\documentclass[useAMS,usenatbib]{mn2e}

\usepackage{graphicx}
\usepackage{amssymb}
\usepackage{amsmath}
\usepackage{lscape}   
\usepackage{threeparttable} 
\usepackage{dcolumn}

\title[Glitches in 36 southern pulsars]{Detection of 107 glitches in
  36 southern pulsars}

\author[Yu et al.]{M. Yu,$^{1,2}$\thanks{E-mail: vela.yumeng@gmail.com}
  R. N. Manchester,$^2$ G. Hobbs,$^2$ S. Johnston,$^2$ V. M. Kaspi,$^{3}$
\newauthor M. Keith,$^{2}$ A. G. Lyne,$^4$
  G. J. Qiao,$^5$ V. Ravi,$^{6,2}$ J. M. Sarkissian,$^2$ R. Shannon,$^{2}$ 
\newauthor and R. X. Xu$^1$ \\ 
$^1$ School of Physics and State Key Laboratory of Nuclear
  Physics and Technology, Peking University, Beijing 100871,
  P. R. China \\ 
$^2$ CSIRO Astronomy and Space Science, Australia
  Telescope National Facility, PO Box 76, Epping, NSW 1710, Australia
  \\ 
$^3$ Department of Physics, Rutherford Physics Building, McGill University, 
3600 University Street, Montreal, Quebec H3A 2T8, Canada \\
$^4$ Jodrell Bank Centre for Astrophysics, The University of
  Manchester, Alan Turing Building, Manchester M13 9PL, UK \\ 
$^5$ Department of Astronomy, School of Physics, Peking University,
  Beijing 100871, P. R. China \\ 
$^6$ School of Physics, University of
  Melbourne, Parkville, VIC 3010, Australia}

\begin{document}
\maketitle
\begin{abstract}
  Timing observations from the Parkes 64-m radio telescope for 165
  pulsars between 1990 and 2011 have been searched for period
  glitches. Data spans for each pulsar ranged between 5.3 years and
  20.8 years. From the total of 1911 years of pulsar rotational
  history, 107 glitches were identified in 36 pulsars. Out of these
  glitches, 61 have previously been reported whereas 46 are new
  discoveries. Glitch parameters, both for the previously known and
  the new glitch detections, were measured by fitting the timing
  residual data. Observed relative glitch sizes $\Delta\nu_{\rm
    g}/\nu$ range between $10^{-10}$ and $10^{-5}$, where $\nu = 1/P$
  is the pulse frequency. We confirm that the distribution of
  $\Delta\nu_{\rm g}/\nu$ is bimodal with peaks at approximately
  $10^{-9}$ and $10^{-6}$. Glitches are mostly observed in pulsars
  with characteristic ages between $10^3$ and $10^5$ years, with large
  glitches mostly occurring in the younger pulsars. Exponential
  post-glitch recoveries were observed for 27 large glitches in 18
  pulsars. The fraction $Q$ of the glitch that recovers exponentially
  also has a bimodal distribution. Large glitches generally have low
  $Q$, typically just a few per cent, but large $Q$ values are
  observed in both large and small glitches. Observed time constants
  for exponential recoveries ranged between 10 and 300 days with some
  tendency for longer timescales in older pulsars. Shorter timescale
  recoveries may exist but were not revealed by our data which
  typically have observation intervals of 2 -- 4 weeks. For most of
  the 36 pulsars with observed glitches, there is a persistent linear
  increase in $\dot\nu$ (i.e., decrease in the slow-down rate
  $|\dot\nu|$) in the inter-glitch interval. Where an exponential
  recovery is also observed, the effects of this are superimposed on
  the linear increase in $\dot\nu$. In some but not all cases, the
  slope of the linear recovery changes at the time of a glitch. The
  $\ddot\nu$ values characterising the linear changes in $\dot\nu$ are
  almost always positive and, after subtracting the magnetospheric
  component of the braking, are approximately proportional to the
  ratio of $|\dot\nu|$ and the inter-glitch interval, as predicted by
  vortex-creep models.
\end{abstract}
\section{Introduction}\label{sect:intro}

Pulsars are thought to be highly-magnetised, rapidly-rotating neutron
stars. They are remarkably stable rotators, which has enabled tests of
general relativity \citep{ksm+06}, searches for gravitational waves
(e.g., \citealt{ych+11}) and the establishment of a pulsar timescale
\citep{hcmc11}.
These results have been being obtained by the technique known as
``pulsar timing''. The pulsar timing technique allows observed pulse
times-of-arrival (ToAs) to be compared with predicted arrival
times. The predicted arrival times are determined using a model of the
pulsar's rotation, position, orbit etc. The differences between the
actual and predicted pulse arrival times are known as ``timing
residuals''. Timing residuals can be induced by an inaccuracy or
omission in the parameters in the timing model or by the timing model
not including all phenomena affecting the propagation of a pulse from
the pulsar to the observer.
The timing residuals for some pulsars are very small. For instance,
PSR J0437$-$4715 has an rms residual of 75\,ns over several years
\citep{mhb+12}. However, most pulsars are not so stable. \citet{hlk10}
analysed the timing residuals of 366 normal and recycled pulsars on
timescales longer than 10\,yr. They found that in most cases the
residuals comprise low-frequency structures. For young pulsars,
the timing residuals were further found to be dominated by recovery
processes from glitch events.

A glitch is an abrupt increase in the pulse frequency $\nu = 1/P$ of a
pulsar, often followed by an exponential recovery toward the
extrapolation of the pre-glitch pulse frequency
\citep{bppr69}. Post-glitch behaviours generally exhibit another
recovery process which is characterised by a linear increase in
$\dot\nu$ or decrease in slow-down rate $|\dot{\nu}|$. This often
extends from the end of the initial exponential recovery until the
next glitch event.  Such ``linear-decay'' processes were first
observed in the Vela pulsar \citep{dow81,lpgc96}, and were
subsequently seen in other sources \citep{ywml10}. The first known
glitch was detected in the Vela pulsar \citep{rm69,rd69}. Since then
more than 350 glitch events have been observed in about 120
pulsars. Glitch databases are now available: the ATNF Pulsar Catalogue
glitch table
\citep{mhth05}\footnote{http://www.atnf.csiro.au/research/pulsar/psrcat/glitchTbl.html}
and the Jodrell Bank Glitch Catalogue
\citep{elsk11}.\footnote{http://www.jb.man.ac.uk/pulsar/glitches.html}
Since the original glitch discovery, the Vela pulsar has been observed
to undergo 15 further glitch events, most of which have
$\Delta\nu_{\rm g}/\nu\sim10^{-6}$. In contrast, the Crab pulsar has
been observed to have $\Delta\nu_{\rm g}/\nu\sim10^{-7}-10^{-9}$ for
most of its glitch events.
Most glitches have been observed in relatively young radio pulsars but
they have also been observed in magnetars \citep{wkt+04,dkgw07} and
even in a millisecond pulsar \citep{cb04}. Observed fractional glitch
sizes range from $\sim10^{-10}$ to $\sim10^{-5}$, but it is important
to note that the low end of this distribution is strongly
limited by observational selection.

The increase of the pulse frequency during a glitch is usually
unresolvable and exponential recoveries typically have timescales of
ten to a few hundred days \citep{wmp+00,ywml10}. However, intensive
observations of glitch events in the Crab and Vela pulsars have shown that
1) the rising edge of the pulse frequency can sometimes be resolved
into multiple components \citep{lsp92} and 2) very short exponential
decays can occur \citep{dml02}.
For two pulsars, sinusoidal oscillations have been observed in timing
residuals after glitch events \citep{mhmk90,ymw+10}.

``Slow glitch'' events have been observed in PSR B1822$-$09 and other
pulsars \citep{zww+04,sha05,sha07,ywml10}. Unlike normal glitches, a
slow glitch builds up over several hundred days, and the increased
pulse frequency is usually maintained until the next event. This
corresponds to a fluctuation in $|\dot{\nu}|$, characterised by an
impulsive decrease followed by a gradual increase. \citet{hlk10} and
\citet{lhk+10} suggested that slow glitches are a manifestation of the
``$\dot\nu$ switching'' observed in some pulsars.

\begin{figure}
\begin{center}
\includegraphics[angle=-90,width=8cm]{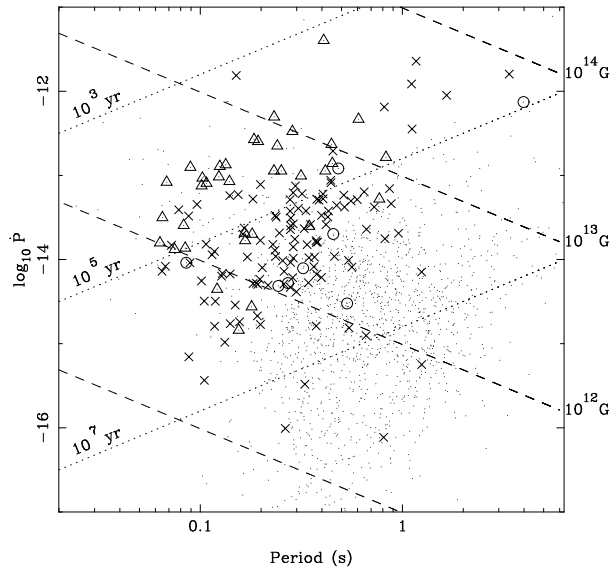} 
\end{center}
\caption{Period--period-derivative ($P$ -- $\dot{P}$) diagram showing
  the pulsars in our sample where no glitch is detected ($\times$), a
  glitch is detected ($\triangle$) and a glitch was detected prior to
  our observations ($\bigcirc$). Other pulsars are marked with a
  dot. Data are from the ATNF Pulsar Catalogue.} \label{fig:ppdot}
\end{figure}

Glitches are thought to be triggered either by the neutron-star
crustquakes (e.g., \citealt{rud91c,rzc98}) or by the sudden transfer
of angular momentum from the faster-rotating crustal neutron
superfluid to the rest of the star \citep[e.g.,][]{ai75,rud76,aaps81}.
The post-event exponential recoveries have been explained as the
re-establishment of an equilibrium between pinning and unpinning in a
vortex-creep region interior to a neutron star
\citep{accp93,lsg00}. Fractional glitch sizes $\Delta\nu_{\rm g}/\nu$
show a bimodal distribution with peaks at $\sim10^{-9}$ and
$\sim10^{-6}$ \citep{lsg00,wmp+00,ywml10}. Using a sample containing
315 glitches, \citet{elsk11} confirmed this bimodal distribution and
also found that the rate of glitch occurrence peaks for pulsars with a
characteristic age ($\tau_{\rm c} \equiv P/(2\dot{P}$) of about
10\,kyr. They also showed that, on average, nearly one per cent of the
spin-down is reversed by glitches for those pulsars with a slow-down
rate $|\dot{\nu}|$ between 10$^{-14}$ to 10$^{-11}$ s$^{-2}$.

\begin{figure}
\begin{center}
\begin{tabular}{c}
\includegraphics[width=0.26\textwidth,angle=-90]{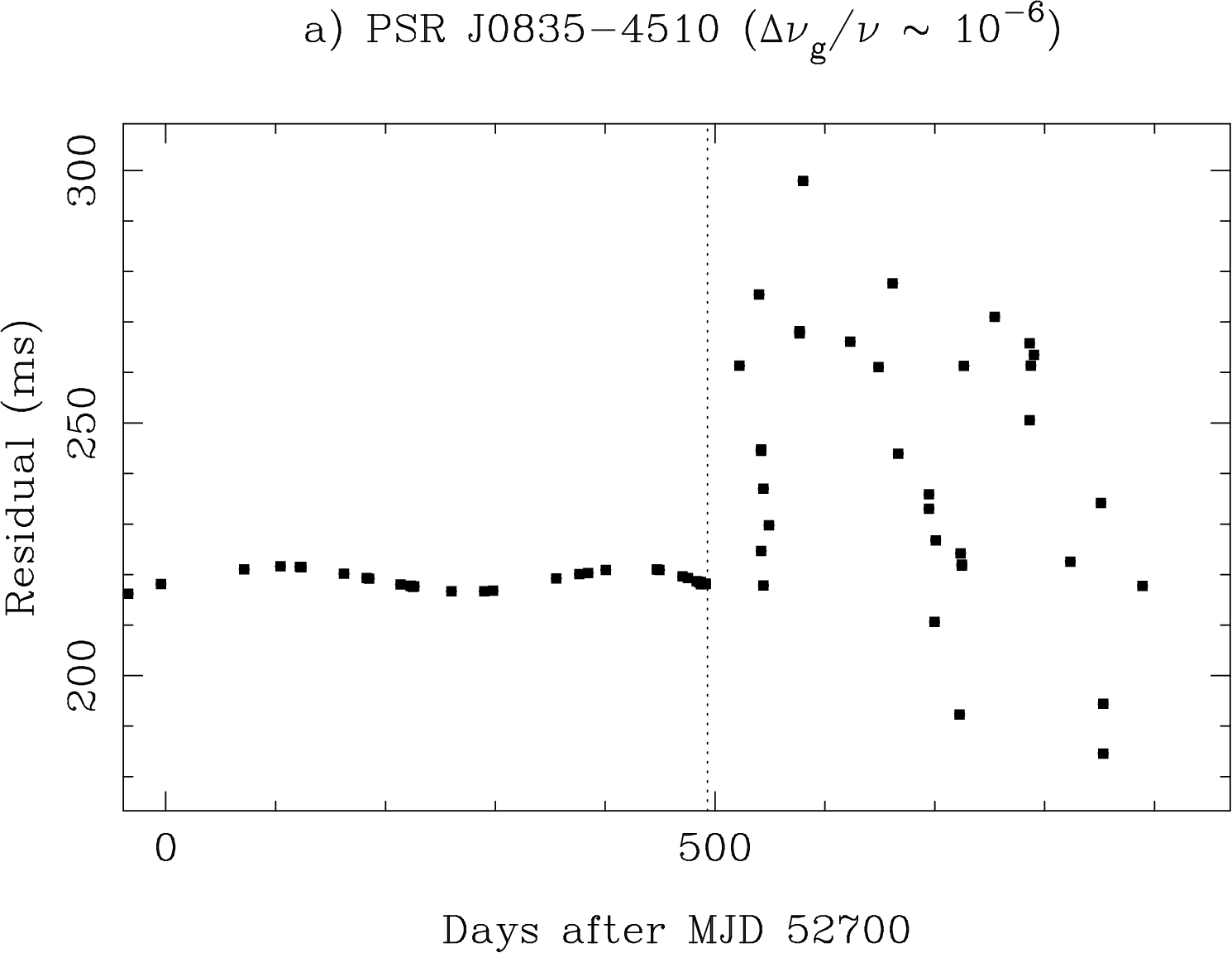}\\
\includegraphics[width=0.26\textwidth,angle=-90]{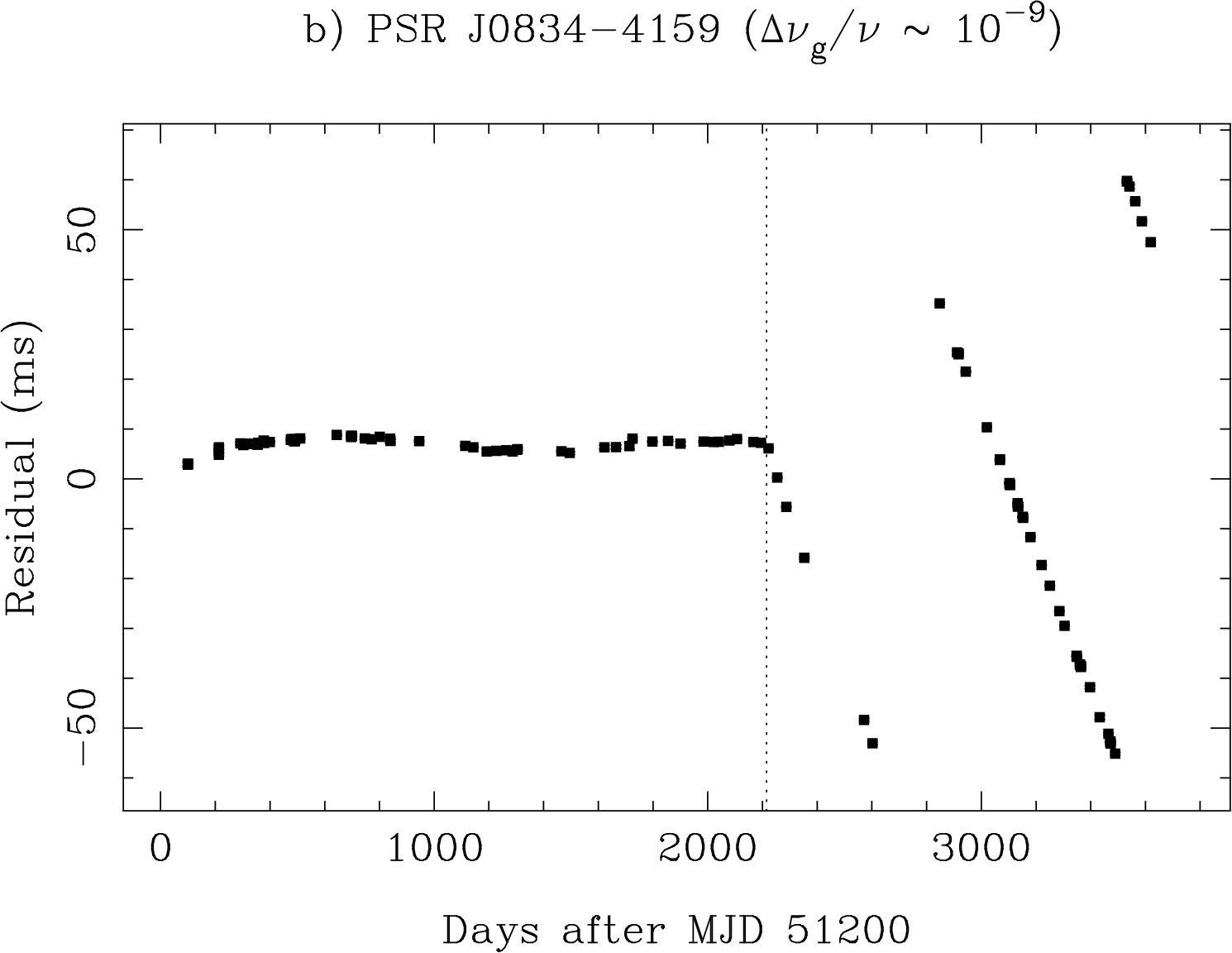}
\end{tabular}
\end{center}
\caption{Illustration of the identification of glitch events with two
  typical sizes $10^{-6}$ and $10^{-9}$. Sub-plot a) shows the effect
  on the timing residuals of the large glitch that occurred in the
  Vela pulsar around MJD 53193.  The pre-glitch solution contains
  $\nu$, $\dot\nu$ and $\ddot\nu$ and phase-coherency is broken by the
  glitch. Sub-plot b) shows timing residuals relative to a pre-glitch
  solution containing $\nu$ and $\dot\nu$ for a small glitch in PSR
  J0834$-$4159 that occurred at MJD $\sim 53415$. In this case, phase
  coherency is maintained after the glitch although there are several
  phase wraps. In each plot the vertical dashed line indicates the
  glitch epoch (from Dodson et al. 2004 for the Vela
  glitch).}\label{fig:illu}
\end{figure}

Even though glitch events and their subsequent recoveries have been
extensively studied, theoretical predictions have been unable to model
fully the timing residuals induced by a glitch event.  Theoretical
models also cannot yet explain why some pulsars exhibit a large number
of glitch events, whereas other pulsars with similar characteristics
have never been observed to glitch. \citet{mpw08} showed that the
waiting-time sequences of the glitches in seven pulsars followed a
constant-rate Poisson process, which suggests that a neutron star
could be a system fluctuating around a self-organised critical state.

For this paper, we searched a total of 1911\,yr of pulsar rotational
history for glitch events. In \S2, we describe our observations. In
\S3, we present our method for determining glitch parameters. Our
results are shown in \S4, and discussed in \S5. We conclude the paper
in \S6.

\section{Observations}\label{sect:obs}

The observations of 165 pulsars analysed in this paper were obtained
using the Parkes 64-m radio telescope between 1990 January and 2011
January. Almost all observations were at radio frequencies near 1400
MHz, in the 20-cm band. For 1990 -- 1994 the ``H-OH'' receiver was
used with an analogue filterbank having $64\times 5$~MHz channels for
each polarisation \citep{jlm+92}. The data were summed and high-pass
filtered before one-bit digitisation. From 1994 -- 2001 most data were
obtained using the ``FPTM'' digital filterbank \citep{sbm+97} with one
(later two) 128 MHz bands for each polarisation. Up until 1997 the
H-OH receiver was used. From 1997 on, most observations used the
centre beam of the 20-cm Multibeam receiver \citep{swb+96} although a
few were at higher frequencies between 1700 and 2200\,MHz using other
receivers. Between 1997 and 2007 observations were usually made with
the analogue filterbank system used for the Parkes Multibeam Pulsar
Survey \citep{mlc+01} which has $96\times3$\,MHz channels for each
polarisation. Intervals between observing sessions were typically 2 --
4 weeks and a ToA was obtained for most of the pulsars each
session. Some pulsars with lower priority were observed less
frequently. Observation times per ToA were normally between one and
ten minutes. 

From 1991 to 2000 these observations were in part used to support the
Energetic Gamma-Ray Experiment Telescope (EGRET) aboard the {\it
  Compton Gamma-Ray Observatory} \citep{tho08}.  Between 2007 and
2011, observations were obtained with the primary goal of supporting
the {\it Fermi Gamma-ray Space Telescope} mission
\citep{sgc+08,wjm+10}. Observing sessions are separated by
approximately four weeks, each lasts for 24\,h, allowing $\sim$170
pulsars to be observed. The centre beam of the Multibeam receiver is
used at 1369\,MHz with a bandwidth of 256\,MHz. Digital filterbank
systems \citep{mhb+12} were used to record the data, with integration
times of 2 to 20\,min for each pulsar to ensure a signal-to-noise
ratio larger than five. A few observations were taken as part of the
PULSE@Parkes project \citep{hhc+09}. These data are available for
download from the Parkes pulsar data
archive\footnote{http://data.csiro.au} \citep{hmm+11}.

In Table~\ref{tab:obs}, we summarise the properties of the 165
pulsars. The pulsar name, pulse period $P$, period derivative
$\dot{P}$, dispersion measure (DM), data span in Modified Julian Day
(MJD) and years, and number of observations are presented. The final
column in this table indicates whether the pulsar has never been
observed to glitch (N), has been detected to glitch during our
observations (Y) or has been reported to glitch prior to our
observations (P). Figure~\ref{fig:ppdot} shows the pulsars in our
sample on the period--period-derivative ($P$ -- $\dot P$) diagram. The
$P$ and $\dot P$ data are from the ATNF Pulsar
Catalogue\footnote{http://www.atnf.csiro.au/research/pulsar/psrcat/;
  catalogue version 1.43} \citep{mhth05}. The identification and
modelling of glitch events will be introduced in detail in the
following section.

\begin{table*}
\caption{Pulsars in our sample observed at the Parkes Observatory with a data span larger than 5 yr.}
\label{tab:obs}
\begin{center}
\begin{threeparttable}
{\tiny
\begin{tabular}{cclrD{.}{.}{4.2}cD{.}{.}{3.2}D{.}{.}{3.0}c}
\hline\\ 
PSR J & PSR B & \multicolumn{1}{c}{$P$} & \multicolumn{1}{c}{$\dot{P}$} & \multicolumn{1}{c}{DM} & Data range & \multicolumn{1}{c}{Data span} & \multicolumn{1}{c}{No. of ToAs} & Glitched? \\
 & & \multicolumn{1}{c}{(s)} & \multicolumn{1}{c}{(10$^{-15}$)} & \multicolumn{1}{c}{(cm$^{-3}$ pc)} & (MJD) & \multicolumn{1}{c}{(yr)} & & \multicolumn{1}{c}{(Y/N/P)$^1$} \\
\hline\\
J0108$-$1431 & - & 0.807565 & 0.08 & 2.38 & 49373 --- 55144 & 15.8 & 135 & N \\
J0401$-$7608 & B0403$-$76 & 0.545253 & 1.54 & 21.60 & 53033 --- 55144 & 5.8 & 58 & N \\
J0536$-$7543 & B0538$-$75 & 1.245856 & 0.56 & 17.50 & 48957 --- 55144 & 16.9 & 103 & N \\
J0630$-$2834 & B0628$-$28 & 1.244419 & 7.12 & 34.47 & 51524 --- 55144 & 9.9 & 62 & N \\
J0729$-$1448 & - & 0.251659 & 113.29 & 92.30 & 51896 --- 55429 & 9.7 & 89 & Y \\\\
J0738$-$4042 & B0736$-$40 & 0.374920 & 1.62 & 160.80 & 52995 --- 55144 & 5.9 & 73 & N \\
J0742$-$2822 & B0740$-$28 & 0.166762 & 16.82 & 73.78 & 49364 --- 55579 & 17.0 & 481 & Y \\
J0834$-$4159 & - & 0.121116 & 4.44 & 240.50 & 51299 --- 55145 & 10.5 & 120 & Y \\
J0835$-$3707 & - & 0.541404 & 9.78 & 112.30 & 50940 --- 53948 & 8.2 & 62 & N \\
J0835$-$4510 & B0833$-$45 & 0.089328 & 125.01 & 67.99 & 49608 --- 55172 & 15.2 & 667 & Y \\\\
J0855$-$4644 & - & 0.064686 & 7.26 & 238.20 & 51158 --- 55144 & 10.9 & 225 & N \\
J0857$-$4424 & - & 0.326774 & 23.34 & 184.43 & 51899 --- 55144 & 8.9 & 94 & N \\
J0901$-$4624 & - & 0.441995 & 87.49 & 198.80 & 50849 --- 55144 & 11.7 & 127 & N \\
J0905$-$5127 & - & 0.346287 & 24.90 & 196.43 & 49363 --- 55145 & 15.8 & 101 & Y \\
J0908$-$4913 & B0906$-$49 & 0.106755 & 15.15 & 180.37 & 48957 --- 55182 & 17.0 & 271 & N \\\\
J0940$-$5428 & - & 0.087545 & 32.87 & 134.50 & 50941 --- 55144 & 11.5 & 163 & N \\
J0942$-$5552 & B0940$-$55 & 0.664367 & 22.85 & 180.20 & 48928 --- 53948 & 13.7 & 181 & N \\
J0954$-$5430 & - & 0.472834 & 43.91 & 200.30 & 50940 --- 55182 & 11.6 & 106 & N \\
J1012$-$5857 & B1011$-$58 & 0.819911 & 17.69 & 383.90 & 50536 --- 53948 & 9.3 & 64 & N \\
J1015$-$5719 & - & 0.139882 & 57.37 & 278.70 & 51215 --- 55182 & 10.9 & 166 & N \\\\
J1016$-$5819 & - & 0.087834 & 0.70 & 252.10 & 50940 --- 55144 & 11.5 & 77 & N \\
J1016$-$5857 & - & 0.107386 & 80.83 & 394.20 & 51299 --- 55429 & 11.3 & 250 & Y \\
J1019$-$5749 & - & 0.162499 & 20.08 & 1039.40 & 51158 --- 55182 & 11.0 & 85 & N \\
J1020$-$6026 & - & 0.140480 & 6.74 & 445.00 & 52854 --- 55182 & 6.4 & 67 & N \\
J1038$-$5831 & B1036$-$58 & 0.661992 & 1.25 & 72.74 & 50536 --- 53948 & 9.3 & 61 & N \\\\
J1043$-$6116 & - & 0.288602 & 10.40 & 449.20 & 51158 --- 55182 & 11.0 & 75 & N \\
J1047$-$6709 & - & 0.198451 & 1.69 & 116.16 & 50538 --- 53948 & 9.3 & 62 & N \\
J1048$-$5832 & B1046$-$58 & 0.123671 & 96.32 & 129.10 & 47910 --- 55183 & 19.9 & 353 & Y \\
J1052$-$5954 & - & 0.180592 & 19.98 & 491.00 & 51411 --- 55460 & 11.1 & 92 & Y \\
J1057$-$5226 & B1055$-$52 & 0.197108 & 5.83 & 30.10 & 49363 --- 55182 & 15.9 & 291 & N \\\\
J1105$-$6107 & - & 0.063193 & 15.83 & 271.01 & 49589 --- 55461 & 16.1 & 297 & Y \\
J1112$-$6103 & - & 0.064962 & 31.46 & 599.10 & 50850 --- 55207 & 11.9 & 178 & Y \\
J1114$-$6100 & B1112$-$60 & 0.880820 & 46.09 & 677.00 & 50538 --- 53948 & 9.3 & 54 & N \\
J1115$-$6052 & - & 0.259777 & 7.23 & 228.20 & 50849 --- 55205 & 11.9 & 97 & N \\
J1119$-$6127 & - & 0.407963 & 4020.22 & 707.40 & 50852 --- 55576 & 12.9 & 348 & Y \\\\
J1123$-$6259 & - & 0.271434 & 5.25 & 223.26 & 50400 --- 55205 & 13.1 & 182 & P \\
J1136$-$5525 & B1133$-$55 & 0.364706 & 8.22 & 85.50 & 51844 --- 53948 & 5.8 & 33 & N \\
J1138$-$6207 & - & 0.117564 & 12.48 & 519.80 & 50849 --- 55205 & 11.9 & 129 & N \\
J1152$-$6012 & - & 0.376570 & 6.68 & 74.00 & 51216 --- 53948 & 7.5 & 54 & N \\
J1156$-$5707 & - & 0.288409 & 26.45 & 243.50 & 51944 --- 55205 & 8.9 & 67 & N \\\\
J1216$-$6223 & - & 0.374047 & 16.82 & 786.60 & 50851 --- 55205 & 11.9 & 61 & N \\
J1224$-$6407 & B1221$-$63 & 0.216476 & 4.95 & 97.47 & 48330 --- 55205 & 18.8 & 661 & N \\
J1248$-$6344 & - & 0.198335 & 16.92 & 433.30 & 51260 --- 55205 & 10.8 & 78 & N \\
J1301$-$6305 & - & 0.184528 & 266.75 & 374.00 & 50941 --- 55104 & 11.4 & 177 & Y \\
J1305$-$6203 & - & 0.427762 & 32.14 & 470.00 & 50940 --- 55205 & 11.7 & 58 & N \\\\
J1316$-$6232 & - & 0.342825 & 5.30 & 983.30 & 49589 --- 53948 & 11.9 & 163 & N \\
J1320$-$5359 & B1317$-$53 & 0.279729 & 9.25 & 97.60 & 50536 --- 55205 & 12.8 & 178 & N \\
J1327$-$6400 & - & 0.280678 & 31.18 & 680.90 & 50940 --- 55205 & 11.7 & 63 & N \\
J1328$-$4357 & B1325$-$43 & 0.532699 & 3.01 & 42.00 & 50738 --- 53948 & 8.8 & 88 & P \\
J1341$-$6220 & B1338$-$62 & 0.193340 & 253.11 & 717.30 & 49540 --- 55461 & 16.2 & 265 & Y \\\\
J1349$-$6130 & - & 0.259363 & 5.12 & 284.60 & 50940 --- 55205 & 11.7 & 71 & N \\
J1359$-$6038 & B1356$-$60 & 0.127501 & 6.34 & 293.71 & 48330 --- 55205 & 18.8 & 661 & N \\
J1412$-$6145 & - & 0.315225 & 98.66 & 514.70 & 50850 --- 55461 & 12.6 & 159 & Y \\
J1413$-$6141 & - & 0.285625 & 333.44 & 677.00 & 50850 --- 55461 & 12.6 & 198 & Y \\
J1420$-$6048 & - & 0.068180 & 83.17 & 358.80 & 51100 --- 55461 & 11.9 & 272 & Y \\\\
J1452$-$5851 & - & 0.386625 & 50.71 & 262.40 & 51088 --- 55205 & 11.3 & 120 & N \\
J1452$-$6036 & - & 0.154991 & 1.45 & 349.70 & 51302 --- 55461 & 11.4 & 93 & Y \\
J1453$-$6413 & B1449$-$64 & 0.179485 & 2.75 & 71.07 & 50669 --- 55205 & 12.4 & 143 & Y \\
J1456$-$6843 & B1451$-$68 & 0.263377 & 0.10 & 8.60 & 48330 --- 55205 & 18.8 & 222 & N \\
J1509$-$5850 & - & 0.088922 & 9.17 & 140.60 & 51214 --- 55205 & 10.9 & 155 & N \\\\
J1512$-$5759 & B1508$-$57 & 0.128694 & 6.85 & 628.70 & 51527 --- 55205 & 10.1 & 85 & N \\
J1513$-$5908 & B1509$-$58 & 0.150658 & 1536.53 & 252.50 & 47913 --- 55205 & 20.0 & 384 & N \\
J1514$-$5925 & - & 0.148796 & 2.88 & 194.10 & 51220 --- 55205 & 10.9 & 66 & N \\
J1515$-$5720 & - & 0.286646 & 6.10 & 482.00 & 51391 --- 55205 & 10.4 & 54 & N \\
J1524$-$5625 & - & 0.078219 & 38.95 & 152.70 & 51214 --- 55205 & 10.9 & 122 & N \\\\
J1524$-$5706 & - & 1.116049 & 356.47 & 833.00 & 51101 --- 55205 & 11.2 & 110 & N \\
J1530$-$5327 & - & 0.278957 & 4.68 & 49.60 & 51013 --- 55205 & 11.5 & 113 & N \\
J1531$-$5610 & - & 0.084202 & 13.74 & 110.90 & 51215 --- 55461 & 11.6 & 161 & Y \\
J1538$-$5551 & - & 0.104675 & 3.21 & 603.00 & 51300 --- 55205 & 10.7 & 81 & N \\
J1539$-$5626 & B1535$-$56 & 0.243392 & 4.85 & 175.88 & 49358 --- 55205 & 16.0 & 261 & P \\\\
J1541$-$5535 & - & 0.295838 & 75.02 & 428.00 & 51300 --- 55205 & 10.7 & 69 & N \\
J1543$-$5459 & - & 0.377119 & 52.02 & 345.70 & 50941 --- 55205 & 11.7 & 113 & N \\
J1548$-$5607 & - & 0.170934 & 10.74 & 315.50 & 50941 --- 55205 & 11.7 & 127 & N \\
J1549$-$4848 & - & 0.288347 & 14.11 & 55.98 & 49358 --- 55205 & 16.0 & 239 & N \\
J1551$-$5310 & - & 0.453394 & 195.13 & 493.00 & 51099 --- 55205 & 11.2 & 152 & N \\\\
J1557$-$4258 & - & 0.329187 & 0.33 & 144.50 & 50538 --- 53948 & 9.3 & 54 & N \\
J1559$-$5545 & B1555$-$55 & 0.957242 & 20.48 & 212.90 & 49359 --- 53948 & 12.6 & 117 & N \\
J1600$-$5044 & B1557$-$50 & 0.192601 & 5.06 & 260.56 & 50618 --- 55205 & 12.6 & 128 & N \\
J1601$-$5335 & - & 0.288457 & 62.37 & 194.60 & 50941 --- 55205 & 11.7 & 118 & N \\
J1602$-$5100 & B1558$-$50 & 0.864227 & 69.58 & 170.93 & 47913 --- 55205 & 20.0 & 246 & N \\\\
J1611$-$5209 & B1607$-$52 & 0.182492 & 5.17 & 127.57 & 51526 --- 55205 & 10.1 & 80 & N \\
J1614$-$5048 & B1610$-$50 & 0.231694 & 494.94 & 582.80 & 47910 --- 55461 & 20.7 & 413 & Y \\
J1623$-$4949 & - & 0.725732 & 42.09 & 183.30 & 50851 --- 53975 & 8.5 & 57 & N \\
J1626$-$4807 & - & 0.293928 & 17.48 & 817.00 & 50941 --- 55205 & 11.7 & 71 & N \\
J1627$-$4706 & - & 0.140746 & 1.73 & 456.10 & 52807 --- 55205 & 6.6 & 84 & N \\\\
\end{tabular}}
\end{threeparttable}
\end{center}
\end{table*}

\addtocounter{table}{-1}

\begin{table*}
\caption{--- {\it continued}}
\begin{center}
\begin{threeparttable}
{\tiny
\begin{tabular}{cclrD{.}{.}{4.2}cD{.}{.}{3.2}D{.}{.}{3.0}c}
\hline\\ 
PSR J & PSR B & \multicolumn{1}{c}{$P$} & \multicolumn{1}{c}{$\dot{P}$} & \multicolumn{1}{c}{DM} & Data range & \multicolumn{1}{c}{Data span} & \multicolumn{1}{c}{No. of ToAs} & Glitched? \\
 & & \multicolumn{1}{c}{(s)} & \multicolumn{1}{c}{(10$^{-15}$)} & \multicolumn{1}{c}{(cm$^{-3}$ pc)} & (MJD) & \multicolumn{1}{c}{(yr)} & & \multicolumn{1}{c}{(Y/N/P)} \\
\hline\\
J1632$-$4757 & - & 0.228564 & 15.07 & 578.00 & 51216 --- 55205 & 10.9 & 92 & N \\
J1632$-$4818 & - & 0.813453 & 650.42 & 758.00 & 50852 --- 55182 & 11.9 & 135 & N \\
J1637$-$4553 & B1634$-$45 & 0.118771 & 3.19 & 193.23 & 50669 --- 55205 & 12.4 & 132 & N \\
J1637$-$4642 & - & 0.154027 & 59.20 & 417.00 & 51393 --- 55205 & 10.4 & 86 & N \\
J1638$-$4417 & - & 0.117802 & 1.61 & 436.00 & 51633 --- 55205 & 9.8 & 82 & N \\\\
J1638$-$4608 & - & 0.278137 & 51.50 & 424.30 & 51089 --- 55205 & 11.3 & 75 & N \\
J1640$-$4715 & B1636$-$47 & 0.517405 & 42.03 & 591.70 & 51528 --- 55205 & 10.1 & 39 & N \\
J1643$-$4505 & - & 0.237383 & 31.83 & 484.00 & 52738 --- 55205 & 6.8 & 44 & N \\
J1644$-$4559 & B1641$-$45 & 0.455060 & 20.09 & 478.80 & 47913 --- 55101 & 19.7 & 298 & P \\
J1646$-$4346 & B1643$-$43 & 0.231603 & 112.75 & 490.40 & 47913 --- 55273 & 20.2 & 305 & Y \\\\
J1648$-$4611 & - & 0.164950 & 23.75 & 392.90 & 51216 --- 55205 & 10.9 & 73 & N \\
J1649$-$4653 & - & 0.557019 & 49.74 & 332.00 & 51089 --- 55205 & 11.3 & 93 & N \\
J1650$-$4502 & - & 0.380870 & 16.06 & 319.70 & 50941 --- 55205 & 11.7 & 82 & N \\
J1650$-$4921 & - & 0.156399 & 1.82 & 229.90 & 52983 --- 55205 & 6.1 & 69 & N \\
J1702$-$4128 & - & 0.182136 & 52.34 & 367.10 & 51089 --- 55205 & 11.3 & 87 & N \\\\
J1702$-$4310 & - & 0.240524 & 223.78 & 377.00 & 51223 --- 55461 & 11.6 & 125 & Y \\
J1705$-$1906 & B1702$-$19 & 0.298987 & 4.14 & 22.91 & 51901 --- 55206 & 9.0 & 75 & N \\
J1705$-$3950 & - & 0.318941 & 60.60 & 207.10 & 51217 --- 55205 & 10.9 & 63 & N \\
J1709$-$4429 & B1706$-$44 & 0.102459 & 92.98 & 75.69 & 47910 --- 55507 & 20.8 & 395 & Y \\
J1713$-$3949 & - & 0.392451 & - & 342.00 & 51557 --- 54504 & 8.1 & 84 & N \\\\
J1715$-$3903 & - & 0.278481 & 37.69 & 313.10 & 51217 --- 55205 & 10.9 & 111 & N \\
J1718$-$3718 & - & 3.378574 & 1613.59 & 371.10 & 51244 --- 54859 & 9.9 & 100 & N \\
J1718$-$3825 & - & 0.074670 & 13.22 & 247.40 & 50878 --- 55507 & 12.7 & 164 & Y \\
J1721$-$3532 & B1718$-$35 & 0.280424 & 25.19 & 496.00 & 51879 --- 55205 & 9.1 & 96 & N \\
J1722$-$3712 & B1719$-$37 & 0.236173 & 10.85 & 99.50 & 49363 --- 55205 & 16.0 & 197 & N \\\\
J1723$-$3659 & - & 0.202722 & 8.01 & 254.20 & 50851 --- 55205 & 11.9 & 102 & N \\
J1726$-$3530 & - & 1.110132 & 1216.75 & 727.00 & 50681 --- 55205 & 12.4 & 147 & N \\
J1730$-$3350 & B1727$-$33 & 0.139460 & 84.83 & 259.00 & 50539 --- 55507 & 13.6 & 182 & Y \\
J1731$-$4744 & B1727$-$47 & 0.829829 & 163.63 & 123.33 & 48184 --- 55507 & 20.0 & 228 & Y \\
J1733$-$3716 & B1730$-$37 & 0.337586 & 15.05 & 153.50 & 51893 --- 55205 & 9.1 & 80 & N \\\\
J1734$-$3333 & - & 1.169008 & 2278.98 & 578.00 & 50686 --- 55205 & 12.4 & 136 & N \\
J1735$-$3258 & - & 0.350963 & 26.08 & 754.00 & 51393 --- 55205 & 10.4 & 64 & N \\
J1737$-$3137 & - & 0.450432 & 138.76 & 488.20 & 51157 --- 55507 & 11.9 & 83 & Y \\
J1737$-$3555 & B1734$-$35 & 0.397585 & 6.12 & 89.41 & 52003 --- 53948 & 5.3 & 21 & N \\
J1738$-$2955 & - & 0.443398 & 81.86 & 223.40 & 51158 --- 55205 & 11.1 & 63 & N \\\\
J1739$-$2903 & B1736$-$29 & 0.322882 & 7.88 & 138.56 & 50739 --- 55205 & 12.2 & 156 & P \\
J1739$-$3023 & - & 0.114368 & 11.40 & 170.00 & 51879 --- 55205 & 9.1 & 98 & N \\
J1740$-$3015 & B1737$-$30 & 0.606887 & 466.12 & 152.15 & 50669 --- 55507 & 13.2 & 190 & Y \\
J1745$-$3040 & B1742$-$30 & 0.367429 & 10.67 & 88.37 & 51901 --- 55205 & 9.0 & 121 & N \\
J1752$-$2806 & B1749$-$28 & 0.562558 & 8.13 & 50.37 & 47911 --- 55083 & 19.6 & 177 & N \\\\
J1756$-$2225 & - & 0.404980 & 52.69 & 326.00 & 51217 --- 54564 & 9.2 & 47 & N \\
J1757$-$2421 & B1754$-$24 & 0.234101 & 12.92 & 179.45 & 51529 --- 55205 & 10.1 & 96 & N \\
J1759$-$2205 & B1756$-$22 & 0.460974 & 10.87 & 177.16 & 51529 --- 53975 & 6.7 & 28 & N \\
J1801$-$2154 & - & 0.375297 & 16.00 & 387.90 & 51218 --- 55205 & 10.9 & 61 & N \\
J1801$-$2304 & B1758$-$23 & 0.415827 & 112.93 & 1073.90 & 47911 --- 55507 & 20.8 & 411 & Y \\\\
J1801$-$2451 & B1757$-$24 & 0.124924 & 127.91 & 289.00 & 48957 --- 55507 & 17.9 & 331 & Y \\
J1803$-$2137 & B1800$-$21 & 0.133667 & 134.36 & 233.99 & 50669 --- 55530 & 13.3 & 182 & Y \\
J1806$-$2125 & - & 0.481789 & 121.40 & 750.40 & 51155 --- 55206 & 11.1 & 74 & P \\
J1809$-$1917 & - & 0.082747 & 25.54 & 197.10 & 50782 --- 55530 & 13.0 & 134 & Y \\
J1812$-$1910 & - & 0.430991 & 37.74 & 892.00 & 51804 --- 55206 & 9.3 & 41 & N \\\\
J1814$-$1744 & - & 3.975905 & 744.70 & 792.00 & 51212 --- 54505 & 9.0 & 50 & P \\
J1815$-$1738 & - & 0.198436 & 77.85 & 728.00 & 51157 --- 55205 & 11.1 & 96 & N \\
J1820$-$1529 & - & 0.333243 & 37.91 & 772.00 & 51244 --- 55206 & 10.8 & 41 & N \\
J1821$-$1419 & - & 1.656010 & 894.50 & 1123.00 & 51410 --- 54505 & 8.5 & 53 & N \\
J1824$-$1945 & B1821$-$19 & 0.189335 & 5.23 & 224.65 & 51844 --- 55206 & 9.2 & 99 & N \\\\
J1825$-$0935 & B1822$-$09 & 0.769006 & 52.50 & 19.38 & 51844 --- 55073 & 8.8 & 85 & Y \\
J1825$-$1446 & B1822$-$14 & 0.279187 & 22.68 & 357.00 & 51844 --- 55205 & 9.2 & 96 & N \\
J1826$-$1334 & B1823$-$13 & 0.101487 & 75.25 & 231.00 & 50749 --- 55530 & 13.1 & 174 & Y \\
J1828$-$1057 & - & 0.246328 & 20.70 & 245.00 & 51805 --- 55206 & 9.3 & 70 & N \\
J1828$-$1101 & - & 0.072052 & 14.81 & 607.40 & 51214 --- 55206 & 10.9 & 42 & N \\\\
J1830$-$1059 & B1828$-$11 & 0.405043 & 60.03 & 161.50 & 51133 --- 55206 & 11.1 & 212 & N \\
J1831$-$0952 & - & 0.067267 & 8.32 & 247.00 & 51301 --- 55206 & 10.7 & 77 & N \\
J1832$-$0827 & B1829$-$08 & 0.647293 & 63.88 & 300.87 & 51844 --- 55206 & 9.2 & 80 & N \\
J1833$-$0827 & B1830$-$08 & 0.085284 & 9.17 & 411.00 & 50748 --- 55206 & 12.2 & 132 & P \\
J1834$-$0731 & - & 0.512980 & 58.20 & 295.00 & 51632 --- 55206 & 9.8 & 63 & N \\\\
J1835$-$0643 & B1832$-$06 & 0.305830 & 40.46 & 472.90 & 51529 --- 55206 & 10.1 & 91 & N \\
J1835$-$1106 & - & 0.165907 & 20.61 & 132.68 & 51945 --- 55530 & 9.8 & 105 & Y \\
J1837$-$0604 & - & 0.096294 & 45.17 & 462.00 & 51089 --- 55206 & 11.3 & 91 & N \\
J1838$-$0549 & - & 0.235303 & 33.43 & 274.00 & 51691 --- 55206 & 9.6 & 49 & N \\
J1839$-$0905 & - & 0.418969 & 26.03 & 348.00 & 51410 --- 55206 & 10.4 & 60 & N \\\\
J1841$-$0524 & - & 0.445749 & 233.72 & 289.00 & 52150 --- 55507 & 9.2 & 134 & Y \\
J1842$-$0905 & - & 0.344643 & 10.49 & 343.30 & 51460 --- 55206 & 10.3 & 48 & N \\
J1843$-$0355 & - & 0.132314 & 1.04 & 797.60 & 51159 --- 55206 & 11.1 & 42 & N \\
J1843$-$0702 & - & 0.191614 & 2.14 & 228.10 & 51692 --- 55206 & 9.6 & 61 & N \\
J1844$-$0256 & - & 0.272963 & - & 820.20 & 51559 --- 55206 & 10.0 & 109 & N \\\\
J1844$-$0538 & B1841$-$05 & 0.255699 & 9.71 & 412.80 & 51844 --- 55206 & 9.2 & 75 & N \\
J1845$-$0743 & - & 0.104695 & 0.37 & 281.00 & 51633 --- 55206 & 9.8 & 50 & N \\
J1847$-$0402 & B1844$-$04 & 0.597769 & 51.71 & 141.98 & 51844 --- 55206 & 9.2 & 75 & N \\
J1853$-$0004 & - & 0.101436 & 5.57 & 438.20 & 51411 --- 55206 & 10.4 & 37 & N \\
J1853$+$0011 & - & 0.397882 & 33.54 & 568.80 & 51148 --- 55183 & 11.0 & 26 & N \\\\
\hline \\
\end{tabular}}
\begin{tablenotes}
\item[$^1$] Y: Glitch detected in this work; N: No glitch detection; P: Previously known glitch before data span.
\end{tablenotes}
\end{threeparttable}
\end{center}
\end{table*}

\section{Data analysis}\label{sect:ana}

Off-line data reduction used the \textsc{psrchive} pulsar data
analysis system \citep{hvm04}. Each observation was summed in time,
frequency and polarisation to form a total intensity pulse profile. In
order to determine the pulse time-of-arrival (ToA), each of the total
intensity profiles was cross-correlated with a high signal-to-noise
ratio ``standard'' profile. Timing residuals were formed using the
pulsar timing software package \textsc{Tempo2} \citep{hem06,ehm06},
with the Jet Propulsion Laboratories (JPL) planetary ephemeris DE405
\citep{sta98b} to correct the local ToAs to the solar-system
barycentre. Each observed ToA was first referred to terrestrial time
as realised by International Atomic Time and subsequently to
Barycentric Coordinate Time.
For each pulsar, \textsc{Tempo2} was used to find a set of parameters
that provided a phase-connected timing solution. The solution contains
the pulse frequency $\nu$ and its first derivative $\dot{\nu}$. The
pulse frequency second derivative was only fitted when a cubic
structure in timing residuals could be seen after fitting for $\nu$
and $\dot{\nu}$. In some cases, particularly for pulsars that had
glitched or have large amounts of timing noise, it was not possible to
obtain a phase-connected timing solution across the entire data
span. In such cases, multiple timing solutions were required.

In order to obtain precise and accurate timing solutions (including
glitch parameters), it is essential to have well-determined pulsar
positions. For some pulsars, positions from the ATNF Pulsar Catalogue
were insufficiently accurate and we therefore determined positions
from our data. Initially, we obtained timing residuals using the
positions (and proper motions) provided by the Catalogue. We fitted
for these parameters using the ``Cholesky'' method that accounts for
the effects of correlated noise \citep{chc+11}. For each fit, we used
the longest data span in which no glitch event was observed. The
resulting positions were held fixed in subsequent processing.

Glitch events are recognised by a sudden discontinuity in the timing
residuals relative to a solution based on earlier data. For glitches
with a fractional size $\Delta\nu_{\rm g}/\nu$ larger than
$\sim10^{-6}$, the residuals change by a large fraction of the pulse
period in a few days and phase connection is normally lost as
illustrated in sub-plot a) of Figure \ref{fig:illu}. By analysing
short sections of post-glitch data, phase coherence can normally be
recovered and an approximate value for the frequency glitch
$\Delta\nu_{\rm g}$ determined. As shown in sub-plot b) of Figure
\ref{fig:illu}, glitches with smaller fractional sizes, typically
$\sim10^{-9}$, have no loss of phase coherence over several hundred
days and the post-glitch pulse frequency is easy to determine. Even
smaller glitches, with $\Delta\nu_{\rm g}/\nu \sim10^{-10}$, are often
hard to distinguish from irregular timing noise and so the observed
sample of these is incomplete.
For each pulsar data set exhibiting a possible glitch, we used the
\textsc{glitch} plug-in of \textsc{Tempo2} to determine the variations
of the pulse frequency and its first time derivative as a function of
time. The \textsc{glitch} plug-in realises this by carrying out a
sequence of local fits for these two parameters to the timing
residuals. Typically, we included five or six observations in each fit
(spanning from about two to six months). After completing the sequence
of local fits, a list of dates, pulse frequencies and pulse frequency
derivatives are obtained for each glitching pulsar.

In \textsc{Tempo2}, the additional pulse phase induced by a glitch is
described by Equation (121) in \citet{ehm06}\footnote{Corrigendum for
  Equation (121) in \citet{ehm06}: the fourth term of the right hand
  side of the equation should be $[1-e^{-(t_{\rm e}^{\rm psr}-t_{\rm
      g})/\tau}]\Delta\nu_t\;\tau$, rather than $[1-e^{-(t_{\rm e}^{\rm
      psr}-t_{\rm g})/\tau}]\Delta\nu_t(t_{\rm e}^{\rm psr}-t_{\rm
    g})$.}:
\begin{eqnarray}
\label{eq:gltmod}
\phi_{\rm g}&=&\Delta\phi+\Delta\nu_{\rm p}(t-t_{\rm
  g})+\frac{1}{2}\Delta\dot{\nu}_{\rm p}(t-t_{\rm g})^2+\\ \nonumber &
&[1-e^{-(t-t_{\rm g})/\tau_{\rm d}}]\Delta\nu_{\rm d}\tau_{\rm d}
\end{eqnarray}
where the glitch event is modelled by an offset in pulse
phase $\Delta\phi$ and the permanent increments in the pulse frequency
$\Delta\nu_{\rm p}$ and first frequency derivative
$\Delta\dot{\nu}_{\rm p}$, in addition to a transient frequency
increment $\Delta\nu_{\rm d}$ which decays exponentially to zero with
a timescale $\tau_{\rm d}$. 
The phase offset $\Delta\phi$ is needed to allow for uncertainty in
the glitch epoch $t_{\rm g}$. An initial estimate of $t_{\rm g}$ was
taken to be halfway between the last pre-glitch observation and the
first post-glitch observation. Initial estimates of $\Delta\nu_{\rm
  p}$ and $\Delta\dot{\nu}_{\rm p}$ were given by the \textsc{glitch}
plug-in. Improved values were obtained by including the glitch model
in the timing model and subsequently using \textsc{Tempo2} to fit for
the glitch parameters.
For our work, we extended the Taylor series in equation
(\ref{eq:gltmod}) to include $\Delta\ddot{\nu}_{\rm p}$ to
characterise the long-term variations in $\ddot{\nu}$. These
parameters and their corresponding uncertainties were obtained from a
\textsc{Tempo2} least-squares-fit to a segment of data typically
spanning $\sim$200\,d to $\sim$3000\,d across the glitch event, with
the glitch epoch around the centre of the data range. The long-term
variations in pulse frequency were described by a truncated Taylor
series, $\phi(t)=\phi_0+\nu t
+ \frac{1}{2}\dot{\nu}t^2+\frac{1}{6}\ddot{\nu}t^3$. Fits including
$\tau_{\rm d}$ are more complicated. As \textsc{Tempo2} 
implements only a linear fitting algorithm, it is necessary to have a good
initial estimate for $\tau_{\rm d}$. The estimate can be realised by
two steps. In the first step, an estimate for $\tau_{\rm d}$ was
obtained by eye by inspecting the post-glitch $\dot{\nu}$
variations. In the second step, the first-step value was introduced
into the fitting. By increasing or decreasing $\tau_{\rm d}$, one can
eventually find a $\tau_{\rm d}$ which minimises the post-fit
$\chi^2$. This $\tau_{\rm d}$ was determined as the estimate and
subsequently included as part of the \textsc{Tempo2} fit. We note
that, when a fit included an exponential recovery, the post-glitch
data range was selected to be larger than the recovery timescale,
$\tau_{\rm d}$.
The changes in the pulse frequency and its first derivative at the
glitch are then described as
\begin{equation}
\Delta\nu_{\rm g}=\Delta\nu_{\rm p}+\Delta\nu_{\rm d} \label{eq:dnug}
\end{equation} 
and
\begin{equation}
\Delta\dot{\nu}_{\rm g}=\Delta\dot{\nu}_{\rm p}-\frac{\Delta\nu_{\rm
    d}}{\tau_{\rm d}}, \label{eq:dnudotg}
\end{equation}
with their uncertainties obtained using standard error propagation
equations. In addition, a factor $Q\equiv\Delta\nu_{\rm
  d}/\Delta\nu_{\rm g}$ can be defined, describing the fraction of
glitch recovery.
In a few cases, after following this procedure the timing residuals
revealed a shorter-timescale exponential recovery. In these cases, the
second exponential was fitted, initially holding the parameters of the
first recovery fixed, and then finally fitting for all parameters
of both recoveries. 

For some glitches, the glitch epoch could be determined by requiring
that the pulse phase was continuous over the glitch, i.e., that
$\Delta\phi =0$. However, a unique solution is only possible when both
the amplitude of the glitch and the interval between the last
pre-glitch observation and the first post-glitch observation are
small, such that $\Delta\phi$ is less than one period between the
bounding observations. For situations in which this was not possible,
we checked the literature to determine whether a precise glitch epoch
had already been published. If so, then we used the published epoch
for the rest of the analysis. If not, the glitch epoch $t_{\rm g}$ was
kept at halfway between the last pre-glitch observation and the first
post-glitch observation, with an uncertainty of half the observation
gap. To take account of this uncertainty for the glitch parameters, we
assume a linear dependence on the epoch for each of the glitch
parameters. The fitting routine was carried out again with $t_{\rm g}$
close to the epoch of the first post-glitch observation. A difference
between the original and the new values for each parameter could then
be obtained. The final uncertainty was then the quadrature sum of the
parameter difference and its original uncertainty. For glitches that
have a large epoch uncertainty and/or large exponential recoveries,
the epoch uncertainty term generally dominates the final parameter
uncertainties.

Slow glitches are difficult to recognise from timing residuals alone
and are best identified in plots of $\dot\nu$ versus time. Their
identification is somewhat subjective and they cannot be fitted with
standard glitch analyses. In this paper (in \S\ref{sect:J1539}) we
describe slow glitches detected in one pulsar, PSR J1539$-$5626.

\section{Results}\label{sect:res}

The data sets for the 165 pulsars in the sample were processed, and 36
pulsars were observed to have glitched (indicated with a ``Y'' in the
last column in Table~\ref{tab:obs}). A total of 107 glitches were
detected, among which 46 are new detections. We identified exponential
recoveries for 27 glitches. A total of 22 previously published
glitches are within our data span, but we were unable to identify
these events. This is mainly because the sampling of our observations
is often insufficient, such that glitches with a small fractional size
($\Delta\nu_{\rm g}/\nu < 10^{-9}$) are hard to detect. For the
same reason, only those exponential recoveries with a timescale
between a few tens to a few hundred days are detectable; any
exponential recoveries with a timescale shorter than a few tens of
days are likely to have been missed.

\begin{table*}
\caption{Position and proper motion parameters for 36 glitching pulsars.}\label{tab:pos}
\begin{threeparttable}
\begin{center}
\begin{tabular}{cllcccc}
\hline\\
PSR J & \multicolumn{1}{c}{R. A.} & \multicolumn{1}{c}{Dec.} & \multicolumn{1}{c}{Position epoch} & $\mu_\alpha$ & $\mu_\delta$ & References \\
 & \multicolumn{1}{c}{(h:m:s)} & \multicolumn{1}{c}{($^{\circ}$ $'$ $''$)} & \multicolumn{1}{c}{(MJD)} & (mas yr$^{-1}$) & (mas yr$^{-1}$) & \\
\hline\\ 
J0729$-$1448 & 07:29:16.45(2) & $-$14:48:36.8(8) & 51367 & - & - & 1 \\
J0742$-$2822 & 07:42:49.058(2) & $-$28:22:43.76(4) & 49326 & $-$29(2) & 4(2) & 2,3 \\
J0834$-$4159 & 08:34:17.815(8) & $-$41:59:36.01(9) & 52347 & - & - & This work \\
J0835$-$4510 & 08:35:20.61149(2) & $-$45:10:34.8751(3) & 51544 & $-$49.68(6) & 29.9(1) & 4 \\
J0905$-$5127 & 09:05:51.94(5) & $-$51:27:54.0(4) & 54072 & - & - & This work \\\\
J1016$-$5857 & 10:16:21.16(1) & $-$58:57:12.1(1) & 52717 & - & - & 5 \\
J1048$-$5832 & 10:48:12.2(1) & $-$58:32:05.8(8) & 50889 & - & - & 6 \\
J1052$-$5954 & 10:52:38.11(7) & $-$59:54:44.1(5) & 51683 & - & - & 7 \\
J1105$-$6107 & 11:05:26.17(4) & $-$61:07:51.4(3) & 50794 & - & - & 6 \\
J1112$-$6103 & 11:12:14.81(4) & $-$61:03:31.1(6) & 51055 & - & - & 8 \\\\
J1119$-$6127 & 11:19:14.30(2) & $-$61:27:49.5(2) & 51485 & - & - & 9 \\
J1301$-$6305 & 13:01:45.76(14) & $-$63:05:33.9(12) & 51206 & - & - & 8 \\
J1341$-$6220 & 13:41:42.63(8) & $-$62:20:20.7(5) & 50859 & - & - & 6 \\
J1412$-$6145 & 14:12:07.69(5) & $-$61:45:28.8(6) & 51186 & - & - & 8 \\
J1413$-$6141 & 14:13:09.87(9) & $-$61:41:13(1) & 51500 & - & - & 7 \\\\
J1420$-$6048 & 14:20:08.237(16) & $-$60:48:16.43(15) & 51600 & - & - & 10 \\
J1452$-$6036 & 14:52:51.898(8) & $-$60:36:31.35(6) & 51630 & - & - & 7 \\
J1453$-$6413 & 14:53:32.684(8) & $-$64:13:15.81(7) & 52608 & $-$16(1) & $-$21.3(8) & This work \\
J1531$-$5610 & 15:31:27.91(1) & $-$56:10:55.0(1) & 51448 & - & - & 7 \\
J1614$-$5048 & 16:14:11.29(3) & $-$50:48:03.5(5) & 50853 & - & - & 6 \\\\
J1646$-$4346 & 16:46:50.8(3) & $-$43:45:48(8) & 52792 & - & - & This work \\
J1702$-$4310 & 17:02:26.94(5) & $-$43:10:40(2) & 51597 & - & - & 7 \\
J1709$-$4429 & 17:09:42.728(2) & $-$44:29:08.24(6) & 50042 & - & - & 6 \\
J1718$-$3825 & 17:18:13.565(4) & $-$38:25:18.06(15) & 51184 & - & - & 8 \\
J1730$-$3350 & 17:30:32.28(6) & $-$33:50:28(4) & 53826 & - & - & This work \\\\
J1731$-$4744 & 17:31:42.17(7) & $-$47:44:37(2) & 54548 & - & - & This work \\
J1737$-$3137 & 17:37:04.29(4) & $-$31:37:21(3) & 51234 & - & - & 1 \\
J1740$-$3015 & 17:40:33.82(1) & $-$30:15:43.5(2) & 52200 & - & - & 3 \\
J1801$-$2304 & 18:01:19.829(9) & $-$23:04:44.2(2) & 50809 & - & - & 11 \\
J1801$-$2451 & 18:01:00.016(8) & $-$24:51:27.5(2) & 53348 & $-$11(9) & $-$1(15) & 12 \\\\
J1803$-$2137 & 18:03:51.4105(10) & $-$21:37:07.351(10) & 51544 & 11.6(18) & 14.8(23) & 13 \\
J1809$-$1917 & 18:09:43.132(6) & $-$19:17:40(1) & 54632 & - & - & This work \\
J1825$-$0935 & 18:25:30.629(6) & $-$09:35:22.3(3) & 53300 & $-$13(11) & $-$9(5) & 14,3 \\
J1826$-$1334 & 18:26:13.175(3) & $-$13:34:46.8(1) & 52400 & 23.0(25) & $-$3.9(31) & 14,15 \\
J1835$-$1106 & 18:35:18.41(7) & $-$11:06:15(4) & 53882 & - & - & This work \\\\
J1841$-$0524 & 18:41:49.32(5) & $-$05:24:29.5(12) & 52360 & - & - & 5 \\
\hline\\
\end{tabular}
\begin{tablenotes}
\item[] References for positions and proper motions: 1 -- \citet{mhl+02}; 2 -- \citet{hlk+04}; 3 -- \citet{fgml97}; 4 -- \citet{dlrm03}; 5 -- \citet{hfs+04}; 6 -- \citet{wmp+00}; 7 -- \citet{kbm+03}; 8 -- \citet{mlc+01}; 9 -- \citet{ckl+00}; 10 -- \citet{dkm+01}; 11 -- \citet{fkv93}; 12 -- \citet{zbcg08}; 13 -- \citet{bckf06}; 14 -- \citet{ywml10}; 15 -- \citet{pkb08}.
\end{tablenotes}
\end{center}
\end{threeparttable}
\end{table*}

Table~\ref{tab:pos} gives the positions and proper motions in J2000
coordinate for each glitching pulsar. The positions for 28 pulsars are
from the ATNF Pulsar Catalogue. As described in \S3, we fit for the
positions for a further eight pulsars. All of the proper motions are
from the ATNF Pulsar Catalogue.

Table~\ref{tab:gltspin} lists the pre-, inter- and post-glitch timing
solutions for the glitching pulsars. For each pulsar, the table
contains the pulsar name, the interval relative to glitch number, $\nu$,
$\dot{\nu}$, $\ddot{\nu}$, reference epoch, fitted data span, number
of ToAs, post-fit rms residuals, reduced $\chi^2$ and the number of
degrees of freedom for the least-squares fit. For pulsars with
exponential post-glitch recoveries, to avoid contaminating the
long-term post-glitch parameters, the start of post-glitch data span
is at least two decay timescales from the glitch.  These solutions
include long-term timing noise and so are only valid within the fitted
data range; they cannot be used for extrapolation.

Table~\ref{tab:glt} contains the parameters for each observed
glitch. The second column gives a reference number for each glitch and
the glitch epochs are given in the third column. The fourth column
indicates whether the glitch is new (N) or has been previously
published (P). References for previously published glitches may be
found in the web databases. For each glitch parameter, we give two
uncertainties. The \textsc{Tempo2} 1\,$\sigma$ uncertainties are given
in the first pair of parentheses. If inclusion of effect of the glitch
epoch uncertainty made a significant difference, the final uncertainty
is given in the second pair of parentheses. Note that errors refer to
the last digit quoted. The number of observations, the fitted data
span, the post-fit rms residuals and the reduced $\chi^2$ and degrees
of freedom are listed in columns 11, 12, 13 and 14, respectively.

\begin{figure*}
\begin{tabular}{cc}
\includegraphics[width=5.5cm,angle=-90]{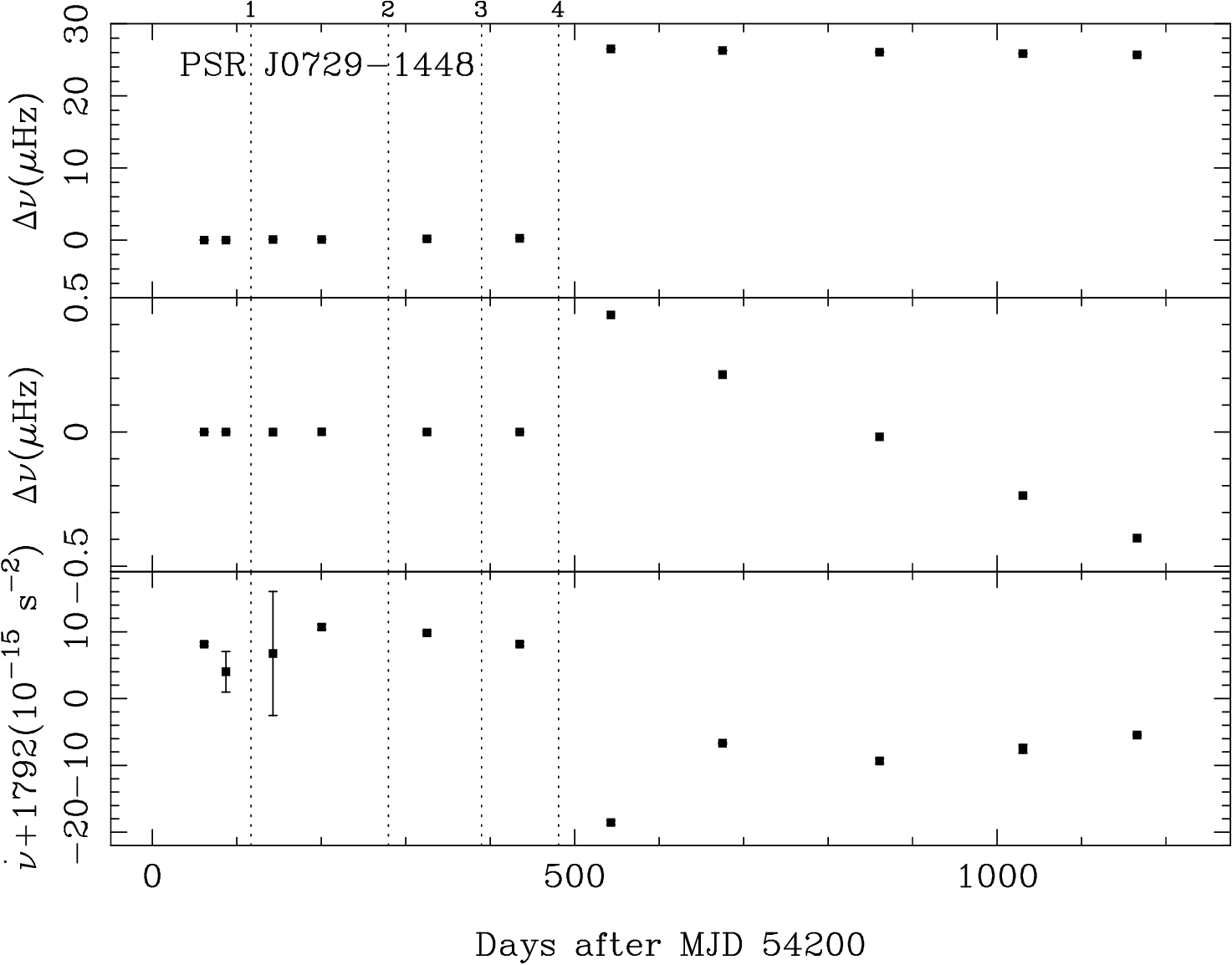} &
\includegraphics[width=5.5cm,angle=-90]{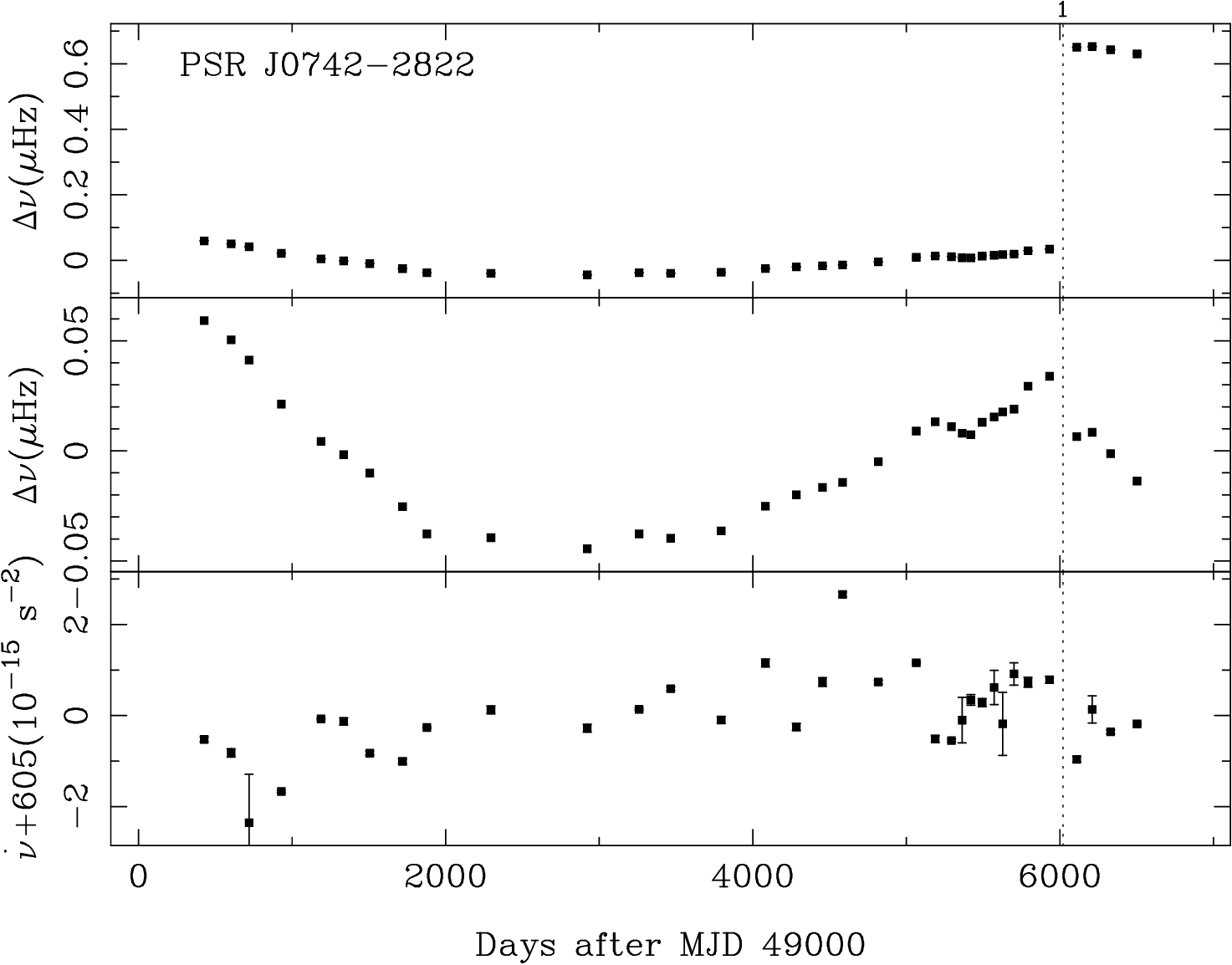} \\ 
\includegraphics[width=5.5cm,angle=-90]{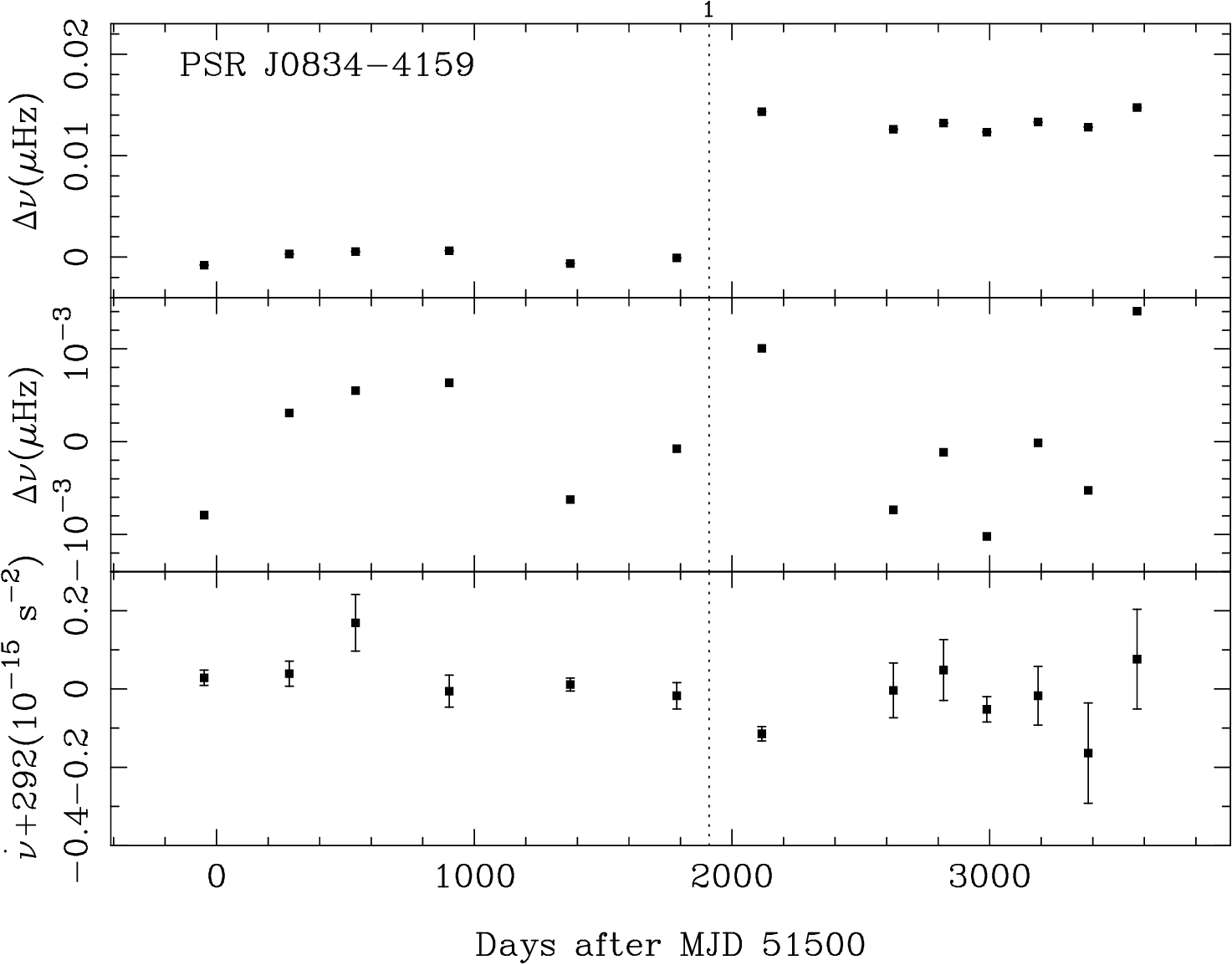} & 
\includegraphics[width=5.5cm,angle=-90]{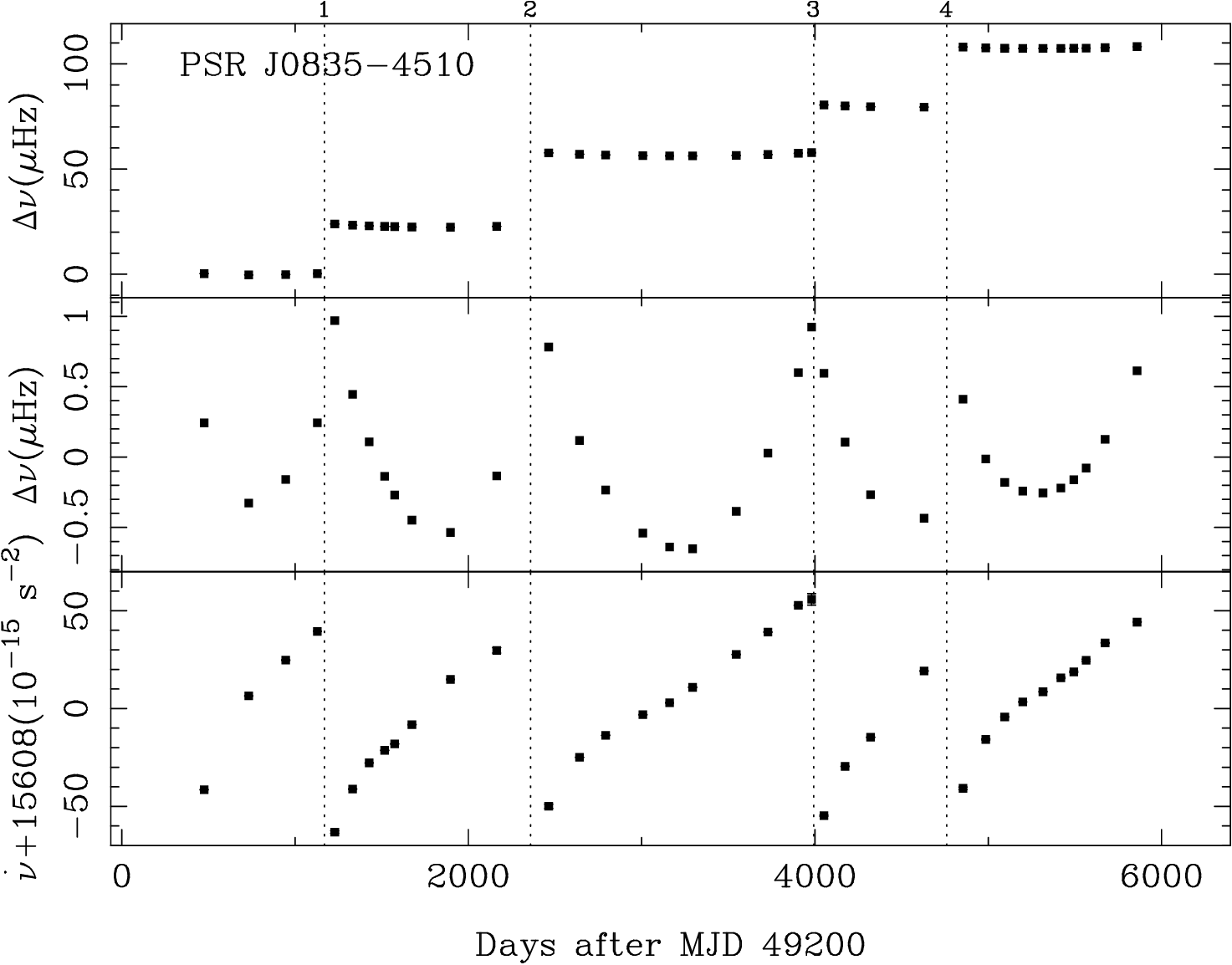} \\
\end{tabular} 
\caption{Glitches in PSRs J0729$-$1448, J0742$-$2822, J0834$-$4159 and
  J0835$-$4510. The top panel shows the
  pulse-frequency residuals $\Delta\nu$, obtained by subtracting the
  (extrapolated) pulse frequency derived from the $\nu$ and $\dot\nu$
  values of the first pre-glitch solution; the middle panel is an
  expanded plot of $\Delta\nu$ where the mean of each inter-glitch (or
  post-glitch) solution is subtracted; the bottom panel shows the
  variations of the pulse-frequency first time derivative
  $\dot{\nu}$. The glitch epochs are indicated by vertical dashed
  lines, and the numbers at the top of each dashed line denote the
  sequence of glitches detected within our data
  span.}\label{fig:glt1}
\end{figure*}

In each of the sub-sections below, we describe the observed glitch
events for each pulsar in more detail. In Figures \ref{fig:glt1} to
\ref{fig:glt9}, for the 36 glitching pulsars, we show the evolution of
pulse frequency and its first time derivative within our data
span. For convenience, observed glitches are numbered as in Table
\ref{tab:glt}.

\subsection{PSR J0729$-$1448}\label{sect:J0729-1448}

A data gap lasting for $\sim 6$\,yr exists in the data set of this
pulsar. During our data span of the recent three years, this pulsar
exhibited four glitches (see Figure~\ref{fig:glt1}). The first three
glitches were small ($\Delta\nu_{\rm g}/\nu\sim10^{-8}$). The fourth
glitch was significantly larger ($\Delta\nu_{\rm
  g}/\nu\sim6\times10^{-6}$). These glitch events have been reported
by \citet{wjm+10} and \citet{elsk11}. The Parkes data are
unfortunately not well sampled. The three small glitches were
identified with prior knowledge from \citet{elsk11}. For the same
reason, it is impossible to evaluate the permanent change in
$\dot{\nu}$ and other long-term parameters. \citet{wjm+10} reported
the large glitch that occurred at MJD 54711(21). Our analysis provides
a more precise epoch of MJD 54681(9) which is consistent with
\citet{elsk11} result of MJD 54687(3).

\subsection{PSR J0742$-$2822 (PSR B0740$-$28)}\label{sect:J0742-2822}

In total, seven glitch events have been reported for this pulsar
\citep{dmk+93,js06,elsk11}. In Figure~\ref{fig:glt1}, we present our
17-yr data span. No new glitches were detected. A glitch at MJD $\sim
55020$ can clearly be seen. However, we were unable to detect the four
small previously reported glitches covered by the data set. For the
observed glitch, \citet{elsk11} gave
$\Delta\dot{\nu}/\dot{\nu}=-0.372(96)$, corresponding to
$\Delta\dot{\nu}=225(58)\times10^{-15}$\,s$^{-2}$. Our measurement of
$\dot{\nu}$ presented in Figure~\ref{fig:glt1} and Table~\ref{tab:glt}
shows $\Delta\dot{\nu}=-1.3(3)\times10^{-15}$\,s$^{-2}$, despite the
evident noise.

\subsection{PSR J0834$-$4159}\label{sect:J0834-4159}

This pulsar was not previously known to glitch, but we identify a
small glitch at MJD $\sim 53415$. Figure~\ref{fig:glt1} shows the
10-yr evolution of pulse frequency and pulse-frequency derivative of
this source observed at Parkes. Both of the measured $\Delta\nu$ and
$\dot{\nu}$ exhibit noise. $\dot{\nu}$ shows a small permanent change
at the glitch event (Table~\ref{tab:glt}). Our observations do not
reveal any post-glitch relaxation process.

\subsection{PSR J0835$-$4510 (PSR B0833$-$45)}\label{sect:J0835-4510}

The Vela pulsar has undergone 16 known glitch events over a period of
$\sim38$\,yr (for a complete list of these glitches, see the ATNF
Pulsar Catalogue glitch table or the Jodrell Bank Glitch
Catalogue). Thirteen have a fractional glitch size larger than
$10^{-6}$. In Figure~\ref{fig:glt1}, we present the variations of
pulse frequency and its first derivative spanning the last
$\sim15$\,yr. Four glitches were detected. These events have been
reported and analysed by \citet{fla96,wmp+00,dmc00,dml02,dbr+04} and
\citet{fb06}.

As Figure~\ref{fig:glt1} shows, these glitches are large, with
$\Delta\nu_{\rm g}/\nu>2\times10^{-6}$. Each of the post-glitch
behaviours exhibits both exponential and linear recoveries. We
attempted to model each of the glitches including both of the
exponential and linear recoveries. Each of the glitches is discussed
in more detail as below.

For glitch 1, \citet{wmp+00} reported an exponential recovery with a
time constant 916(48)\,d. However, the post-event $\dot{\nu}$
variations shown in Figure~\ref{fig:glt1} indicate that the
exponential recovery completes within $\sim200$\,d and the long-term
evolution exhibits a linear recovery. Fitting the timing phase
residuals showed that the exponential timescale is 186(12)\,d, with
$Q=0.030(4)$ (Table~\ref{tab:glt}). Glitch 2 was captured with high
time resolution by \citet{dml02}. Four short-term exponential decays
were identified with the smallest timescale just
$\sim1.2$\,min. Parkes data, however, are not sufficient to resolve
these short-term recoveries. However an exponential recovery that
completes in $\sim100$\,d can be seen in our data. Fitting gave $Q
\sim 0.02$ and $\tau_{\rm d}\sim 125$\,d. For glitch 3, the
exponential recovery is characterised by $Q\sim 0.009$ and $\tau_{\rm
  d}\sim 37$\,d. The most recent glitch 4 also exhibits an exponential
recovery. Our fitting showed $Q\sim 0.0119$ and $\tau_{\rm
  d}\sim 73$\,d.

At least in the long term, the post-glitch behaviour is dominated by
the linear recovery of $\dot\nu$. This is superimposed on the
shorter-term exponential decays and persists until the next glitch. As
Table~\ref{tab:gltspin} shows, the observed values of $\ddot\nu$
representing the slope of this long-term linear recovery are
relatively large for the Vela pulsar and also that they change
signficantly after each glitch. Fitting for $\Delta\ddot{\nu}_{\rm p}$
along with the other glitch parameters was generally difficult. The
superimposed timing noise and the shorter data spans used for the
glitch fitting often led to values somewhat different to those
obtained by differencing the long-term fits for $\ddot\nu$ given in
Table~\ref{tab:gltspin} or to an insignficant value. For example,
glitch 3 has a fitted value of $304(23)\times10^{-24}$\,s$^{-3}$ but
the difference between the post- and pre-glitch values of $\ddot\nu$
in Table~\ref{tab:gltspin} is $491(7)\times10^{-24}$\,s$^{-3}$.

\begin{figure*} 
\begin{tabular}{cc}
\includegraphics[width=5.5cm,angle=-90]{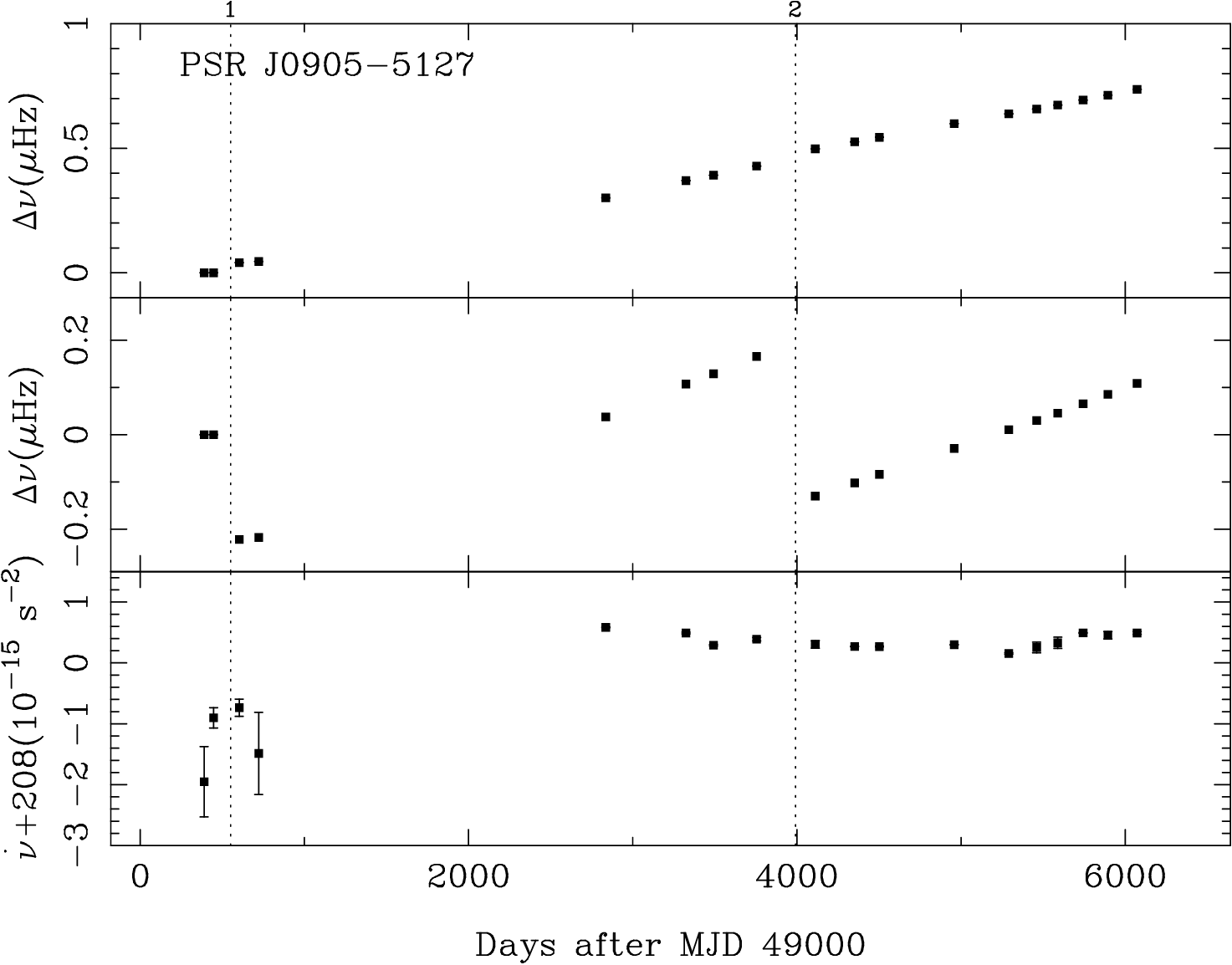} &
\includegraphics[width=5.5cm,angle=-90]{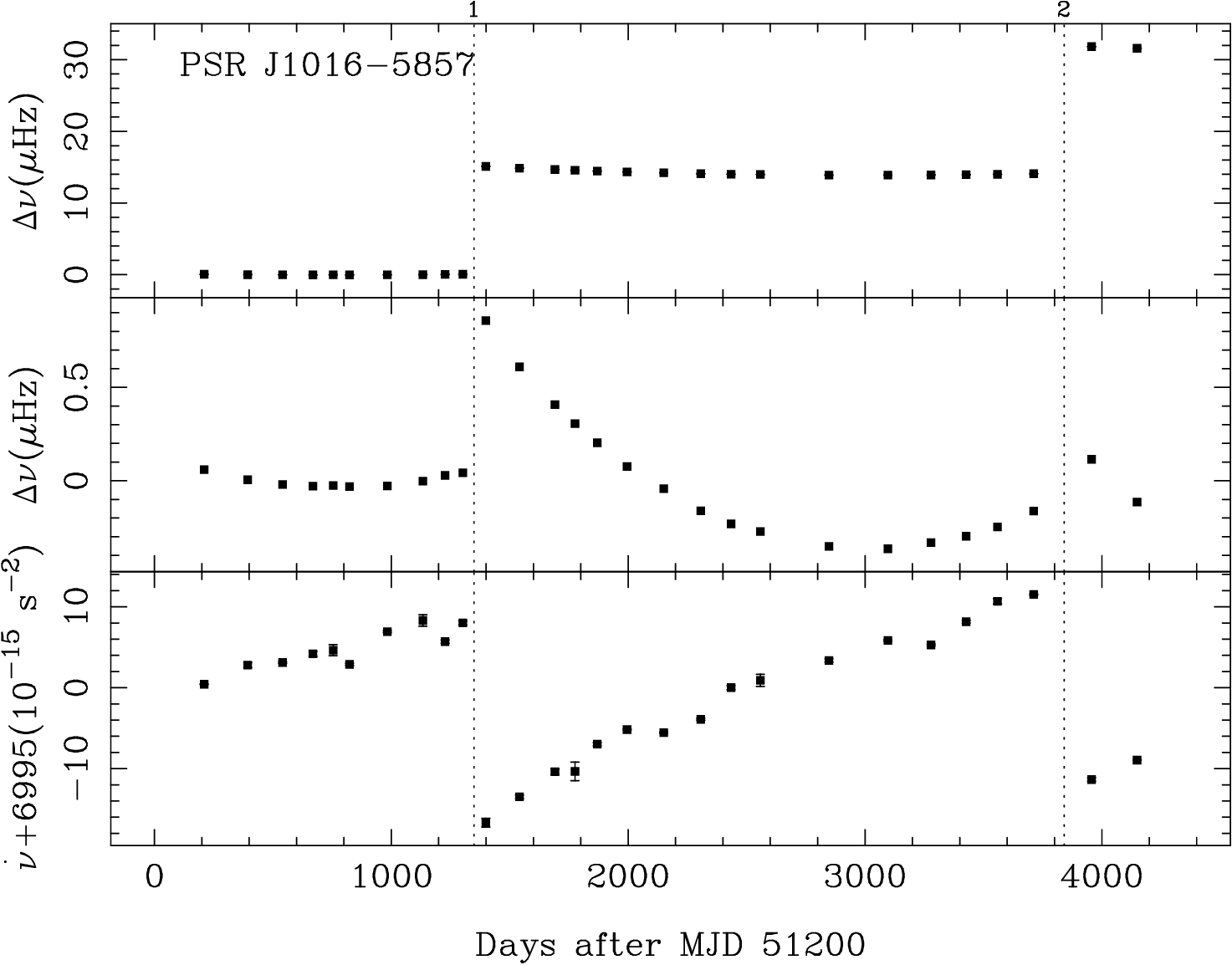} \\ 
\includegraphics[width=5.5cm,angle=-90]{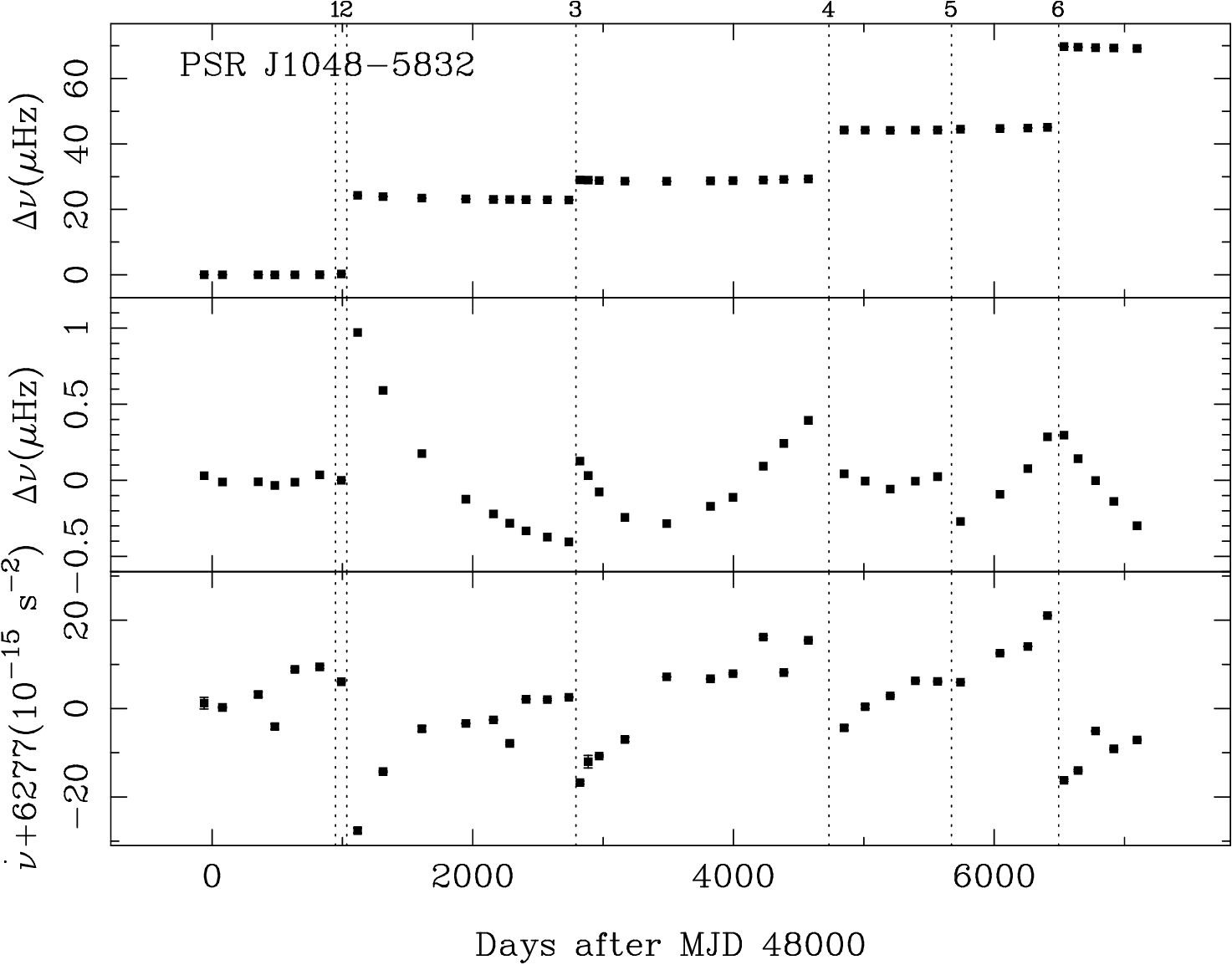} & 
\includegraphics[width=5.5cm,angle=-90]{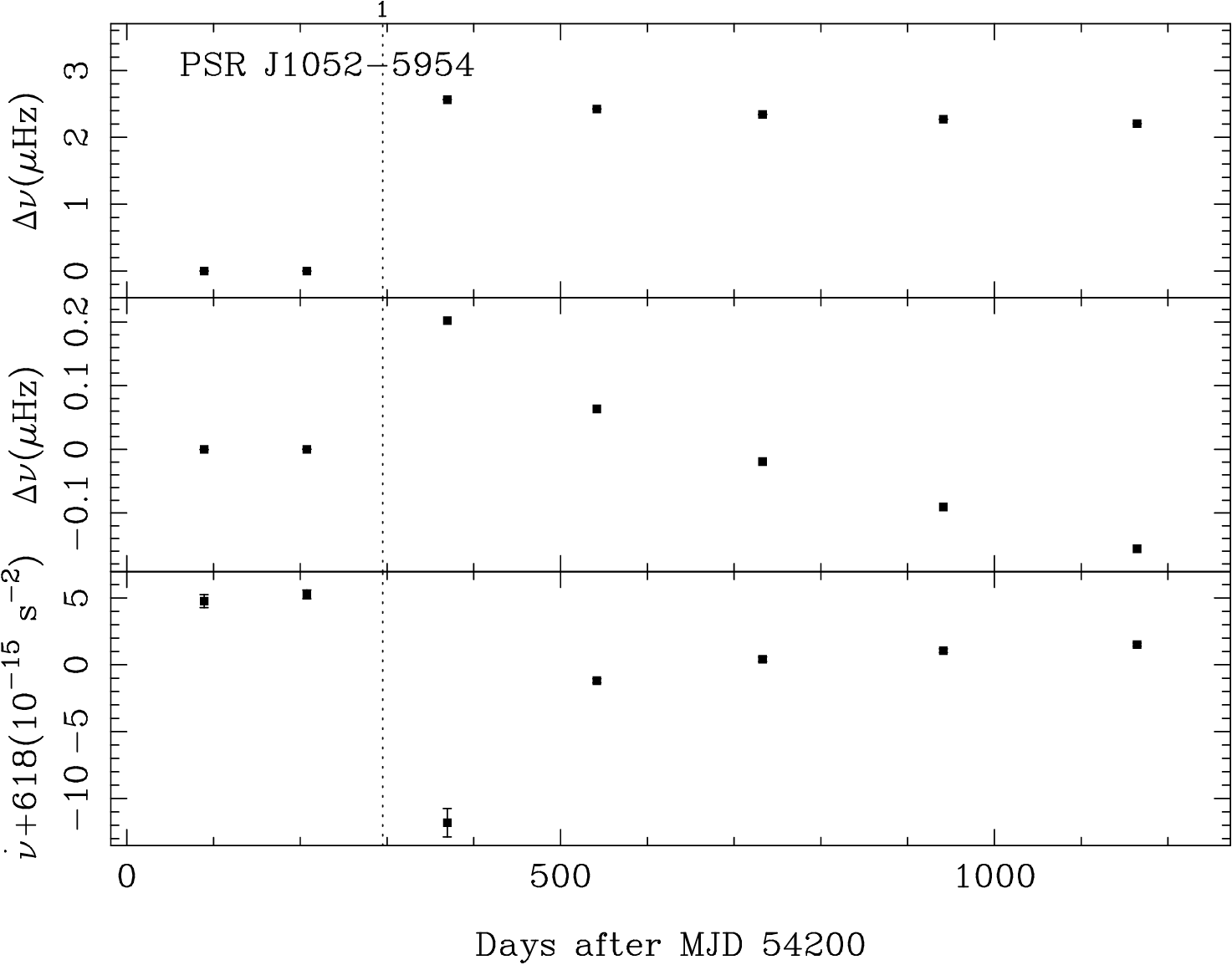} \\
\end{tabular} 
\caption{Glitches in PSRs J0905$-$5127, J1016$-$5857, J1048$-$5832 and
  J1052$-$5954. See Figure~\ref{fig:glt1} for a description
  of each sub-plot.}\label{fig:glt2}
\end{figure*}

\subsection{PSR J0905$-$5127}\label{sect:J0905-5127}

No glitch events have previously been reported for this pulsar.
Figure~\ref{fig:glt2} presents the evolution of $\nu$ and $\dot{\nu}$
observed at Parkes. The entire data span is $\sim16$\,yr, but there
exists a data gap lasting for $\sim4$\,yr. Two glitch events were
detected. Both are small, with a fractional glitch size
$\sim10^{-8}$. The available observations are not sufficient to study
the post-glitch behaviour for glitch 1. For glitch 2, no significant
post-glitch recovery was observed.

\subsection{PSR J1016$-$5857}\label{sect:J1016-5857}

In Figure \ref{fig:glt2}, the variations of $\nu$ and $\dot{\nu}$ of
this pulsar for $\sim11$\,yr are shown. Two glitches were
detected. They are similar with $\Delta\nu_{\rm
  g}/\nu\sim2\times10^{-6}$ and $\Delta\dot{\nu}_{\rm
  g}/\dot{\nu}\sim4\times10^{-3}$. The different slopes of $\dot{\nu}$
before and after glitch 1 imply a permanent change in $\ddot\nu$;
fitting showed that $\Delta\ddot{\nu}_{\rm p}=69(7)\times10^{-24}$
s$^{-3}$, approximately consistent with the $\ddot\nu$ values in
Table~\ref{tab:gltspin}. For glitch 2, the available data are not
sufficient to characterise the long-term post-glitch relaxations.

\subsection{PSR J1048$-$5832 (PSR B1046$-$58)}\label{sect:J1048-5832}

For PSR~J1048$-$5832, the evolution of $\nu$ and $\dot{\nu}$ spanning
20\,yr is shown in Figure \ref{fig:glt2}. \citet{wmp+00} and
\citet{ura02} have published details for glitches 1, 2 and
3. \citet{wjm+10} discovered glitch 6. We report here glitches 4 and 5
as new discoveries. Glitch 4 is large with $\Delta\nu_{\rm
  g}/\nu\sim1.8\times10^{-6}$, whereas glitch 5 is much smaller with a
fractional size $\sim2.5\times10^{-8}$. As shown in
Figure~\ref{fig:glt2}, for these two glitches there is little evidence
for exponential recoveries.

For glitches 2 and 3, \citet{wmp+00} included exponential terms to
model the post-glitch behaviour; the time constants were assumed to be
100\,d and 400\,d, respectively. \citet{ura02} observed glitch 3 with
high observing cadence. Two exponential recoveries were detected; the
timescales are 32(9)\,d and 130(40)\,d, respectively. As shown in
Figure~\ref{fig:glt2}, for both of the glitches 2 and 3, the
post-glitch $\dot{\nu}$ variations exhibit significant noise. For
glitch 2, our fitting for the exponential recovery showed $Q=0.026(6)$
and $\tau_{\rm d}=160(43)$\,d. We note that, for this glitch, because
there is only one pre-glitch measurement of $\dot{\nu}$, so we are
unable to measure the permanent change in $\dot\nu$. For glitch 3, our
fitting showed $Q=0.008(3)$ and $\tau_{\rm d}=60(20)$\,d.
Table~\ref{tab:gltspin} shows signficant values of $\ddot\nu$ for all
except glitch 2, with significant variations from glitch to glitch. On
average, the values are about an order of magnitude smaller than those
for the Vela pulsar.

\subsection{PSR J1052$-$5954}\label{sect:J1052-5954}

The available data set for this pulsar contains a data gap of
$\sim6$\,yr. Figure~\ref{fig:glt2} shows the evolution of $\nu$ and
$\dot{\nu}$ after the data gap. The detected glitch at MJD $\sim
54495$ was reported by \citet{wjm+10}. An exponential relaxation and a
significant permanent increase in spin-down rate $|\dot\nu|$ can be
identified in the post-glitch data. Fitting to the timing residuals
indicated that $\sim0.067$ of the glitch recovered in $\sim46$\,d.

\begin{figure*} 
\begin{tabular}{cc}
\includegraphics[width=5.5cm,angle=-90]{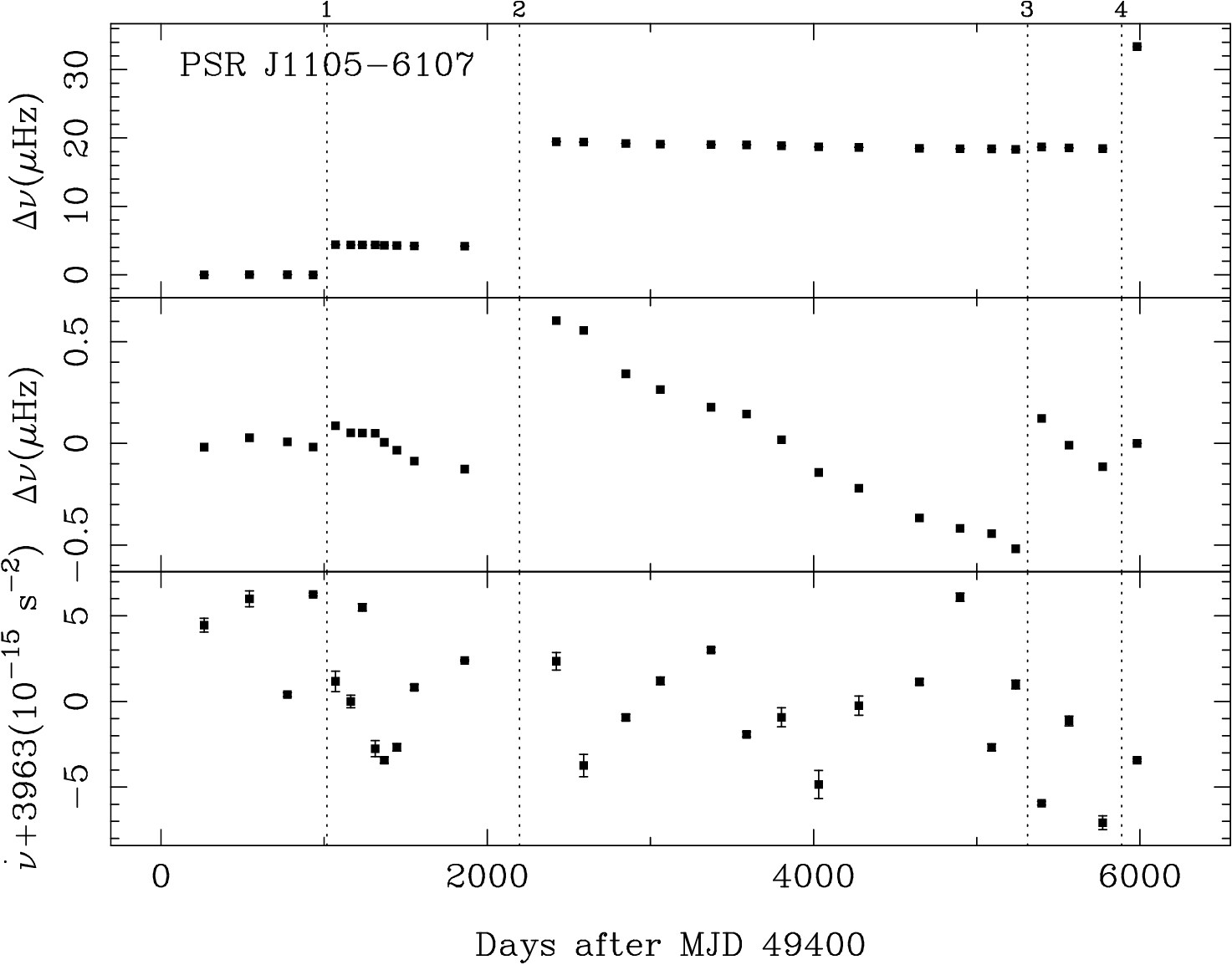} &
\includegraphics[width=5.5cm,angle=-90]{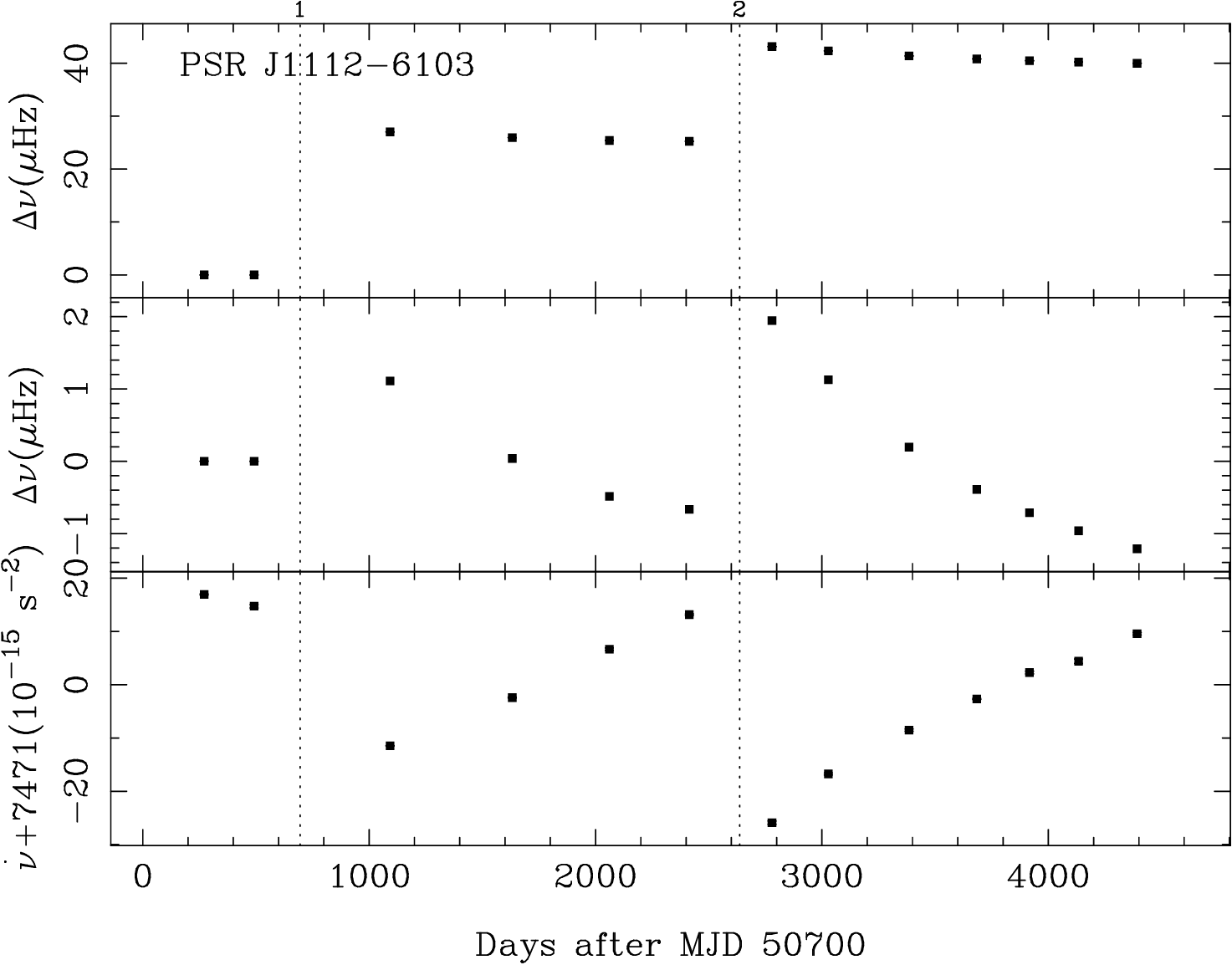} \\ 
\includegraphics[width=5.5cm,angle=-90]{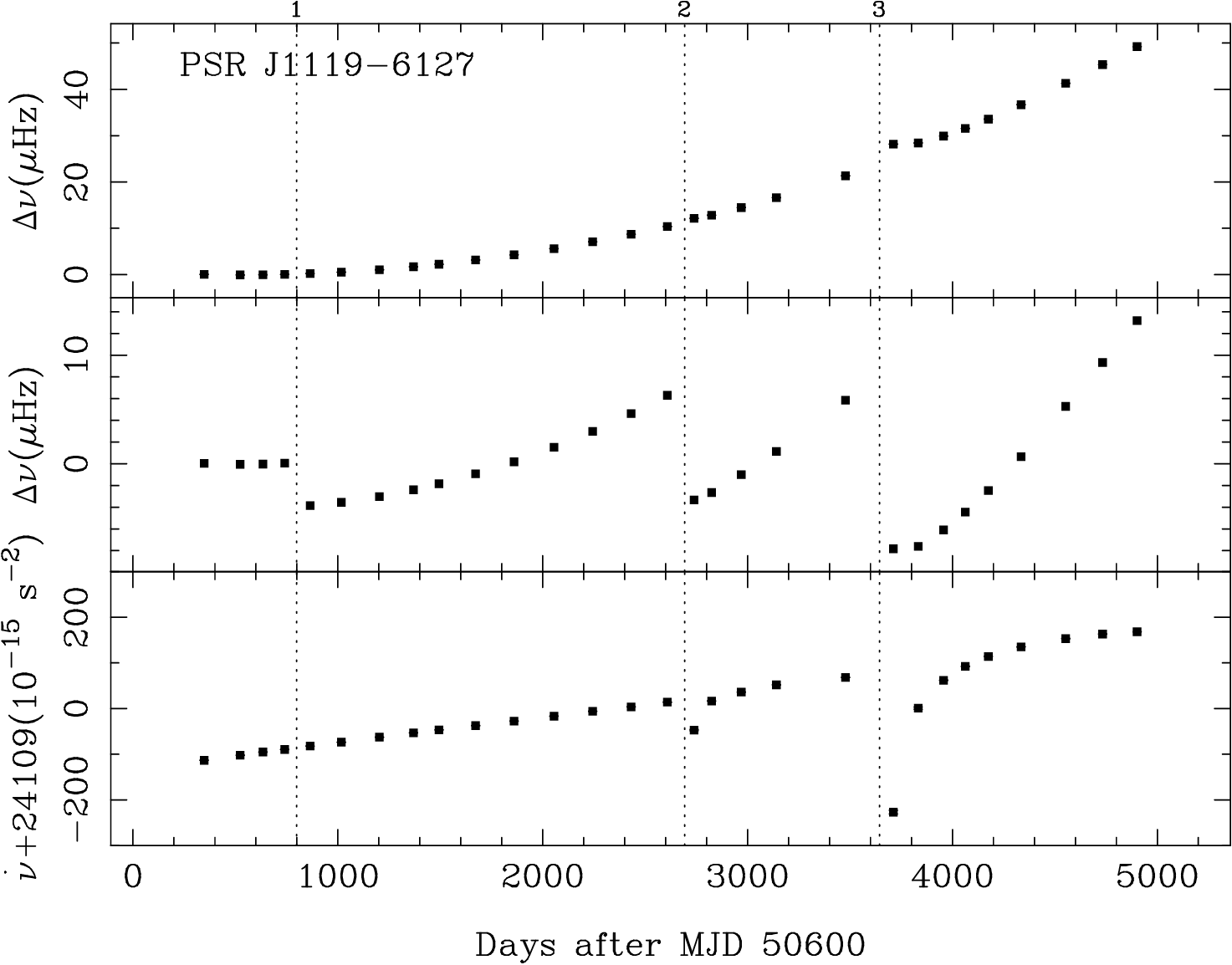} & 
\includegraphics[width=5.5cm,angle=-90]{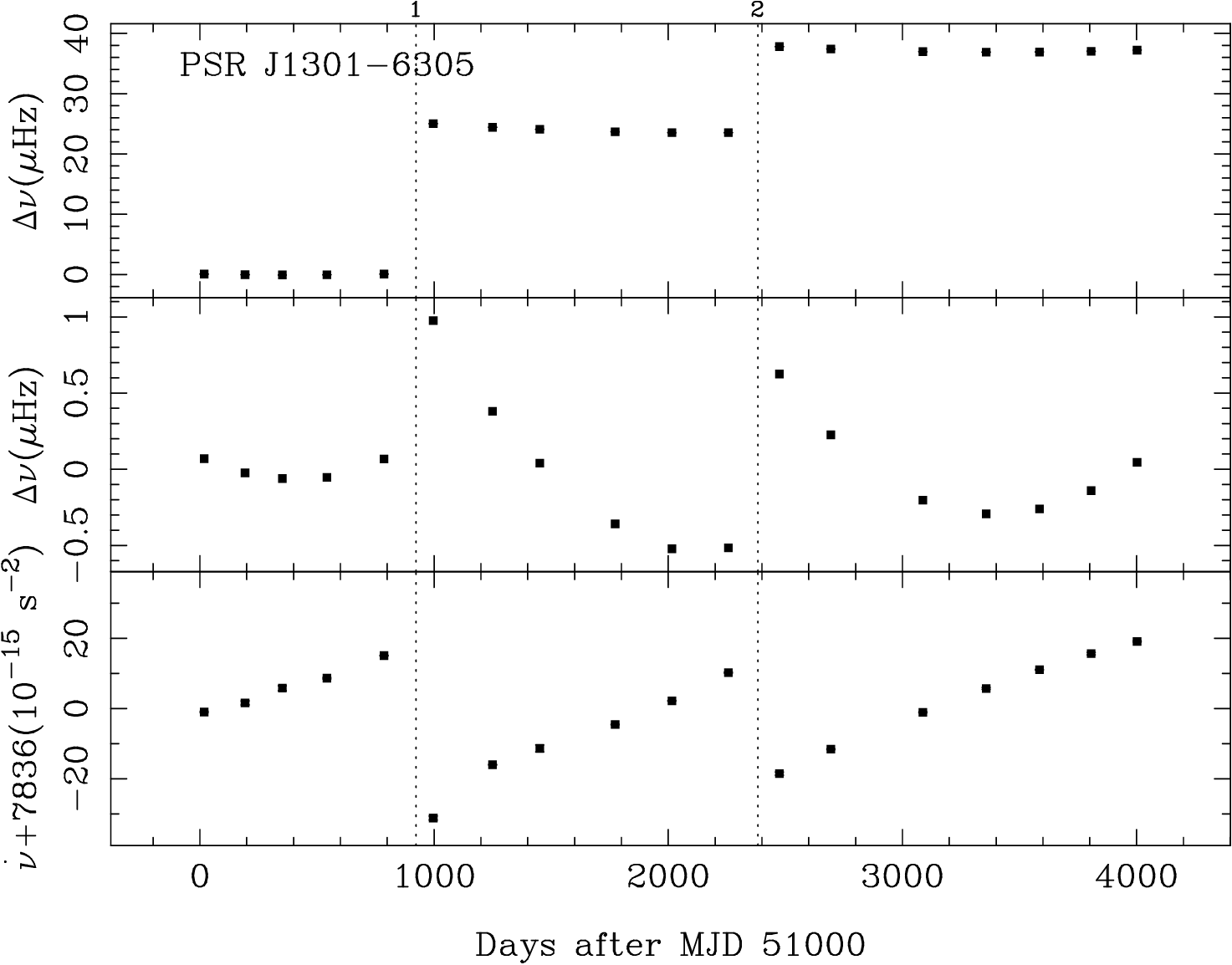} \\
\end{tabular} 
\caption{Glitches in PSRs J1105$-$6107, J1112$-$6103, J1119$-$6127 and
  J1301$-$6305. See Figure~\ref{fig:glt1} for a description
  of each sub-plot.}\label{fig:glt3}
\end{figure*}

\subsection{PSR J1105$-$6107}\label{sect:J1105-6107}

Three glitch events have previously been identified for this source
\citep{wmp+00,wjm+10}. In Figure~\ref{fig:glt3}, we present the
evolution of $\nu$ and $\dot{\nu}$ for $\sim16$\,yr of this pulsar. We
confirm the previously detected glitches 1 and 3
\citep{wmp+00,wjm+10}. \citet{wmp+00} reported a small glitch occurred
at MJD $\sim50610$. However, we found that the timing behaviour of
this source around this epoch is more likely to be dominated by timing
noise. We report new glitch events as glitches 2 and 4. As shown in
Figure~\ref{fig:glt3}, the post-glitch behaviour is noisy and no
exponential recoveries were observed. There appears to be a persistent
increase in $|\dot\nu|$ at the time of each of glitches 1, 2 and
3. For glitch 4, the available data span is not adequate to study the
post-glitch behaviour.

\subsection{PSR J1112$-$6103}\label{sect:J1112-6103}

As shown in Figure~\ref{fig:glt3}, two large glitch events were
detected in this pulsar, with $\Delta\nu_{\rm g}/\nu\sim10^{-6}$. For
glitch 1, the observed variations of $\dot{\nu}$ indicate a large
change in $\ddot{\nu}$; fitting gave $\Delta\ddot{\nu}_{\rm
  p}\sim240\times10^{-24}$ s$^{-3}$. No exponential recovery was
observed for this glitch. For glitch 2, a long-term exponential
relaxation was observed, which is characterised by $Q\sim0.022$ and
$\tau_{\rm d}\sim300$\,d.

\subsection{PSR J1119$-$6127}\label{sect:J1119-6127}

For PSR~J1119$-$6127, Figure~\ref{fig:glt3} shows the evolution of the
observed pulse frequency and its first derivative spanning
$\sim13$\,yr. Three glitches were observed. The first is small with
$\Delta\nu_{\rm g}/\nu\sim4\times10^{-9}$ as was reported by
\citet{ckl+00}. The second and third glitches are much larger and were
studied in detail by \citet{wje11}. Our results are generally
consistent with theirs.

\subsection{PSR J1301$-$6305}\label{sect:J1301-6305}

Figure~\ref{fig:glt3} shows the evolution of $\nu$ and $\dot{\nu}$ for
PSR~J1301$-$6305 over $\sim11$\,yr. We detected two large glitch
events. Glitch 1 has $\Delta\nu_{\rm g}/\nu\sim4.6\times10^{-6}$ and
for glitch 2 the fractional frequency change is about half this. For
glitch 1, an exponential recovery was identified; fitting gave
$Q\sim0.0049$ and $\tau_{\rm d}\sim58$\,d. As shown in Figure
\ref{fig:glt3}, the pre- and post-glitch intervals show clear linear
recoveries. As Table~\ref{tab:gltspin} indicates, the $\ddot{\nu}$
values are all $\sim250\times10^{-24}$~s$^{-3}$. Because of timing
noise, we were unable to fit for the $\Delta\ddot{\nu}_{\rm p}$
changes at the glitches.

\begin{figure*} 
\begin{tabular}{cc}
\includegraphics[width=5.5cm,angle=-90]{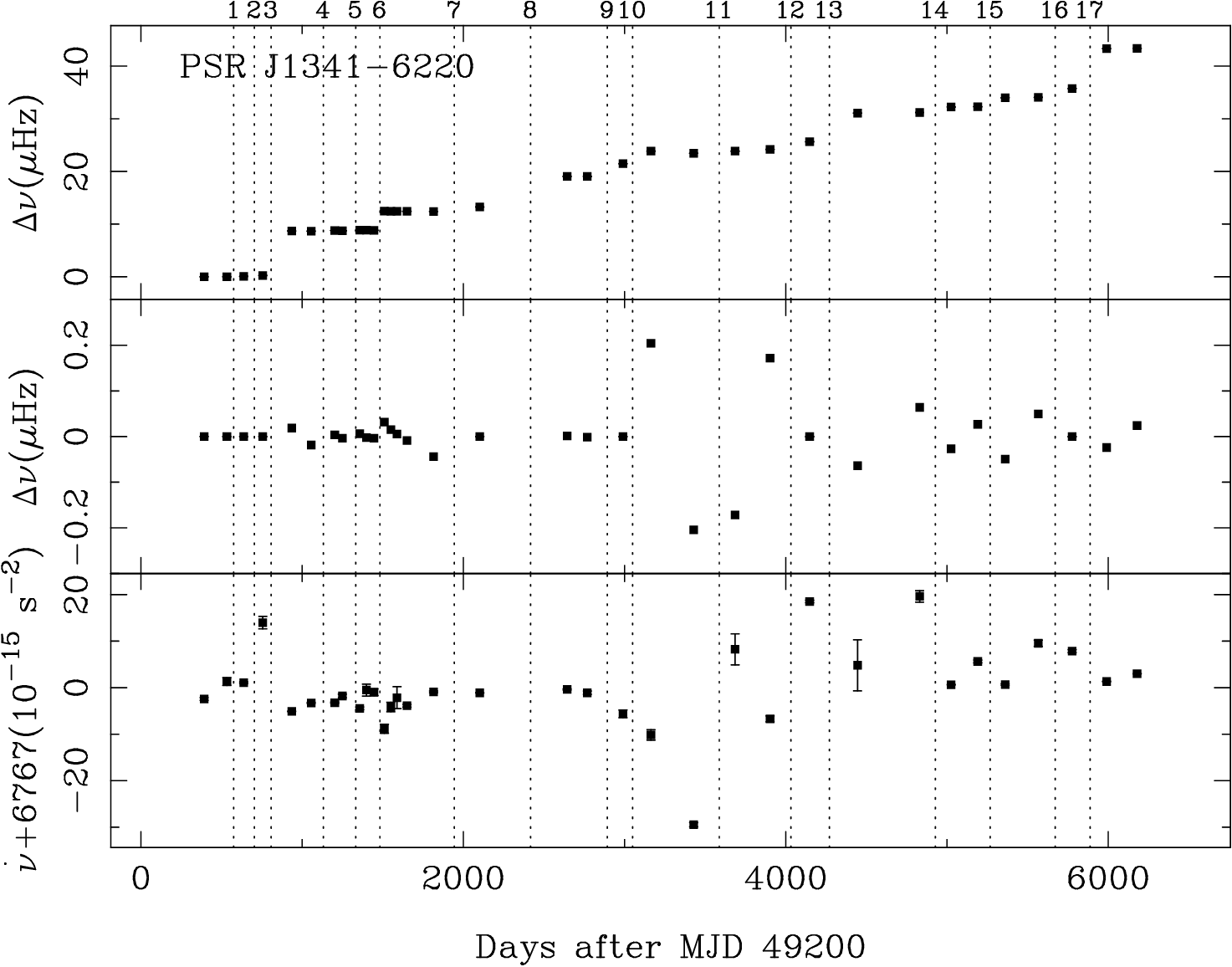} &
\includegraphics[width=5.5cm,angle=-90]{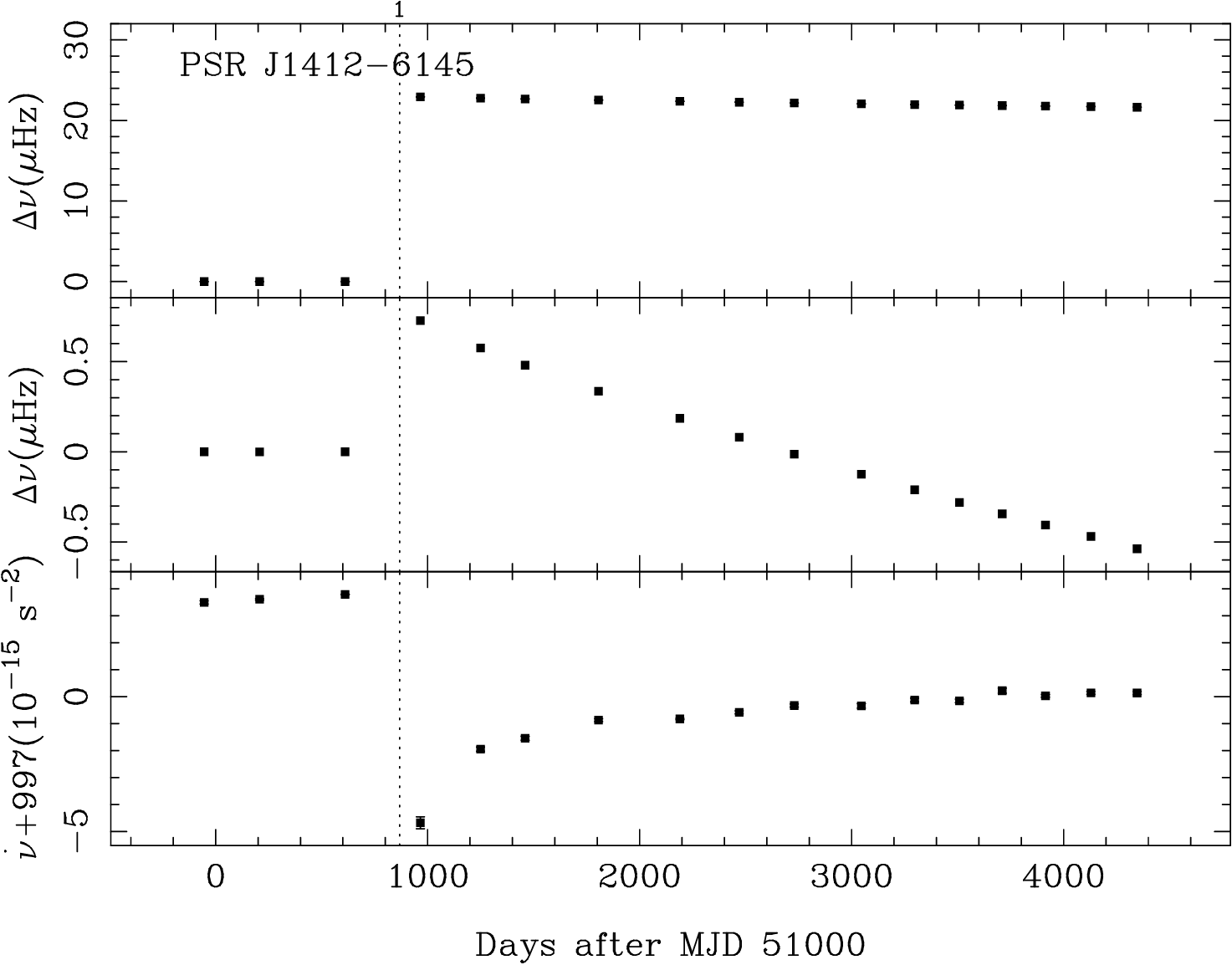} \\ 
\includegraphics[width=5.5cm,angle=-90]{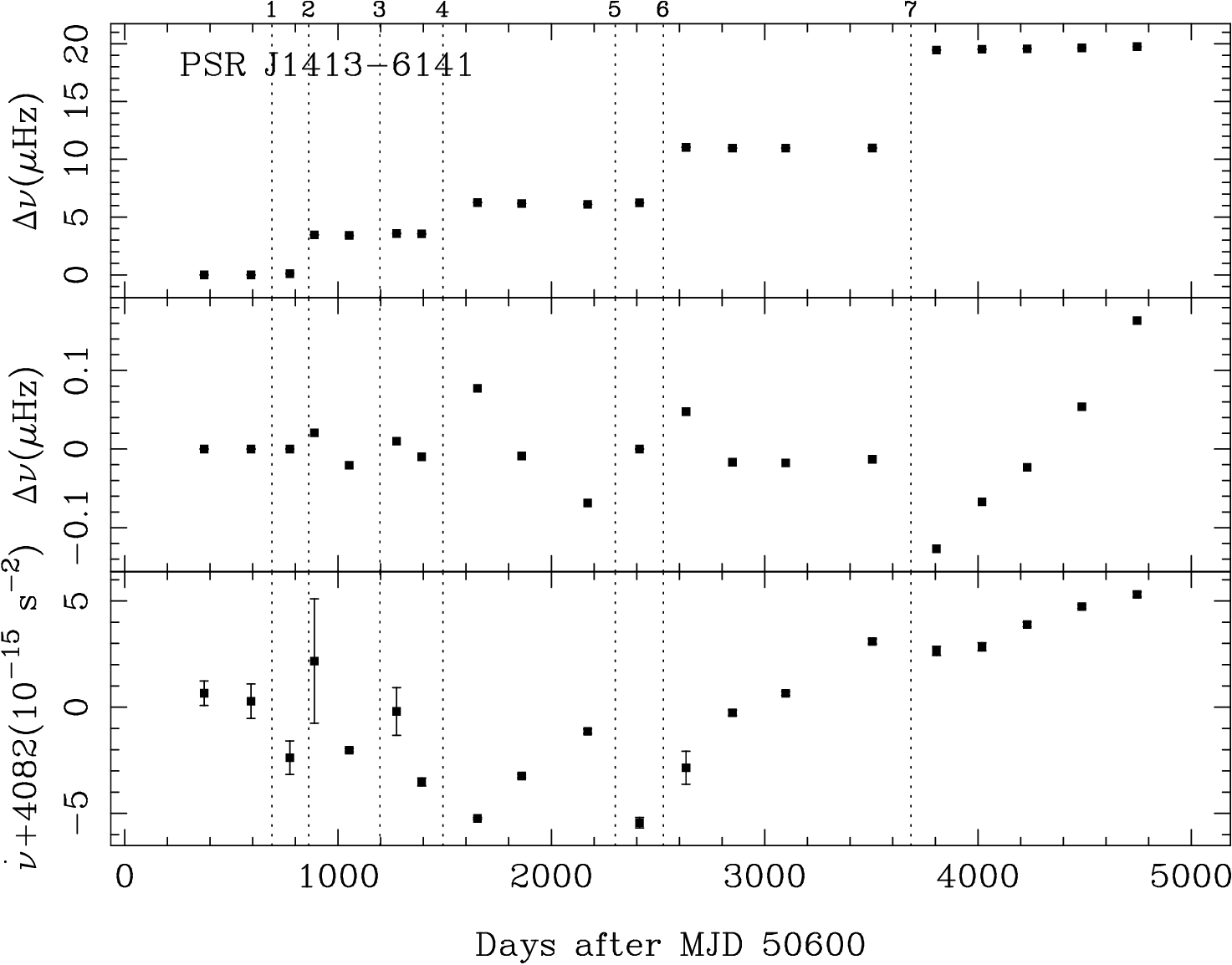} & 
\includegraphics[width=5.5cm,angle=-90]{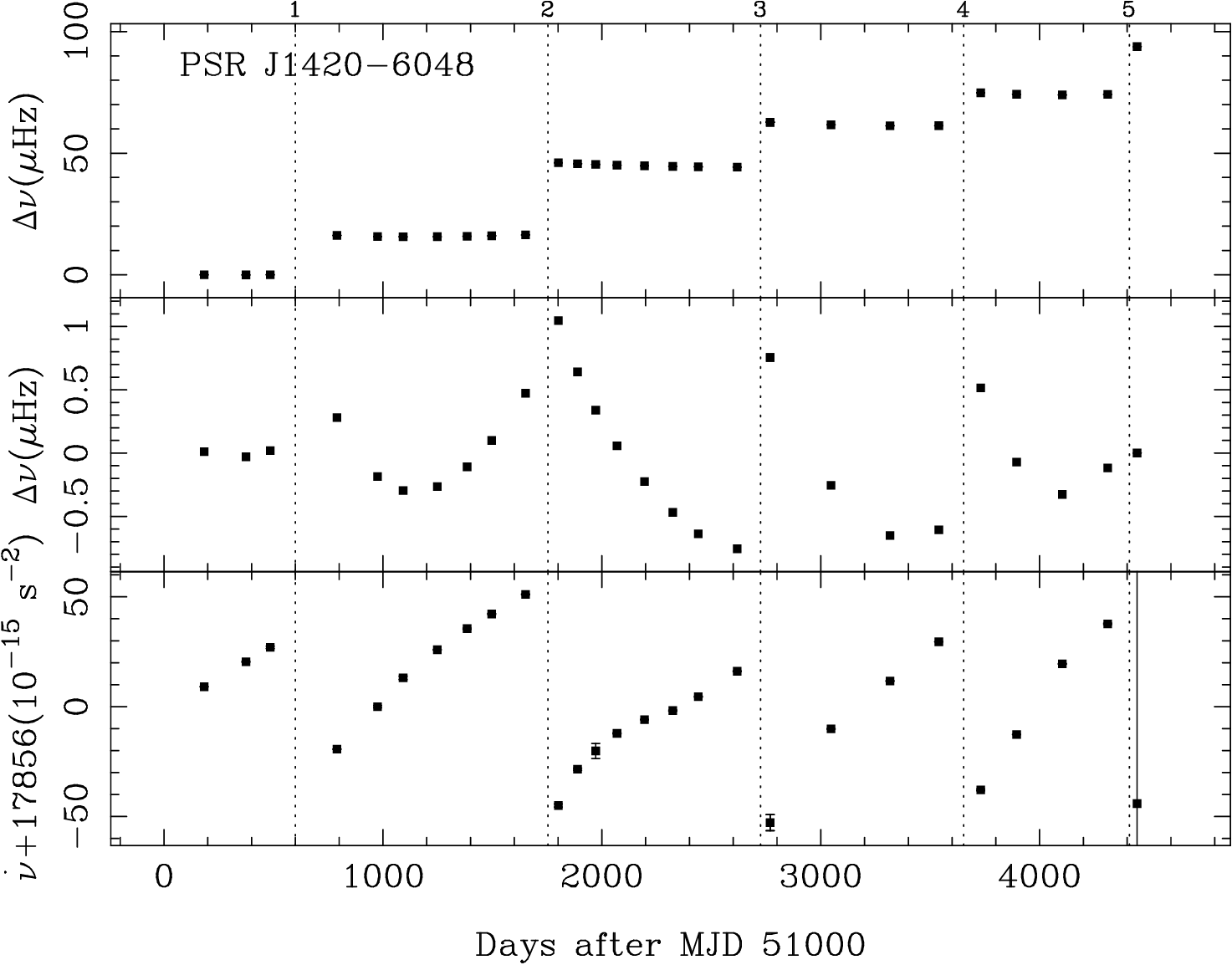} \\
\end{tabular} 
\caption{Glitches in PSRs J1341$-$6220, J1412$-$6145, J1413$-$6141 and
  J1420$-$6048. See Figure~\ref{fig:glt1} for a description
  of each sub-plot.}\label{fig:glt4}
\end{figure*}

\subsection{PSR J1341$-$6220 (PSR B1338$-$62)}\label{sect:J1341-6220}

This pulsar is well known to have frequent glitches --- \citet{wmp+00}
and \citet{wjm+10} have reported 14 glitches. Figure \ref{fig:glt4}
shows the evolution of $\nu$ and $\dot{\nu}$ for $\sim16$\,yr, where a
total of 17 glitches are presented. We report here the new detections
of nine glitch events. For glitch 6, an exponential decay with
$Q\sim0.0112$ and $\tau_{\rm d}\sim24$\,d was detected. Unfortunately,
for the other glitches the observations are insufficient to study the
post-glitch behaviour.

\subsection{PSR J1412$-$6145}\label{sect:J1412-6145}

PSR~J1412$-$6145 has not previously been known to glitch. Here, we
report the discovery of a large glitch with $\Delta\nu_{\rm g}/\nu\sim
7.2\times10^{-6}$ that occurred at MJD $\sim51868$. As shown in Figure
\ref{fig:glt4}, there was a clear exponential recovery with timescale
$\sim60$\,d, a significant increase in $|\dot\nu|$ at the time of the
glitch and a slow linear recovery of part of this increase.

\subsection{PSR J1413$-$6141}\label{sect:J1413-6141}

Figure~\ref{fig:glt4} presents seven new glitch events detected in
this pulsar over a 12.6-yr data span. Among these events, three are
small ($\Delta\nu_{\rm g}/\nu\sim10^{-8}$), while the other four are
larger, with a fractional size $\gtrsim 10^{-6}$. Exponential
post-glitch recoveries are not observed for these
glitches. Significant values of $\ddot{\nu}$ are seen after each of
the latest four glitches (cf. Table~\ref{tab:gltspin}). We were able
to fit for $\Delta\ddot{\nu}_{\rm p}$ for glitch 4, giving a value of
$491(42)\times10^{-24}$\,s$^{-3}$; this is consistent with the
difference between the post- and pre-glitch solutions for
$\ddot{\nu}$, which is $457(40)\times10^{-24}$\,s$^{-3}$.

\subsection{PSR J1420$-$6048}\label{sect:J1420-6048}

Figure~\ref{fig:glt4} shows that five glitch events were observed in
this pulsar. Glitch 4 was first reported by \citet{wjm+10}. All of
these glitches are large, with $\Delta\nu_{\rm g}/\nu\sim10^{-6}$. The
post-glitch slow-down rates exhibit linear decays, and changes in
$\ddot{\nu}$ are observed (Tables \ref{tab:gltspin} and
\ref{tab:glt}). For glitch 2, an exponential recovery was measured,
with obtaining $Q\sim0.008$ and $\tau_{\rm d}\sim99$\,d. For glitch 5,
the available data are not sufficient to study the post-glitch
behaviour.

\begin{figure*} 
\begin{tabular}{cc}
\includegraphics[width=5.5cm,angle=-90]{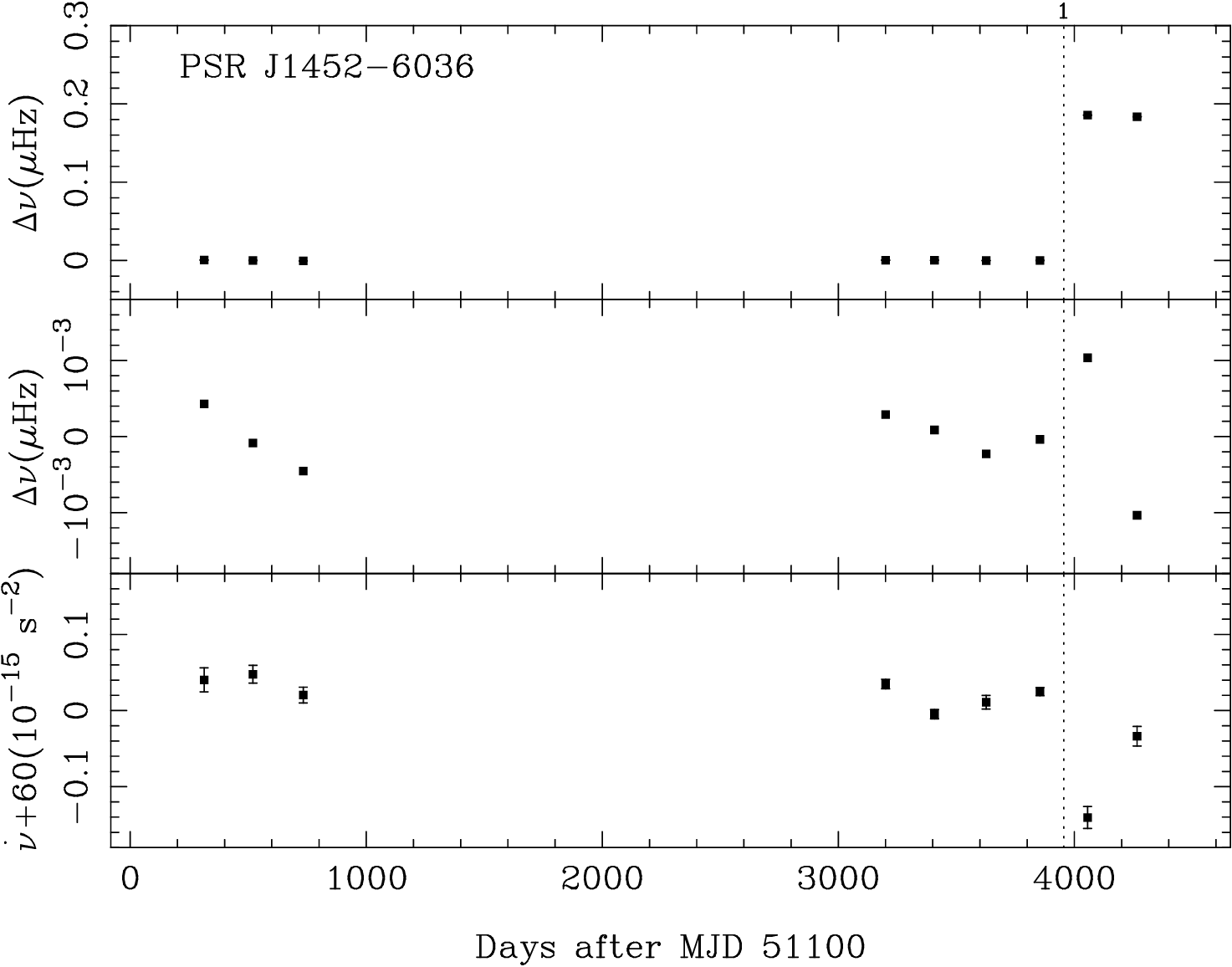} &
\includegraphics[width=5.5cm,angle=-90]{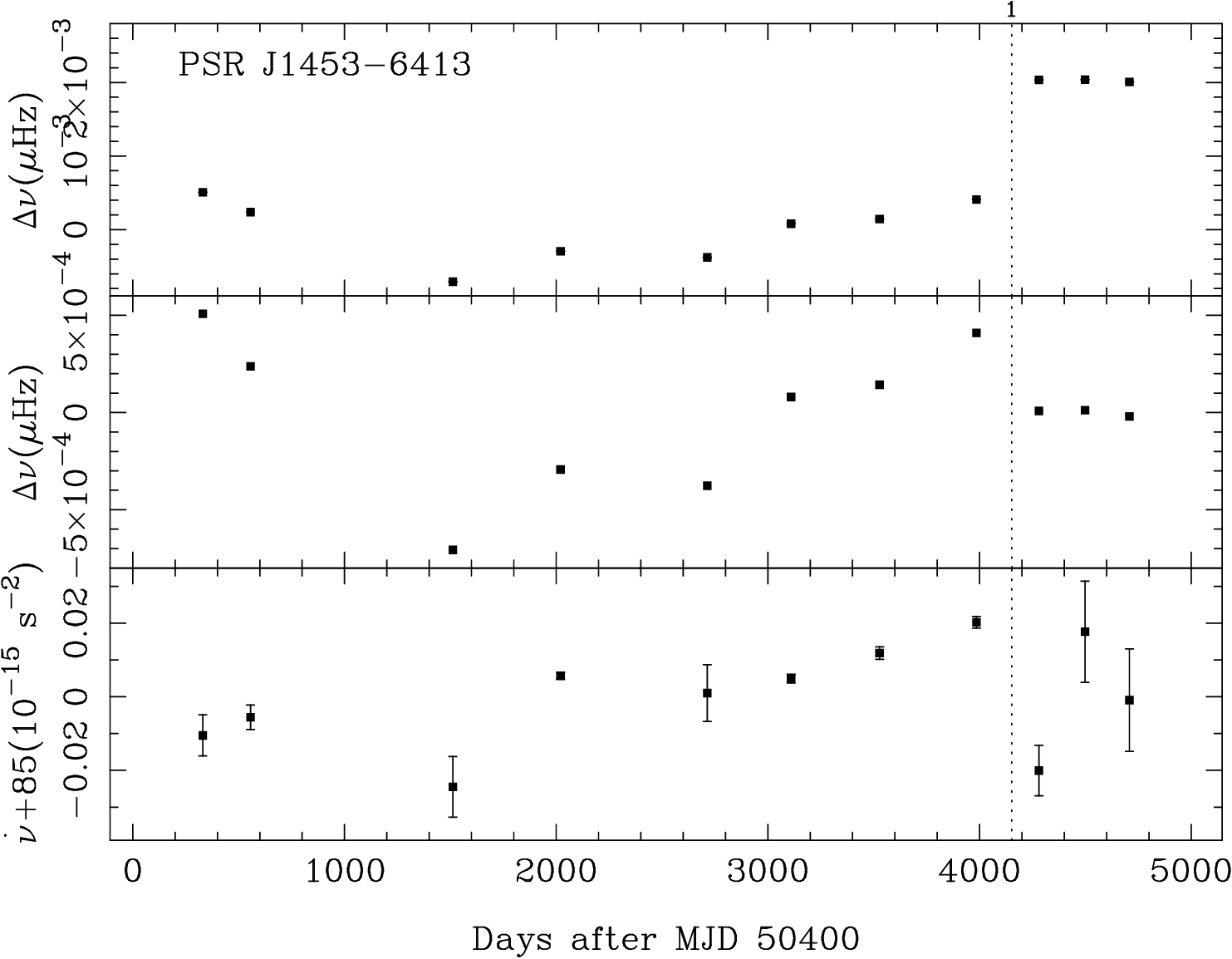} \\ 
\includegraphics[width=5.5cm,angle=-90]{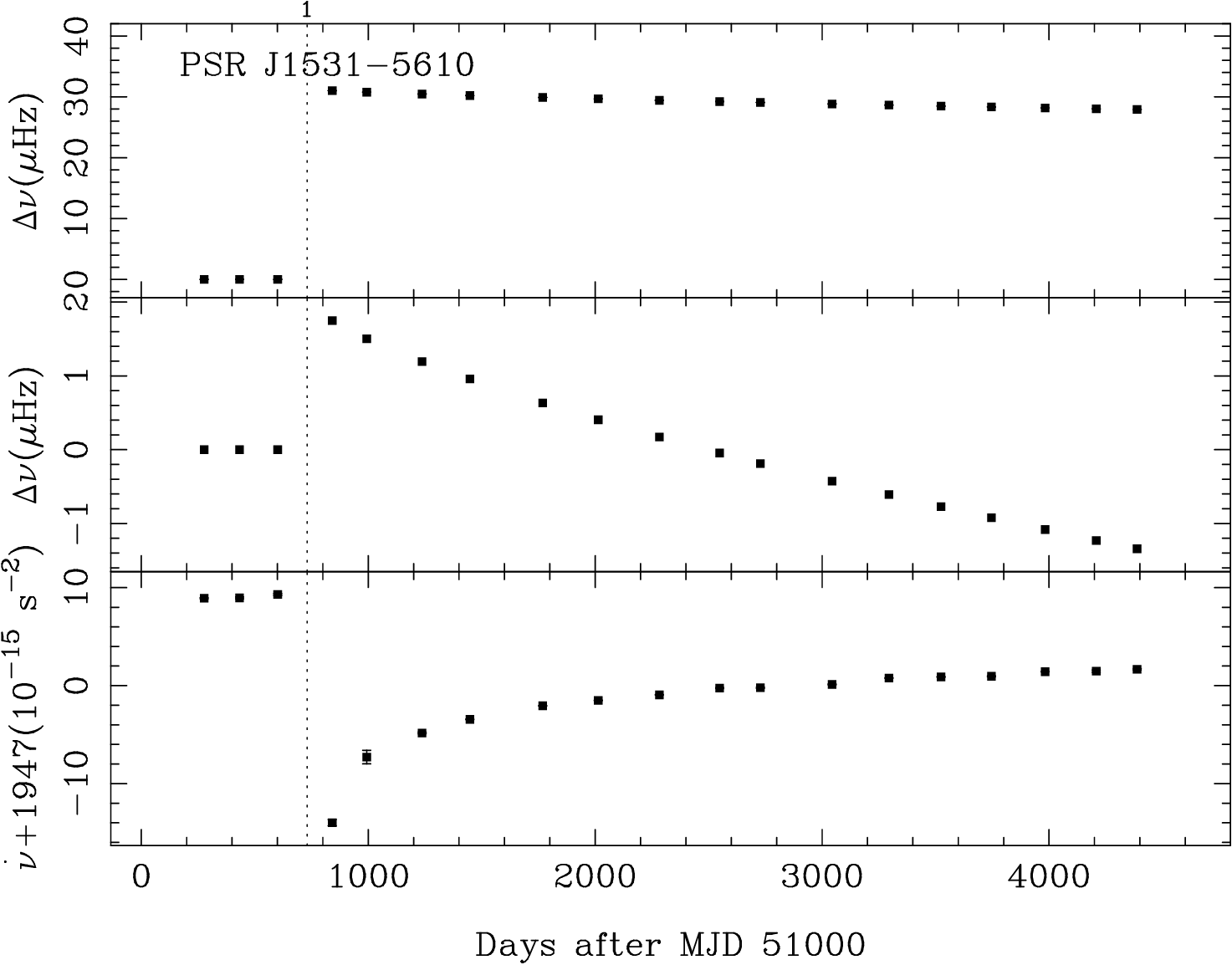} & 
\includegraphics[width=5.5cm,angle=-90]{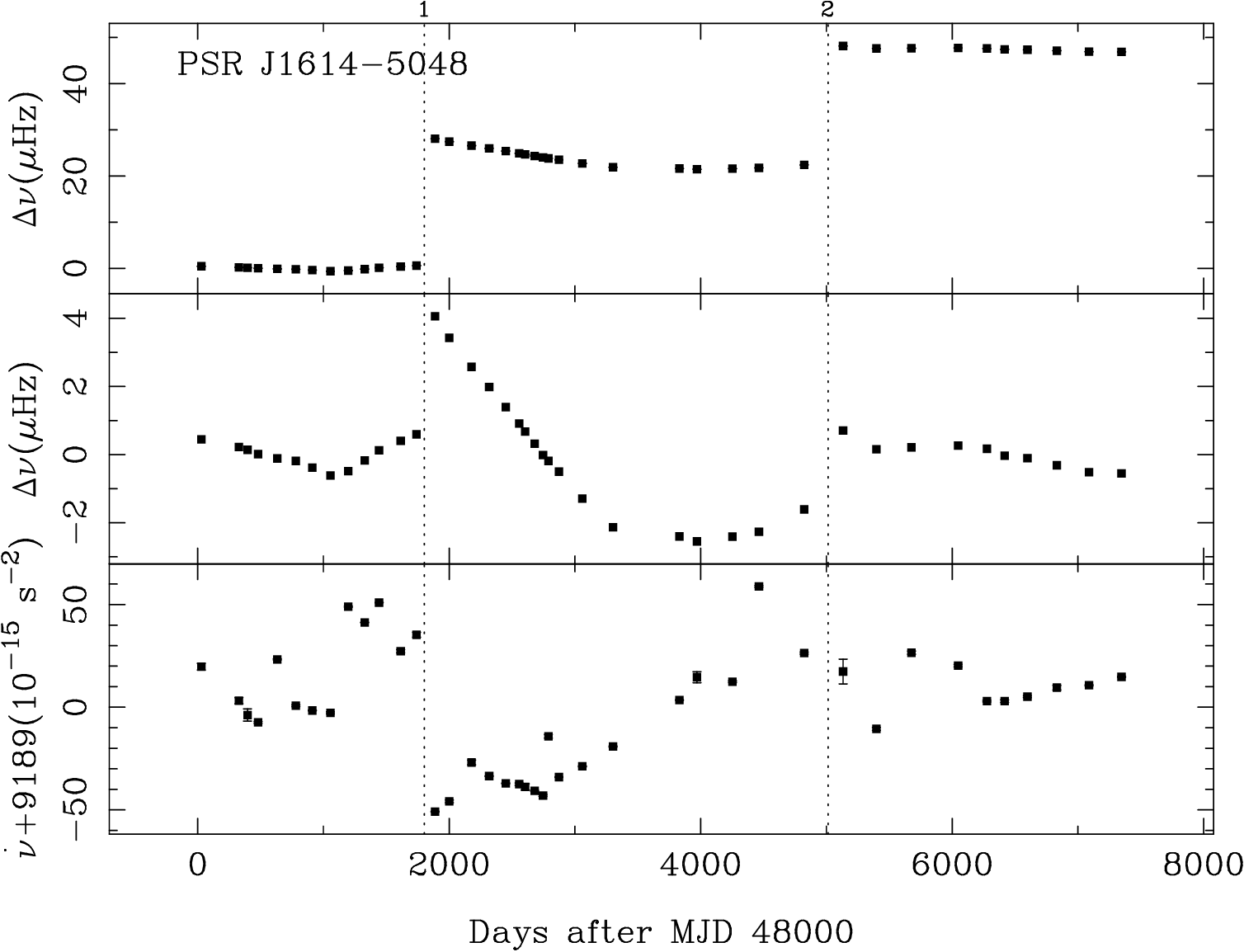} \\
\end{tabular} 
\caption{Glitches in PSRs J1452$-$6036, J1453$-$6413, J1531$-$5610 and
  J1614$-$5048. See Figure~\ref{fig:glt1} for a description
  of each sub-plot.}\label{fig:glt5}
\end{figure*}

\subsection{PSR J1452$-$6036}\label{sect:J1452-6036}

This pulsar was not previously known to glitch. Since the end of the
Multibeam Survey timing \citep{kbm+03}, no observations were made
until the start of the {\it Fermi} project. Hence, a data gap lasting
for $\sim5$\,yr exists. Figure \ref{fig:glt5} shows the evolution of
$\nu$ and $\dot\nu$. A small glitch event with $\Delta\nu_{\rm
  g}/\nu\sim 3\times10^{-8}$ was detected at MJD $\sim 55055$. The
available data are not adequate to study the post-glitch behaviour.

\subsection{PSR J1453$-$6413 (PSR B1449$-$64)}\label{sect:J1453-6413}

No glitch event has previously been reported for this pulsar. Figure
\ref{fig:glt5} shows the evolution of $\nu$ and $\dot{\nu}$ for
$\sim12$\,yr. We detected a very small glitch with $\Delta\nu_{\rm
  g}/\nu\sim3\times10^{-10}$. We cannot comment on the post-glitch
behaviour since the data are insufficient.

\subsection{PSR J1531$-$5610}\label{sect:J1531-5610}

PSR~J1531$-$5610 was not previously known to glitch. Figure
\ref{fig:glt5} shows a large glitch event at MJD $\sim 51730$,
detected by Parkes timing. As in PSR~J1412$-$6145, this glitch has an
exponential recovery, an offset in $\dot\nu$ at the time of the glitch
and a slow linear recovery. Our fitting of the exponential term showed
that $\sim0.007$ of the glitch recovered within a timescale of
$\sim76$\,d and the long-term $\ddot\nu$ is $\sim
20\times10^{-24}$\,s$^{-3}$ (Table~\ref{tab:gltspin}).

\subsection{PSR J1614$-$5048 (PSR B1610$-$50)}\label{sect:J1614-5048}

PSR~J1614$-$5048 has been observed at Parkes for $\sim20$\,yr. As
shown in Figure~\ref{fig:glt5}, two glitches were detected. Both of
the events are large, with $\Delta\nu_{\rm
  g}/\nu>6\times10^{-6}$. This pulsar exhibits remarkable timing
noise; the large-scale fluctuations in $\dot{\nu}$ reflect this. As a
result, phase-connected timing residuals cannot be obtained for the
entire data range between the two glitch events. We thus report the
timing solutions for this data span in two sections (see
Table~\ref{tab:gltspin}). Glitch 1 has previously been reported by
\citet{wmp+00}; our results for $\Delta\nu_{\rm g}/\nu$ and
$\Delta\dot{\nu}_{\rm g}/\dot{\nu}$ are consistent with theirs. Glitch
2 is a new detection. Values of $\ddot\nu$ given in
Table~\ref{tab:gltspin} show significant variations, but these are
likely to be contaminated by the timing noise. Despite the noise,
there does appear to be a signficant linear recovery after glitch 1
with $\ddot\nu \sim 200\times 10^{-24}$~s$^{-3}$.

\begin{figure*} 
\begin{tabular}{cc}
\includegraphics[width=5.5cm,angle=-90]{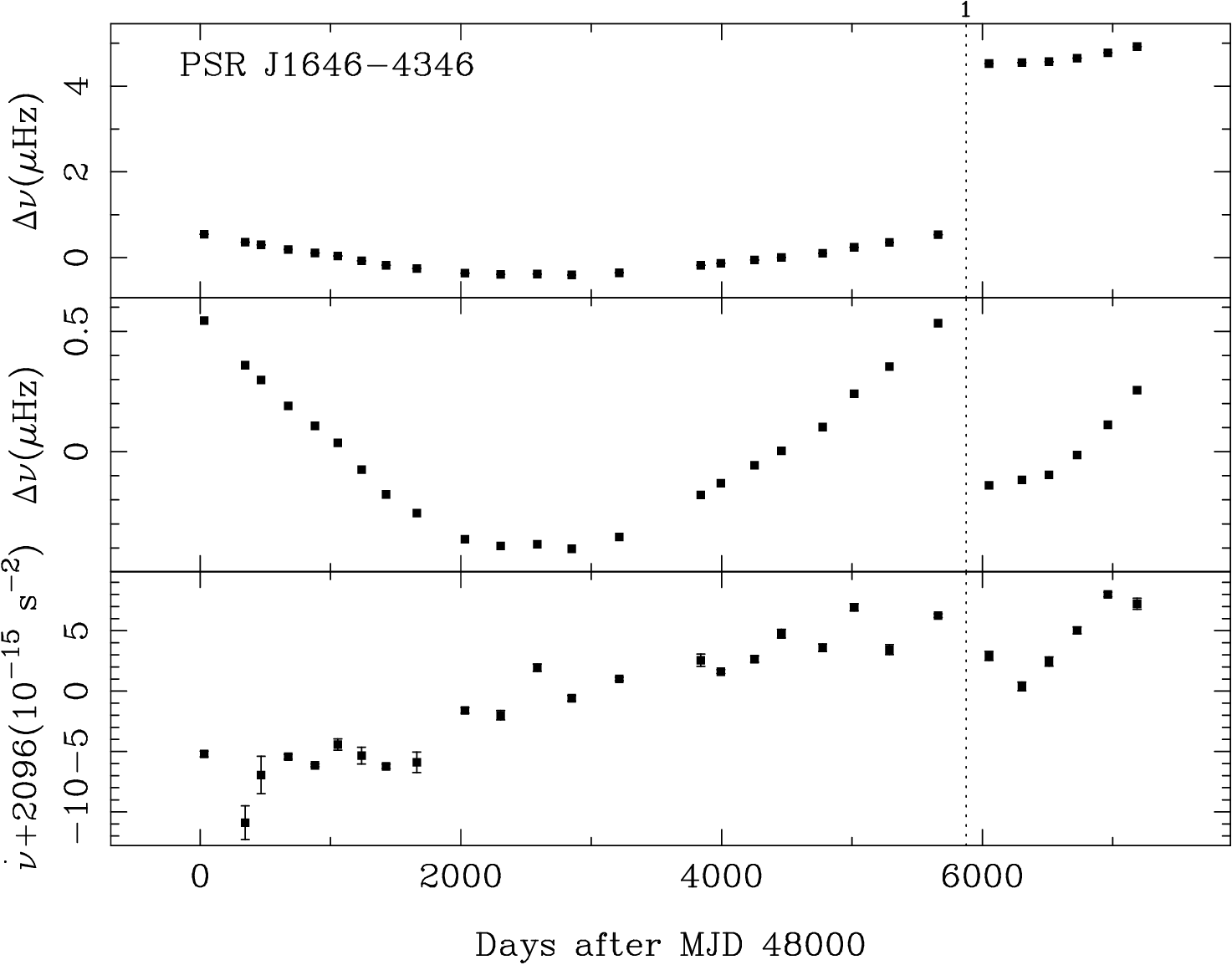} &
\includegraphics[width=5.5cm,angle=-90]{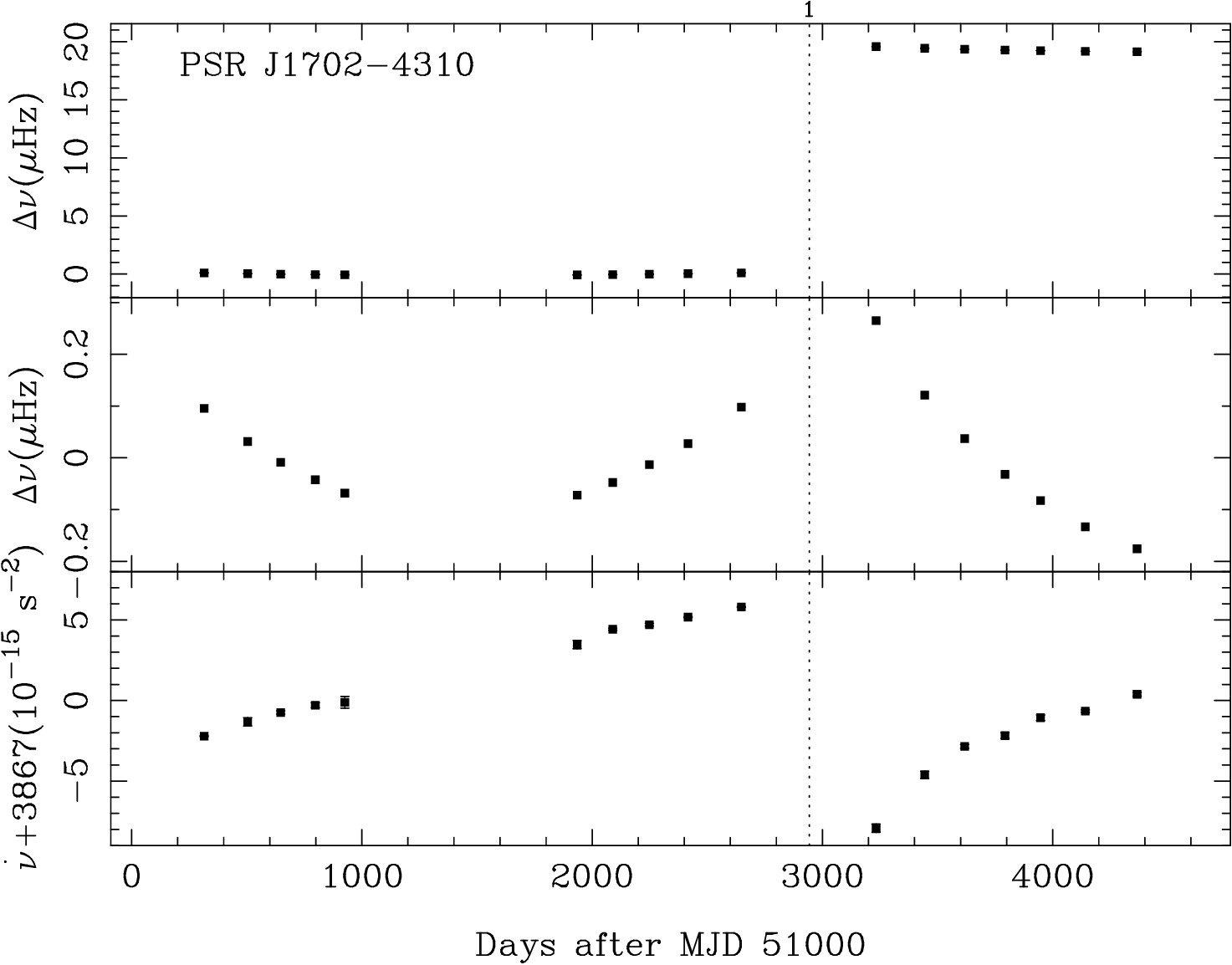} \\ 
\includegraphics[width=5.5cm,angle=-90]{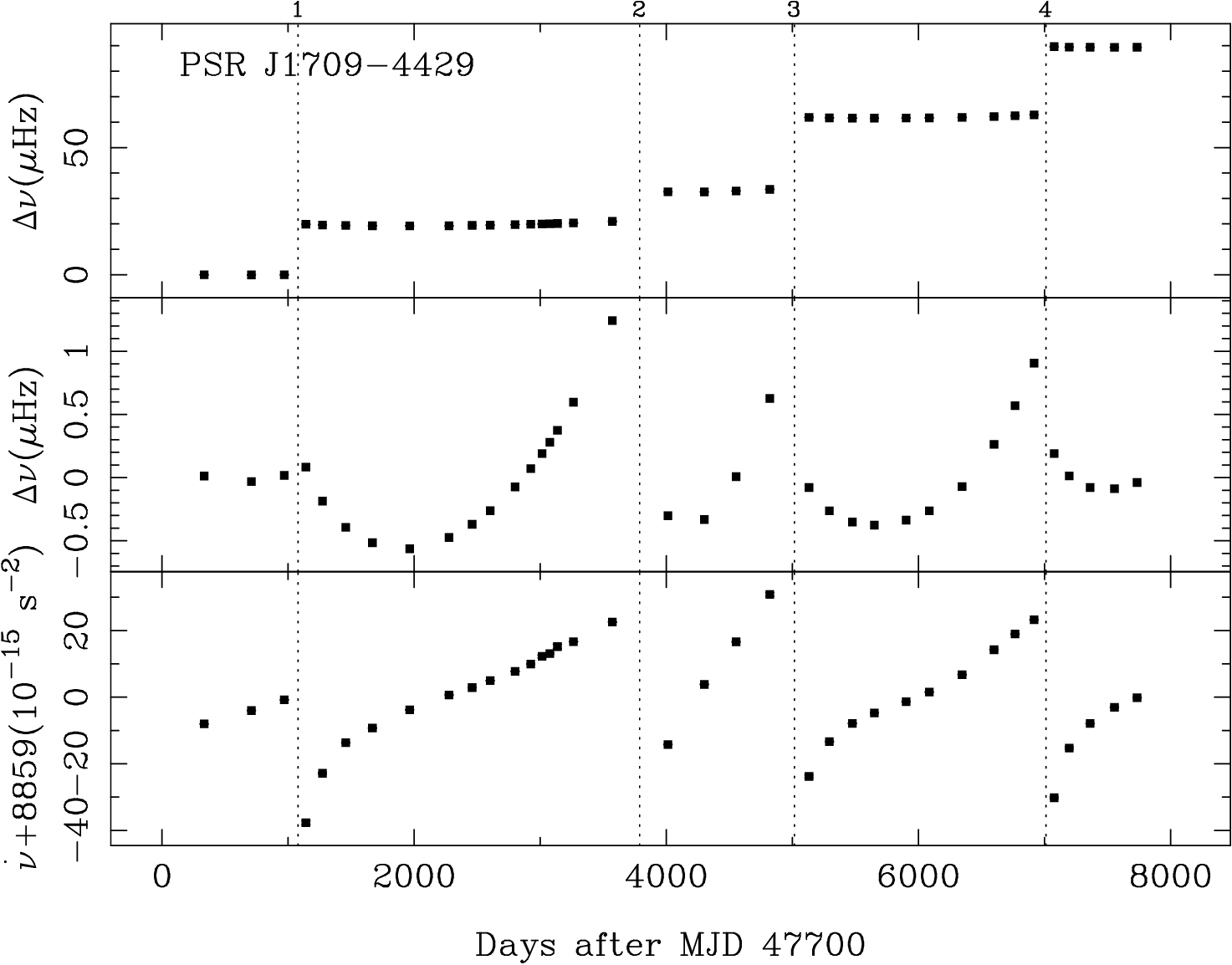} & 
\includegraphics[width=5.5cm,angle=-90]{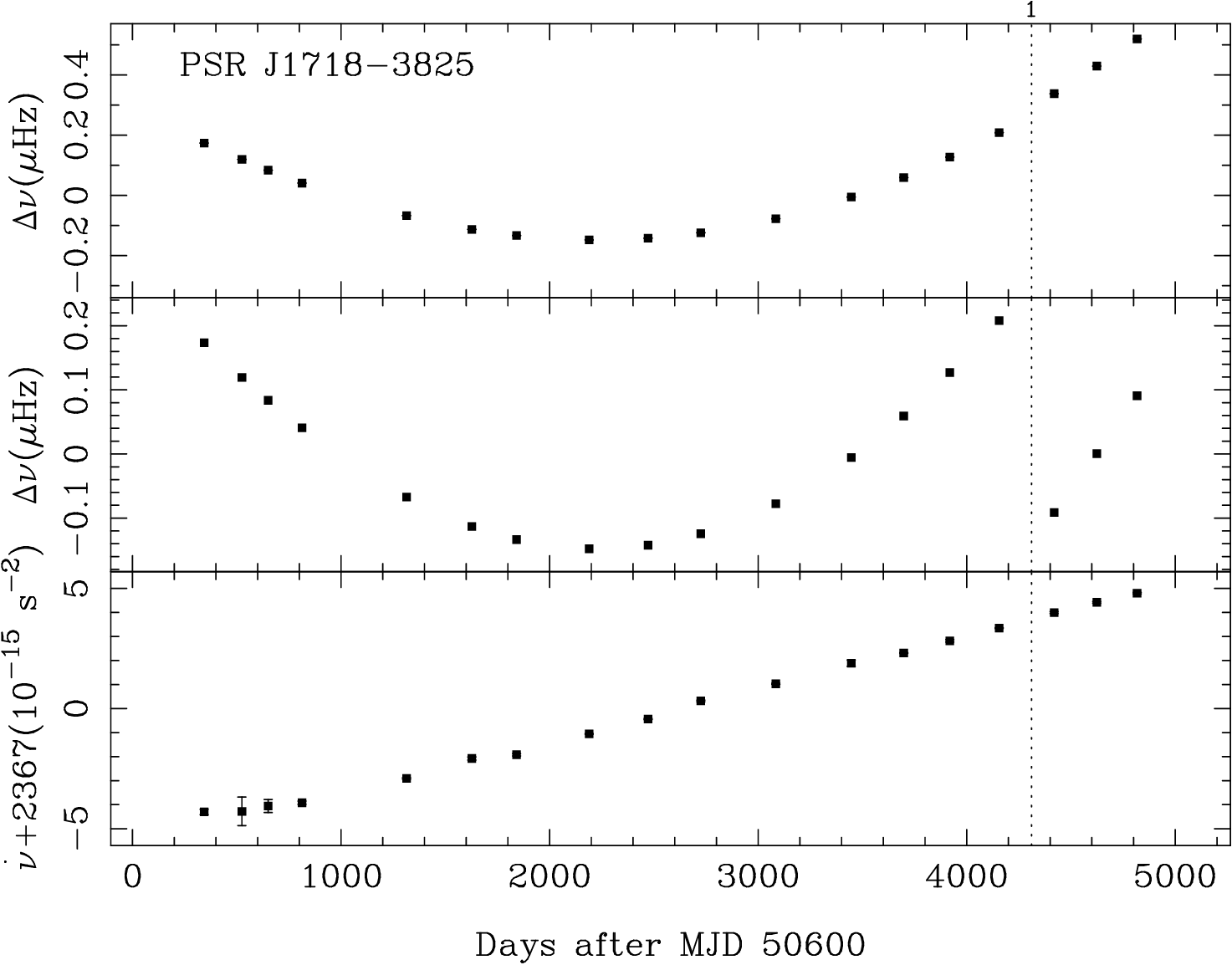} \\
\end{tabular} 
\caption{Glitches in PSRs J1646$-$4346, J1702$-$4310, J1709$-$4429 and
  J1718$-$3825. See Figure~\ref{fig:glt1} for a description
  of each sub-plot.}\label{fig:glt6}
\end{figure*}

\subsection{PSR J1646$-$4346 (PSR B1643$-$43)}\label{sect:J1646-4346}

Figure~\ref{fig:glt6} presents the evolution of $\nu$ and $\dot\nu$
for this pulsar for $\sim16$\,yr. A glitch event was detected at MJD
$\sim 53875$. This is the first reported glitch for this pulsar and it
has a fractional size $\sim8.8\times10^{-7}$. There is a clear linear
recovery from a presumed earlier glitch before the observed glitch. 

\subsection{PSR J1702$-$4310}\label{sect:J1702-4310}

This pulsar was not previously known to glitch. Here, as shown in
Figure~\ref{fig:glt6}, we report our discovery of a glitch event. Our
observations suggest both exponential and linear decays following the
glitch and a linear decay with almost the same slope preceding the
glitch. Fitting for the exponential decay gave $Q\sim0.023$ and
$\tau_{\rm d}\sim96$\,d. A data gap of $\sim3$\,yr exists in the
pre-glitch data span. However, there is little period noise and no
phase ambiguity across this gap.

\subsection{PSR J1709$-$4429 (PSR B1706$-$44)}\label{sect:J1709-4429}

In Figure~\ref{fig:glt6}, we present the evolution of $\nu$ and
$\dot{\nu}$ of this pulsar for $\sim20$\,yr. Four glitches were
detected. These glitches are large, with $\Delta\nu_{\rm
  g}/\nu>1\times10^{-6}$. \citet{jml+95} and \citet{wjm+10} have
reported glitches 1 and 4, but no post-glitch recoveries were
reported. As Figure~\ref{fig:glt6} shows, all four glitches show
significant post-glitch recoveries, most with both exponential and
linear components. Dramatic slope changes in the linear recoveries are
seen after each glitch (Table~\ref{tab:gltspin}). Just a small
fraction of each glitch recovers exponentially, with time constants
$\sim 100$~d (Table~\ref{tab:glt}).

\subsection{PSR J1718$-$3825}\label{sect:J1718-3825}

This pulsar was not previously reported to glitch. Figure
\ref{fig:glt6} shows the evolution of $\nu$ and $\dot\nu$ for
$\sim$13\,yr. A glitch was detected at MJD $\sim 54910$. This event is
small with $\Delta\nu_{\rm g}/\nu\sim2\times10^{-9}$.

\begin{figure*} 
\begin{tabular}{cc}
\includegraphics[width=5.5cm,angle=-90]{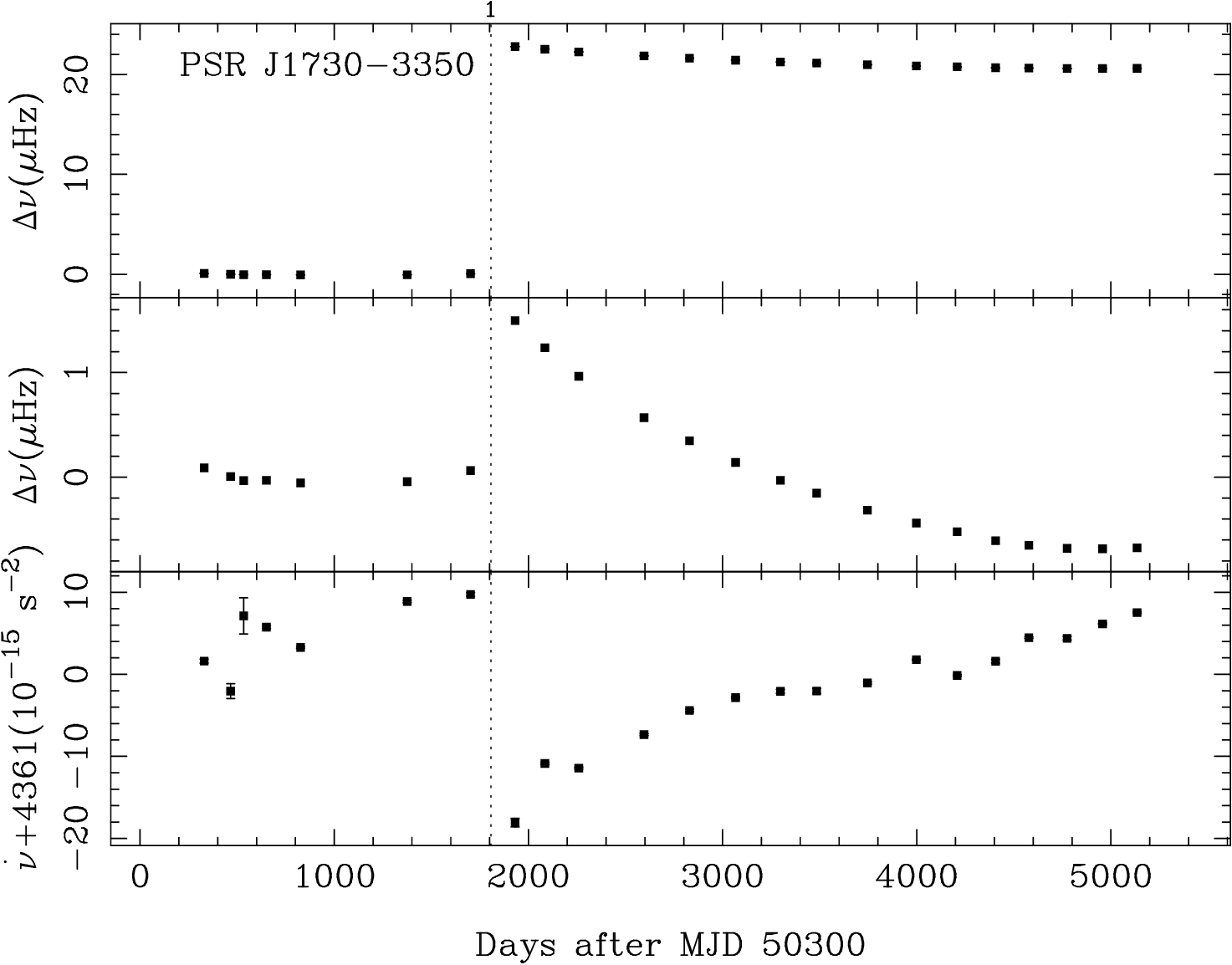} &
\includegraphics[width=5.5cm,angle=-90]{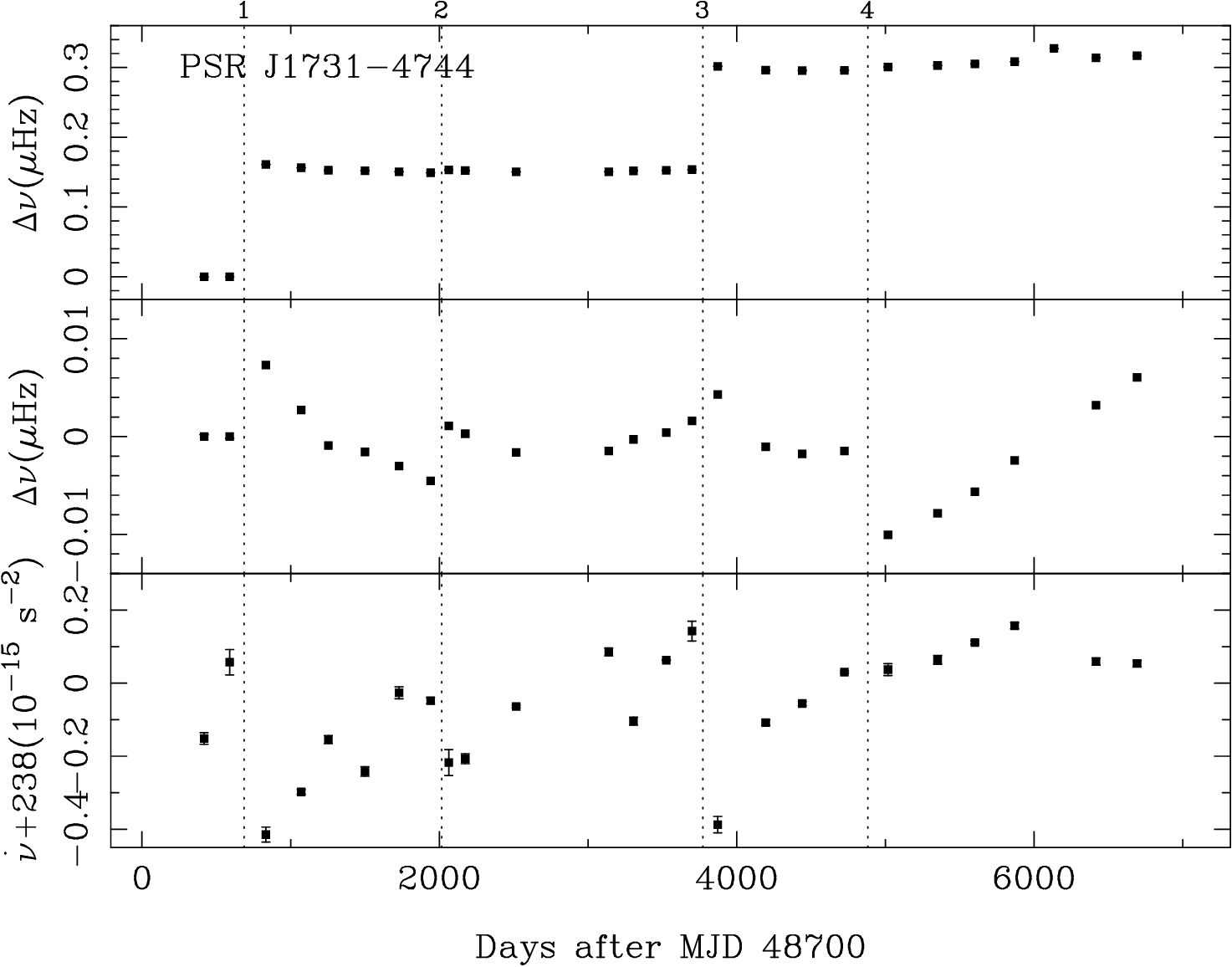} \\ 
\includegraphics[width=5.5cm,angle=-90]{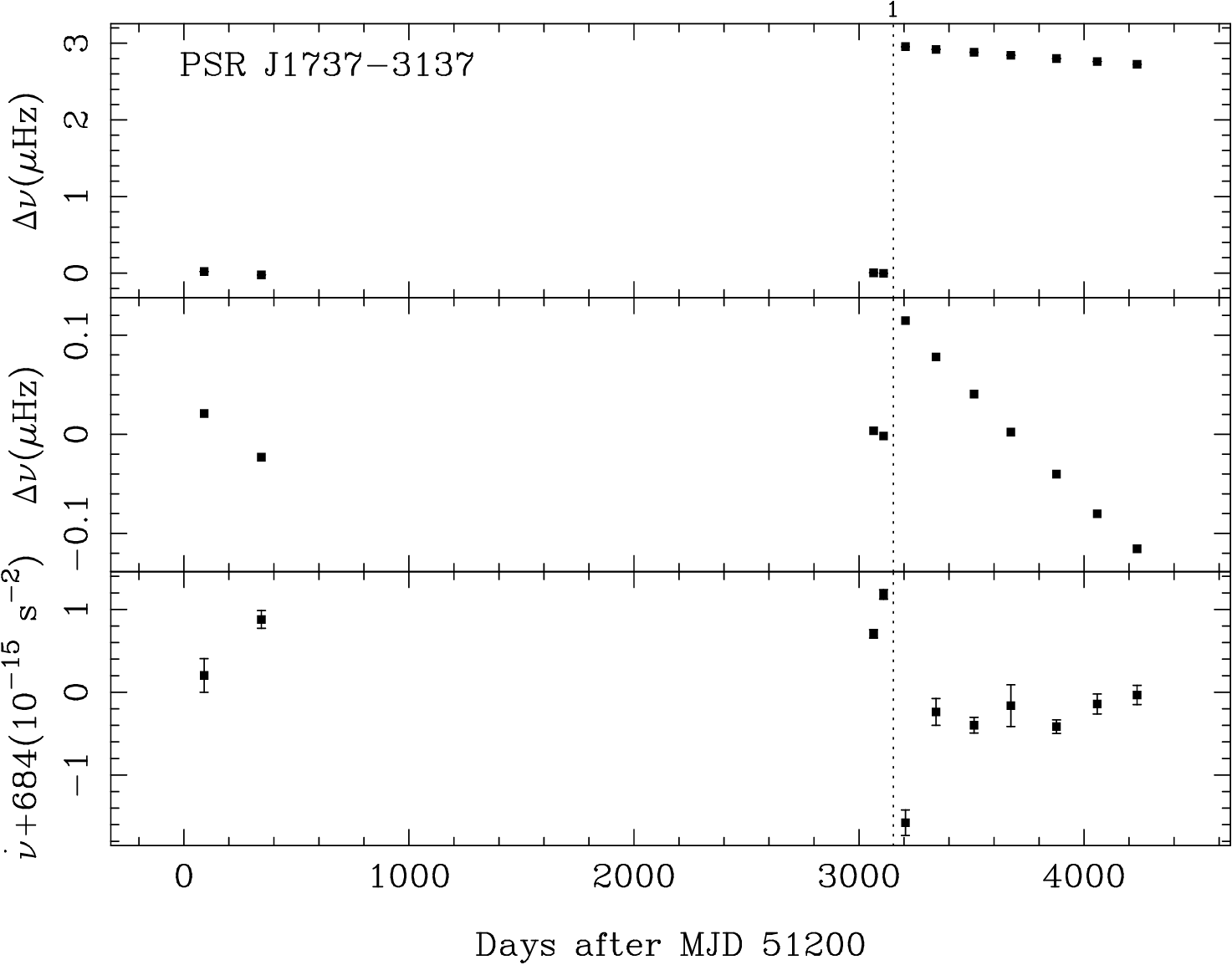} & 
\includegraphics[width=5.5cm,angle=-90]{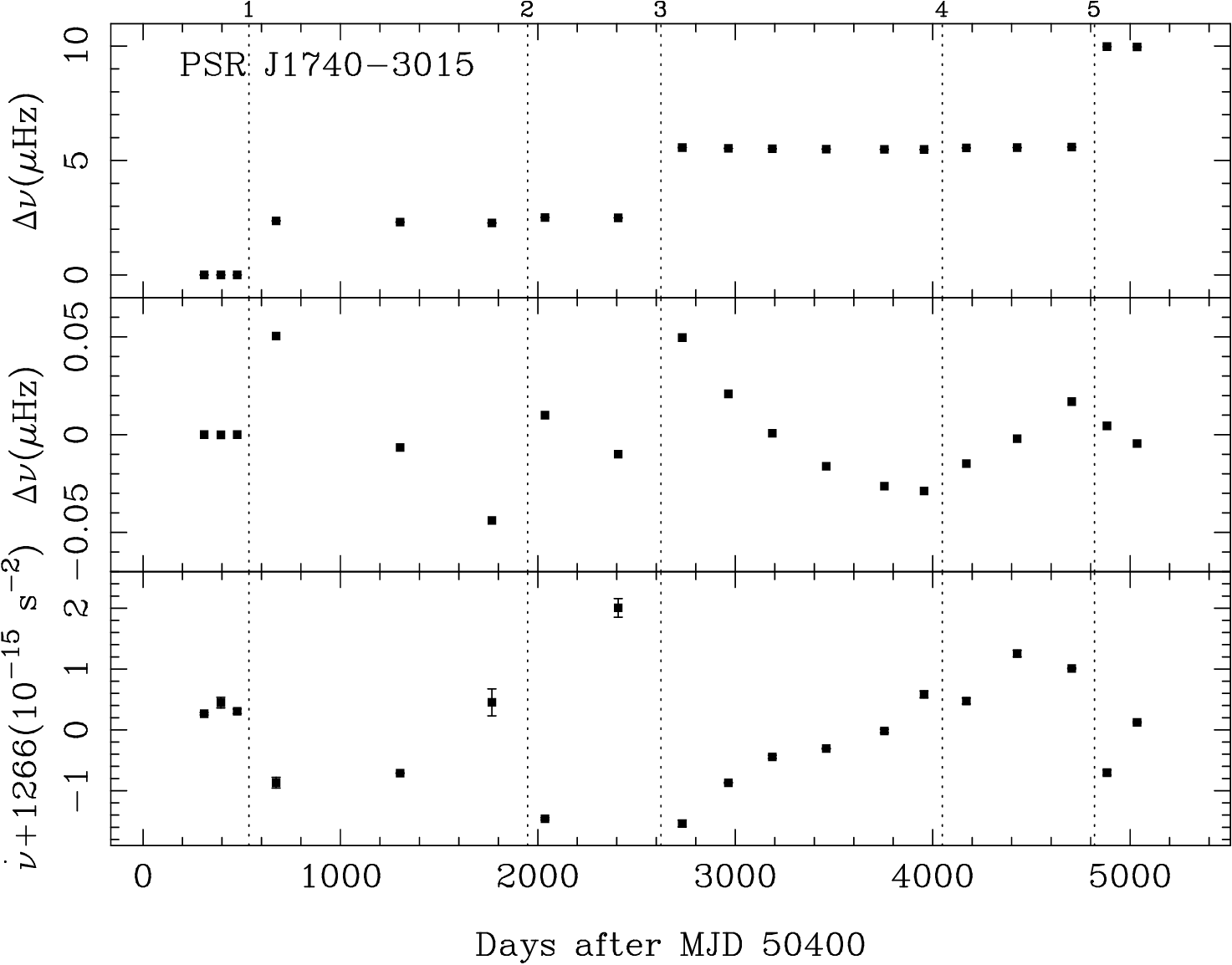} \\
\end{tabular} 
\caption{Glitches in PSRs J1730$-$3350, J1731$-$4744, J1737$-$3137 and
  J1740$-$3015. See Figure~\ref{fig:glt1} for a description
  of each sub-plot.}\label{fig:glt7}
\end{figure*}

\subsection{PSR J1730$-$3350 (PSR B1727$-$33)}\label{sect:J1730-3350}

Two glitch events that occurred at MJDs $\sim48000$ and $\sim52107$
were previously detected for this pulsar \citep{jml+95,elsk11}. Both
are large, with $\Delta\nu_{\rm g}/\nu>3\times10^{-6}$.
Figure~\ref{fig:glt7} shows the MJD $\sim52017$ glitch. There is a
clear linear recovery following the glitch and a small exponential
recovery with $Q\sim 0.01$ and timescale of $\sim 100$~d.

\subsection{PSR J1731$-$4744 (PSR B1727$-$47)}\label{sect:J1731-4744}

Three glitches have been reported for this pulsar
\citep{dm97,wmp+00,elsk11}. Figure~\ref{fig:glt7} shows these numbered
1, 2 and 3 and a newly detected glitch 4. Glitches 1 and 3 have
$\Delta\nu_{\rm g}/\nu\sim1\times10^{-7}$, whereas glitches 2 and 4
have $\Delta\nu_{\rm g}/\nu\sim3\times10^{-9}$. For glitch 3, we
measured an exponential recovery, with obtaining $Q\sim0.073$ and
$\tau_{\rm d}\sim210$\,d. There is some evidence for linear recoveries
following each glitch \citep[cf.,][]{wmp+00}. The measured values of
$\ddot\nu$ are small, ranging from $1\times 10^{-24}$~s$^{-3}$ to
$20\times 10^{-24}$~s$^{-3}$. For glitch 2, \citet{wmp+00} reported
the glitch epoch as MJD 50703(5). We obtained a more accurate epoch
MJD 50715.9(8) by solving for phase continuity across the glitch.

\subsection{PSR J1737$-$3137}\label{sect:J1737-3137}

Since the end of the Multibeam Pulsar Survey timing \citep{mhl+02},
this source was not observed until the commencement of the {\it Fermi}
project, leaving a data gap of $\sim7$\,yr. Three glitches have been
reported for this pulsar \citep{wjm+10,elsk11}. We detected the most
recent glitch (Figure~\ref{fig:glt7}). This glitch is large, with
$\Delta\nu_{\rm g}/\nu\sim10^{-6}$. There is evidence for a signficant
change in $\dot\nu$ at the time of the glitch, and maybe exponential
and linear recoveries, but the available data are insufficient to be
sure. 

\subsection{PSR J1740$-$3015 (PSR B1737$-$30)}\label{sect:J1740-3015}

PSR~J1740$-$3015 is one of the most frequently glitching pulsars
known. Previously, a total of 31 glitches were detected over a 25-yr
data span. Figure~\ref{fig:glt7} presents the evolution of $\nu$ and
$\dot{\nu}$ for $\sim13$\,yr from Parkes timing. During this period,
this pulsar was found to have undergone 17 glitches
\citep{ura02,klgj03,js06,zwm+08,ywml10,wjm+10,elsk11}. However,
because of the frequent glitching and the relatively poor sampling of
our observations, only five major glitches can be detected. The first
four have been published. Glitch 5 is a new detection; it has a
fractional size of $\sim2.7\times10^{-6}$, which is the largest ever
seen in this pulsar. Linear recoveries are seen after most of the
large glitches, including glitch 5, but we have no evidence for
exponential recoveries. If such recoveries exist, they must have
timescales of less than a few tens of days.

\begin{figure*} 
\begin{tabular}{cc}
\includegraphics[width=5.5cm,angle=-90]{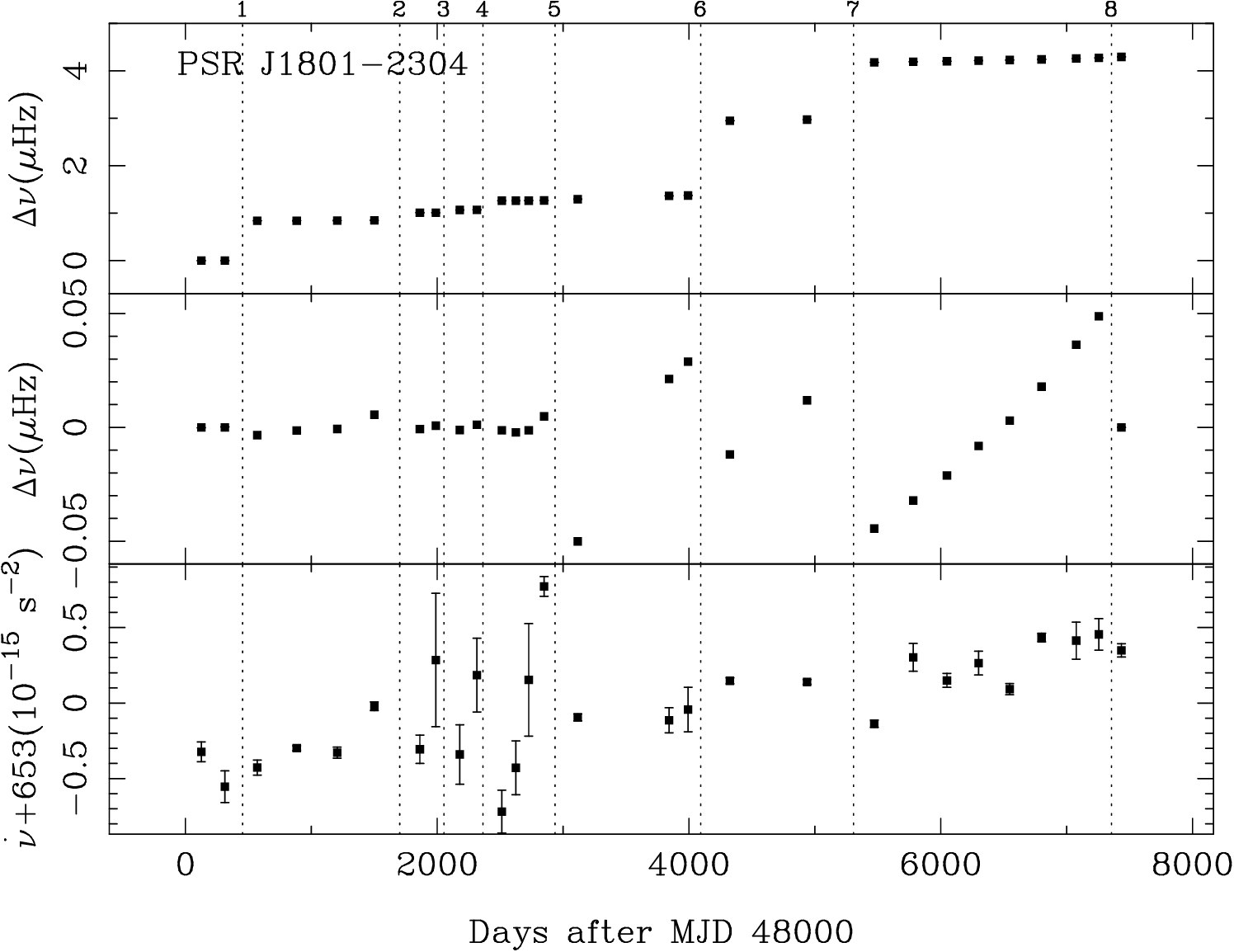} &
\includegraphics[width=5.5cm,angle=-90]{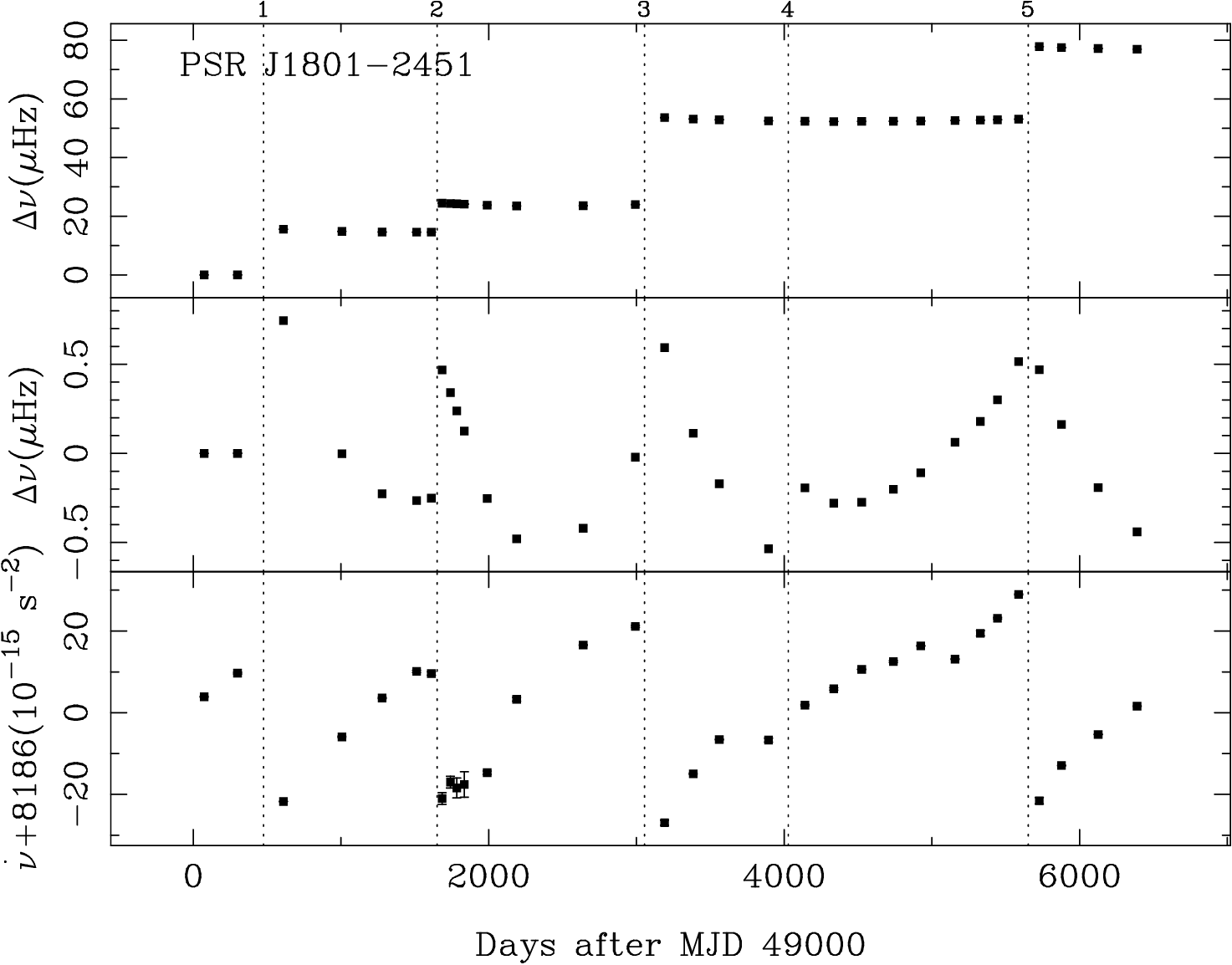} \\ 
\includegraphics[width=5.5cm,angle=-90]{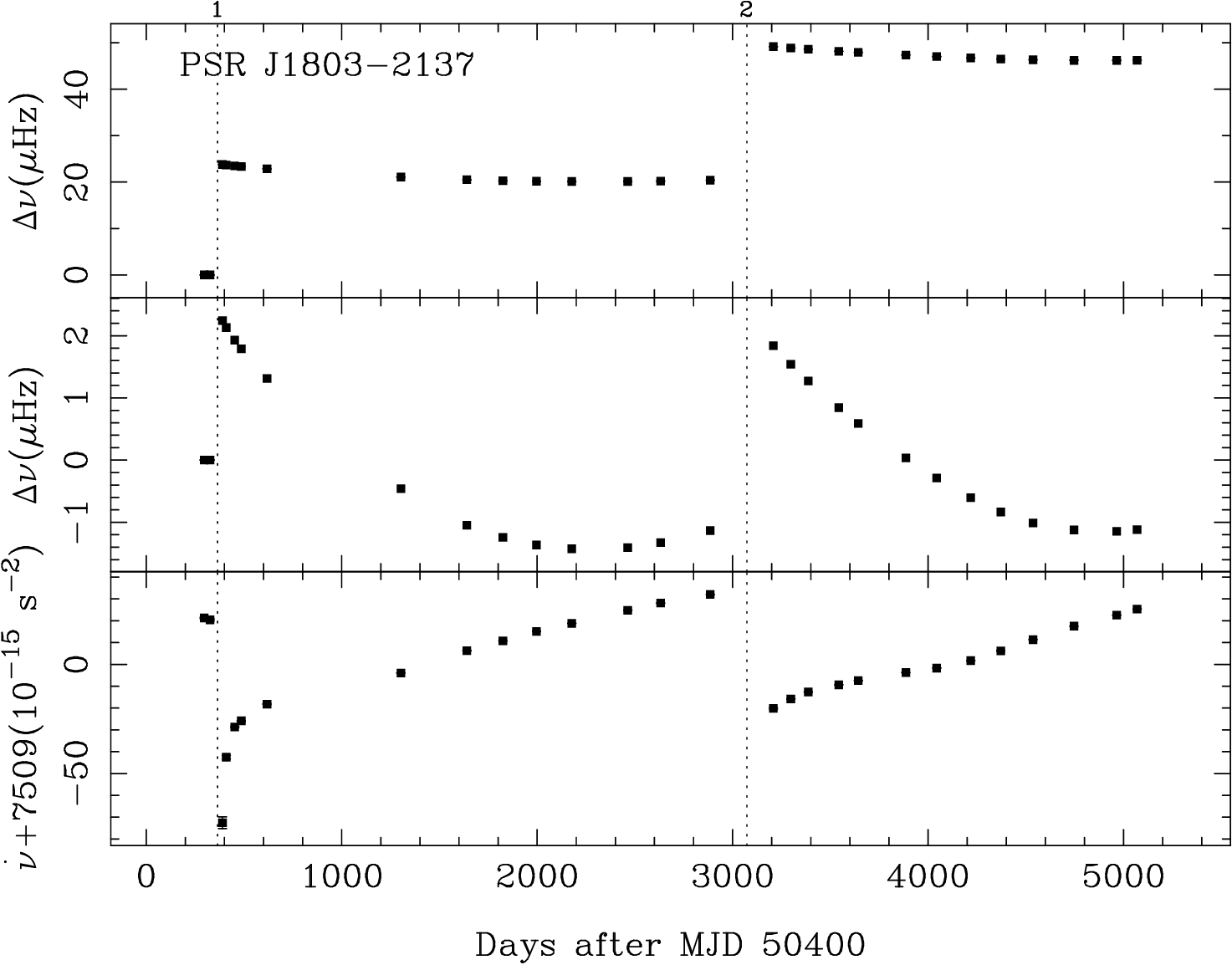} & 
\includegraphics[width=5.5cm,angle=-90]{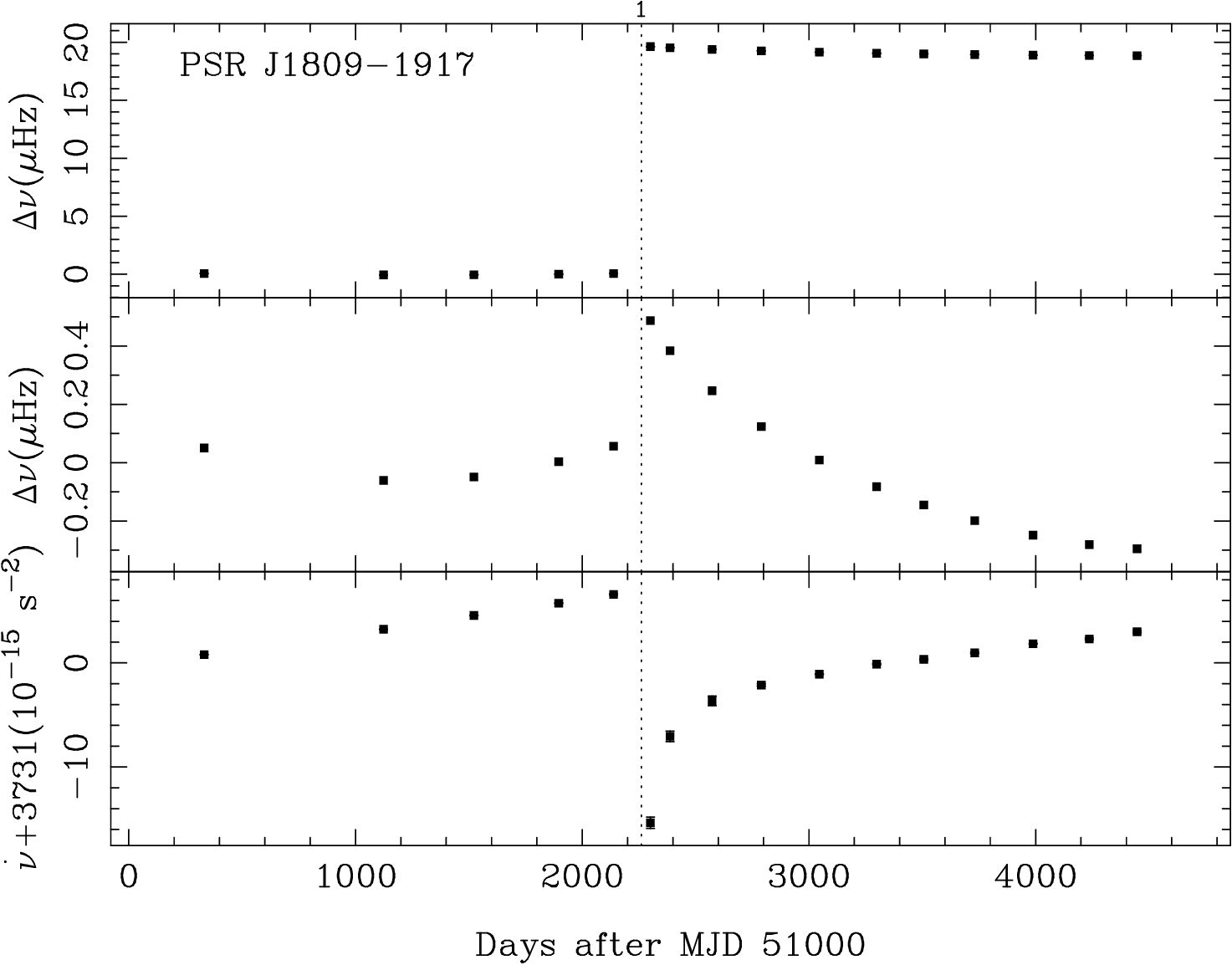} \\
\end{tabular} 
\caption{Glitches in PSRs J1801$-$2304, J1801$-$2451, J1803$-$2137 and
  J1809$-$1917. See Figure~\ref{fig:glt1} for a description
  of each sub-plot.}\label{fig:glt8}
\end{figure*}

\subsection{PSR J1801$-$2304 (PSR B1758$-$23)}\label{sect:J1801-2304}

PSR~J1801$-$2304 has been found to have suffered nine glitches in
$\sim24$\,yr \citep{klm+93,sl96,wmp+00,klgj03,ywml10,elsk11}. Figure
\ref{fig:glt8} presents the variations of $\nu$ and $\dot{\nu}$ for
$\sim20$\,yr, where eight glitches are shown, the last being a new
discovery. This is a small glitch, with a fractional size
$\sim4\times10^{-9}$. For glitch 4, \citet{wmp+00} fit it with an
exponential recovery, assuming the timescale to be 100\,d. However,
as shown in Figure~\ref{fig:glt8}, our observations suggest that a
linear recovery with $\ddot\nu \sim 40\times 10^{-24}$\,s$^{-3}$
(Table~\ref{tab:gltspin}) dominates the post-glitch behaviour.

\subsection{PSR J1801$-$2451 (PSR B1757$-$24)}\label{sect:J1801-2451}

Figure~\ref{fig:glt8} shows the evolution of $\nu$ and $\dot{\nu}$ of
this pulsar for $\sim18$\,yr, where five glitches are shown. These
glitches have been reported previously
\citep{lkb+96,wmp+00,wjm+10,elsk11}. Figure \ref{fig:glt8} suggests
that the post-glitch recoveries are dominated by linear recoveries but
there may be exponential recoveries as well \citep{lkb+96,wmp+00}. In
this work, we fit exponential recoveries for glitches 3 and 5. For
glitch 3, we obtained $Q\sim0.025$ and $\tau_{\rm d}\sim200$~d and for
glitch 5, $Q\sim0.0065$ and $\tau_{\rm d}\sim 25$~d.

\subsection{PSR J1803$-$2137 (PSR B1800$-$21)}\label{sect:J1803-2137}

PSR~J1803$-$2137 is very similar to PSR~J1801$-$2451 in its timing
properties with a characteristic age of 15 kyr. Four glitches have
been detected in this pulsar
\citep{sl96,wmp+00,klgj03,ywml10,elsk11}. Figure~\ref{fig:glt8} shows
the two glitches detected in this work. Both of these show clear
long-term linear recoveries with a very similar slope
(Table~\ref{tab:gltspin}). Clear exponential recoveries with shorter
timescales are also seen. For glitch 1, \citet{wmp+00} fitted a short
timescale ($\sim18$~d) recovery and a $\sim850$~d one to the
longer-term decay. We found evidence in the phase residuals for two
short-term decays with timescales of $\sim12$\,d and $\sim69$\,d;
further relaxation is dominated by the linear recovery. For glitch 2,
\citet{ywml10} fit an exponential decay, obtaining $Q=0.009(2)$ and
$\tau_{\rm d}=120(20)$\,d. We obtained consistent results as $Q=
0.00630(16)$ and $\tau_{\rm d}=133(11)$\,d.

\subsection{PSR J1809$-$1917}\label{sect:J1809-1917}

This pulsar has previously been reported to glitch once
\citep{elsk11}. Figure~\ref{fig:glt8} shows the evolution of $\nu$ and
$\dot{\nu}$ for $\sim12$\,yr. The glitch is large, with $\Delta\nu_{\rm
  g}/\nu\sim1.6\times10^{-6}$. Both exponential and linear recoveries
can clearly be seen in our observations. Fitting showed that the
exponential recovery is characterised by $Q\sim0.006$ and $\tau_{\rm
  d}\sim125$\,d. A linear recovery, presumably from an earlier glitch,
is also seen before the glitch. The recovery after the observed glitch
is slightly less steep, but both have $\ddot{\nu}\sim
35\times10^{-24}$\,s$^{-3}$.

\begin{figure*} 
\begin{tabular}{cc}
\includegraphics[width=5.5cm,angle=-90]{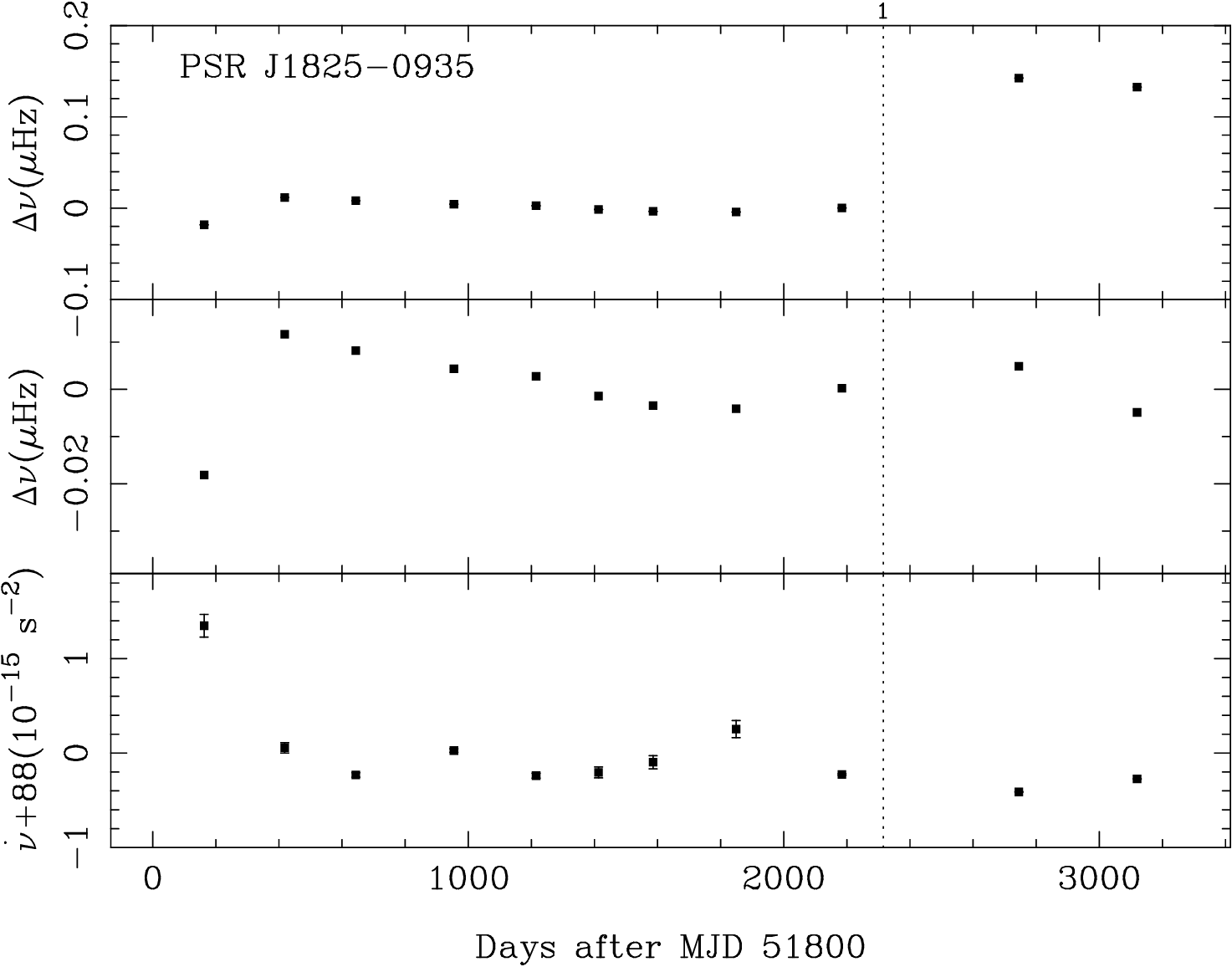} &
\includegraphics[width=5.5cm,angle=-90]{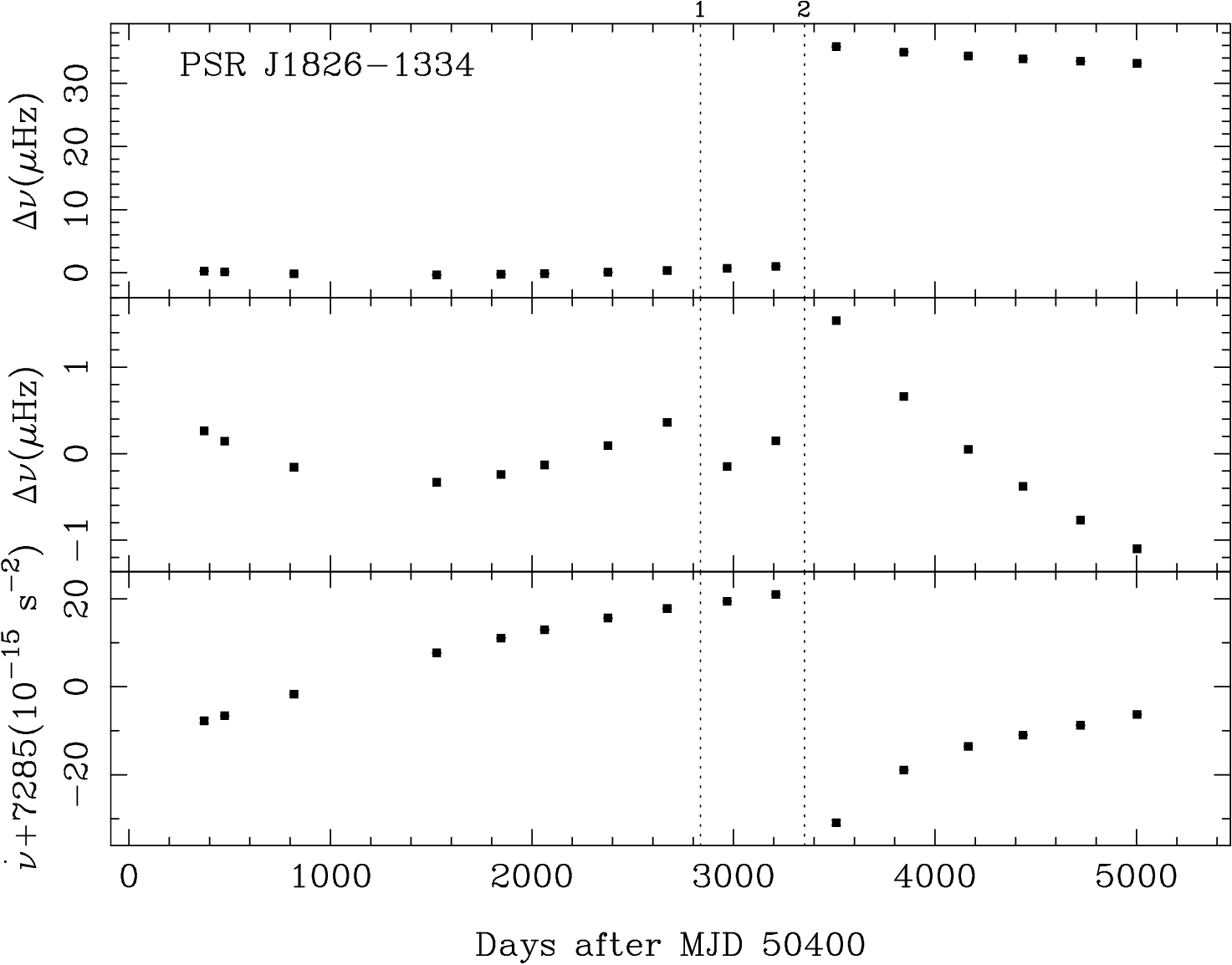} \\ 
\includegraphics[width=5.5cm,angle=-90]{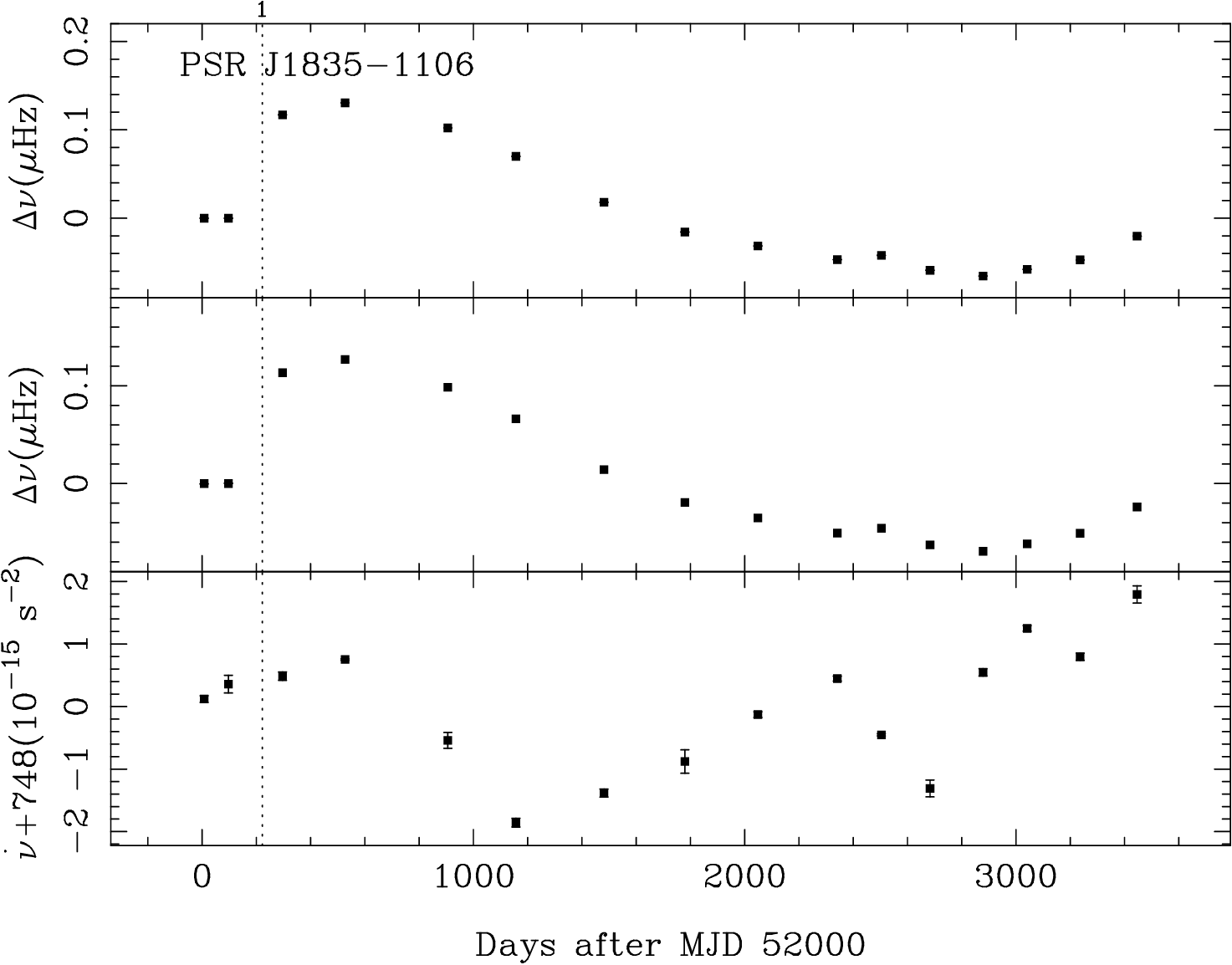} & 
\includegraphics[width=5.5cm,angle=-90]{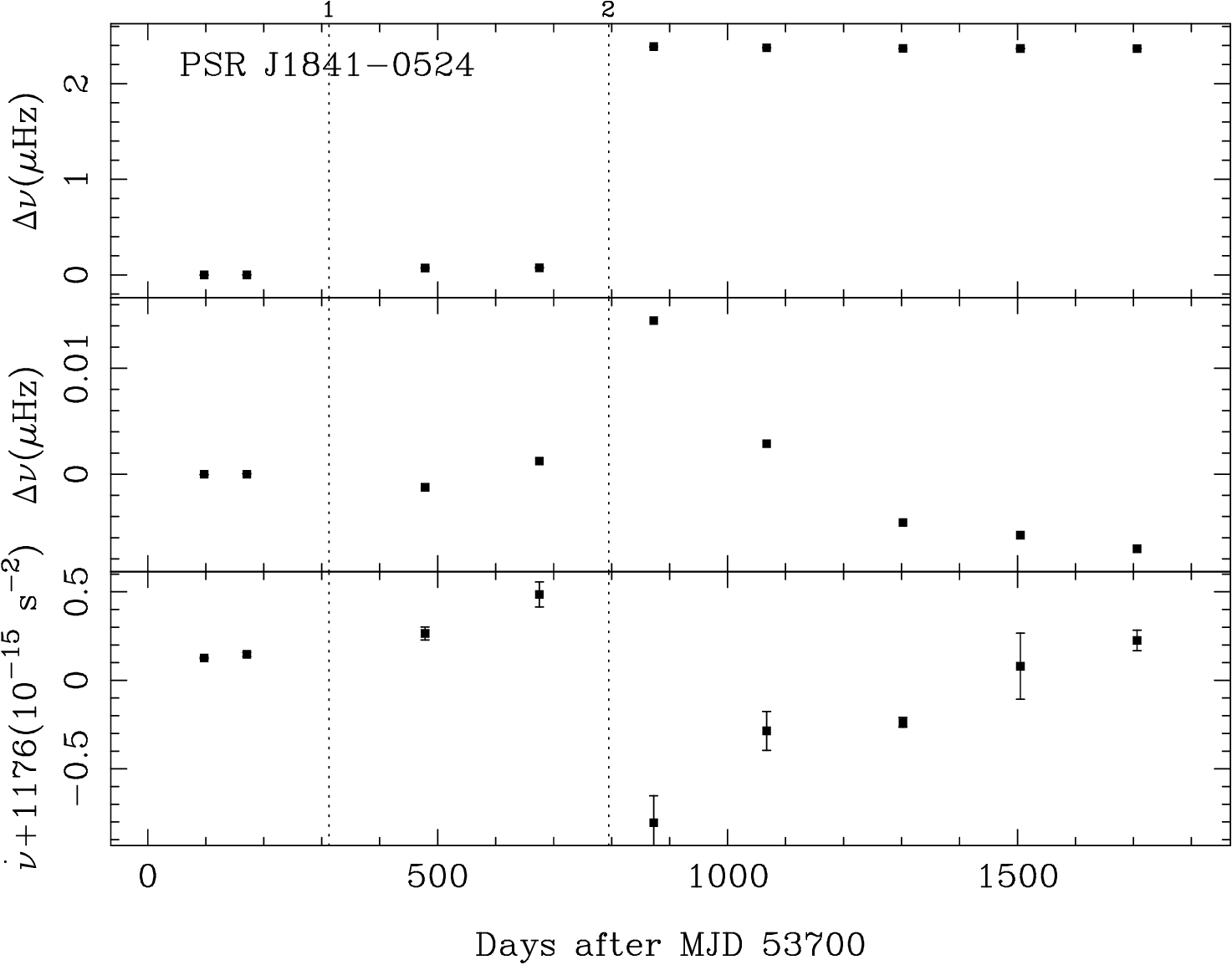} \\
\end{tabular} 
\caption{Glitches in PSRs J1825$-$0935, J1826$-$1334, J1835$-$1106 and
  J1841$-$0524. See Figure~\ref{fig:glt1} for a description
  of each sub-plot.}\label{fig:glt9}
\end{figure*}

\subsection{PSR J1825$-$0935 (PSR B1822$-$09)}\label{sect:J1825-0935}

So far, eight glitches have been detected for this pulsar; six were
identified as ``slow'' glitches
\citep{zww+04,sha07,ywml10,elsk11}. \citet{lhk+10} have shown that
these ``slow'' glitches probably are a manifestation of ``two-state''
magnetospheric switching. Figure~\ref{fig:glt9} presents our
observations of the evolution of $\nu$ and $\dot{\nu}$ for
$\sim9$\,yr. During this period, this pulsar was found to undergo
three slow glitches \citep{zww+04,sha07,ywml10}. Because of
insufficient observations, these slow glitches were missed in our
data. By using the available data, we were able to identify a
``normal'' glitch with $\Delta\nu_{\rm
  g}/\nu\sim1.3\times10^{-7}$. However, a data gap of $\sim500$\,d
exists shortly after this glitch, making it impossible to study the
post-glitch behaviour. This glitch was observed in more detail by the
Xinjiang group \citep{ywml10}.

\subsection{PSR J1826$-$1334 (PSR B1823$-$13)}\label{sect:J1826-1334}

Five glitch events have been reported for this pulsar
\citep{sl96,ywml10,elsk11}. Figure~\ref{fig:glt9} presents the
evolution of $\nu$ and $\dot{\nu}$ for $\sim13$\,yr. For glitch 1, our
last pre-glitch observation was at MJD $\sim53186$, and the first
post-glitch observation was at MJD $\sim53279$. \citet{elsk11} showed
that there are actually two small glitches in this interval. One was
at MJD $\sim53206$, with $\Delta\nu_{\rm g}/\nu\sim0.6\times10^{-9}$
and the other was at MJD $\sim53259$, with $\Delta\nu_{\rm
  g}/\nu\sim3\times10^{-9}$. The much larger glitch 2 was also
observed by both \citet{ywml10} and \citet{elsk11}. We detected an
exponential recovery for this glitch, characterised by $Q\sim0.007$
and $\tau_{\rm d}\sim80$\,d.

\subsection{PSR J1835$-$1106}\label{sect:J1835-1106}

Figure~\ref{fig:glt9} shows the evolution of $\nu$ and $\dot{\nu}$ for
PSR~J1835$-$1106 over about 10\,yr. A small glitch with $\Delta\nu_{\rm
  g}/\nu\sim1.6\times10^{-8}$ was detected at MJD $\sim 52220$. This
event was also observed by \citet{zww+04} and \citet{elsk11}. As shown
in Figure~\ref{fig:glt9}, the post-glitch frequency residuals exhibit
a cubic structure, indicating a measurable $\dddot{\nu}$; fitting
showed that this term is $1.58(13)\times10^{-31}$\,s$^{-4}$. These
signficant higher-order frequency derivatives indicate the presence of
noise processes in the pulsar rotation. It is possible that this
pulsar also has a two-state magnetospheric modulation affecting the
value of $\dot\nu$ \citep{lhk+10}. 

\subsection{PSR J1841$-$0524}\label{sect:J1841-0524}

This pulsar has been found to undergo three glitches
\citep{elsk11}. The first two are small, with $\Delta\nu_{\rm
  g}/\nu\gtrsim 2\times10^{-9}$. The latest one is large, with
$\Delta\nu_{\rm g}/\nu\sim10^{-6}$. Parkes data have a gap, spanning
from MJD $\sim52570$ to MJD $\sim53619$. As a result, the MJD
$\sim53562$ glitch was missed. Figure~\ref{fig:glt9} presents the
evolution of $\nu$ and $\dot{\nu}$ for the past five years. At least
for glitch 2, the post-glitch behaviour exhibits a linear recovery.

\section{Discussion}\label{sect:dis}

\begin{figure}
\begin{center}
\begin{tabular}{c}
\includegraphics[width=5.4cm,angle=-90]{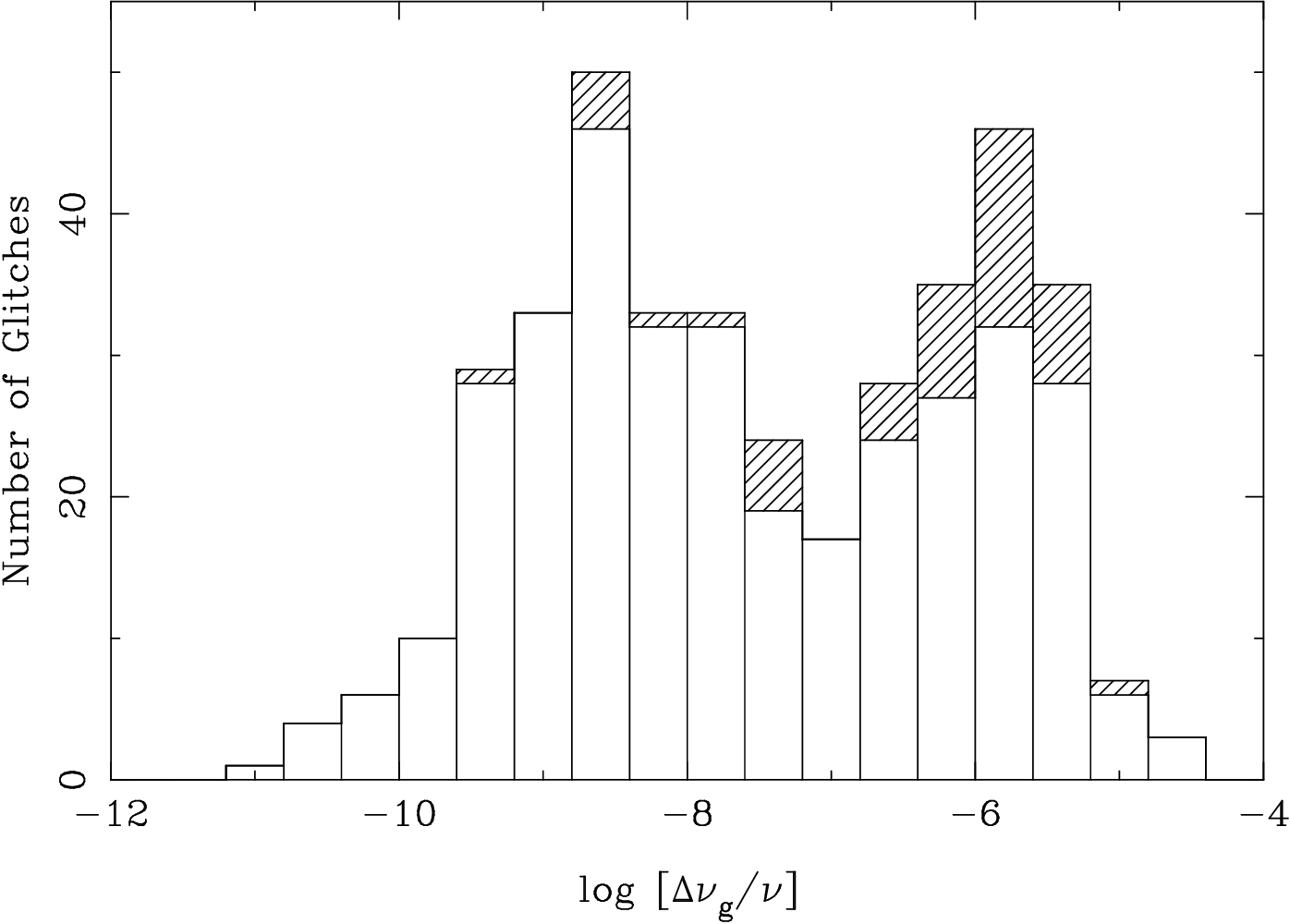} \\
\includegraphics[width=5.5cm,angle=-90]{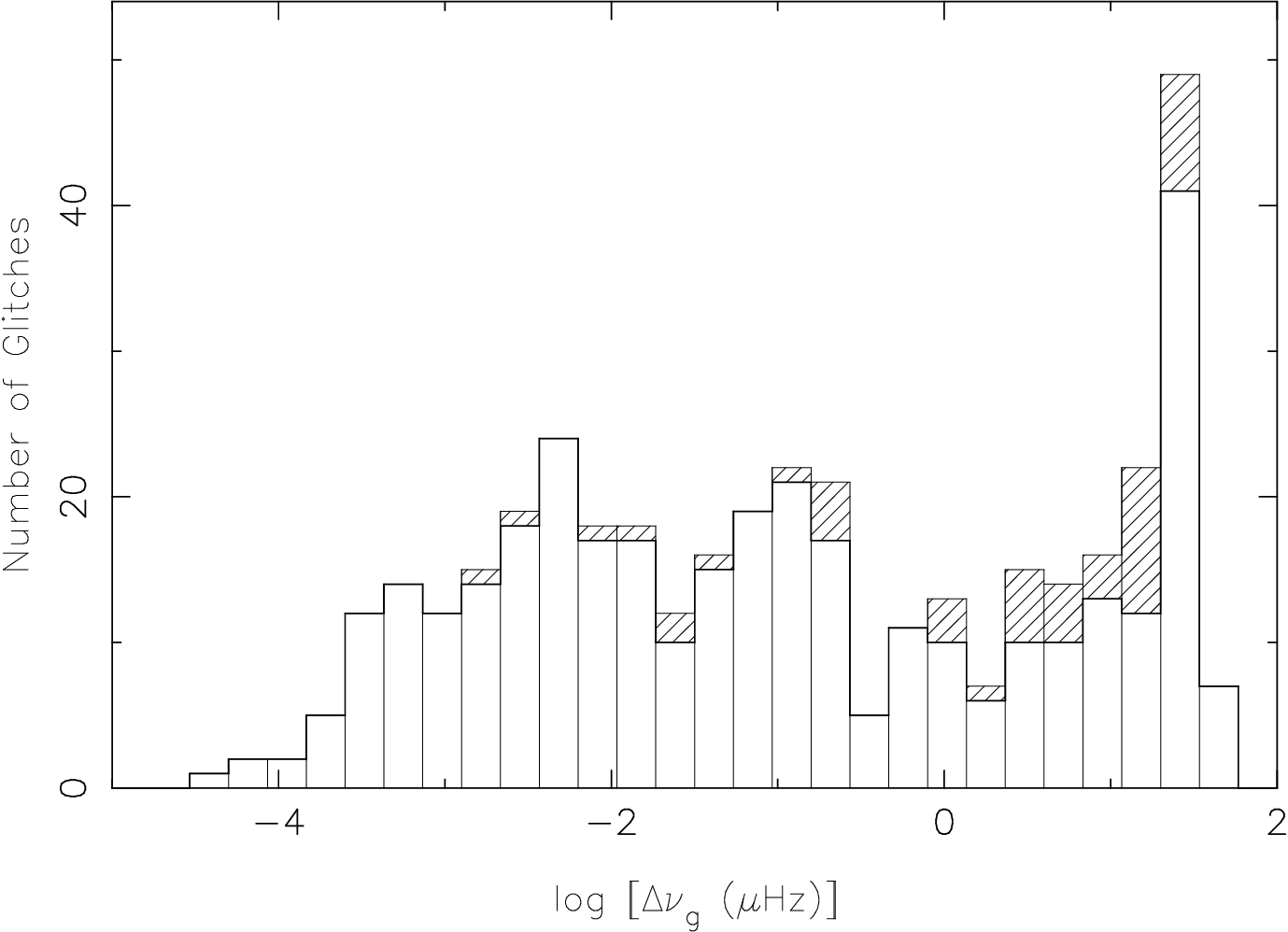}
\end{tabular}
\caption{Histograms of the distributions of glitch fractional size
  $\Delta\nu_{\rm g}/\nu$ (upper panel) and of glitch size
  $\Delta\nu_{\rm g}$ (lower panel). The blank bars show the results
  from the previous work, while the shaded bars show the new
  detections from this work added on top of the previous
  results. For previously reported glitches, we use the ATNF
  glitch database which includes magnetar glitches.}\label{fig:glsize_hist}
\end{center}
\end{figure}

We have searched for glitch events in the timing residuals of 165
pulsars. Out of these, 107 glitches were identified with 46 new
discoveries.  Most of these new discoveries occur for
southern-hemisphere pulsars that cannot be observed by the long-term
monitoring programmes carried out in the northern hemisphere. Because
of the relatively large gaps between many of our observations, there
are 22 known glitches that are undetectable in our data sets. These
missed glitches generally have fractional sizes $\Delta\nu_{\rm
  g}/\nu$ between $10^{-10}$ and $10^{-9}$. The measurement of
$\Delta\dot{\nu}_{\rm g}$ is also very dependent on the data sampling
as short-term transients can easily be missed or underestimated and in
some cases our results differ from those in the literature.

In general, the post-glitch behaviour shows two types of recovery: a
short-term exponential recovery (characterised by $Q$ and $\tau_{\rm
  d}$) and a longer-term linear recovery in $\dot\nu$ (characterised by
$\ddot{\nu}$). Both can be identified by inspecting the evolution of
$\dot{\nu}$. In most cases, we have insufficient observations to study
any exponential recoveries with timescale $\lesssim$20\,d. Most such
short-term exponential recoveries will have been missed. However, we
found that 27 glitches do show measurable post-glitch exponential
recovery with time constants between 12\,d and 300\,d. For
more than 90\% of the observed glitches, values of
post-glitch $\ddot{\nu}$, indicating a linear recovery in $\dot\nu$,
were obtained. For many of these, the $\dot\nu$ slope was similar
before and after the glitch, so the value of $\Delta\ddot{\nu}_{\rm p}$
was not significant. For 13 glitches, the slope change was larger
and a significant value for $\Delta\ddot{\nu}_{\rm p}$ was obtained.

The discovery of 46 new glitches allows further study on the
distribution of glitches and their post-glitch behaviour. The
discussion on these two aspects is presented in \S\ref{sect:glDistrn}
and \S\ref{sect:glRec} respectively.

\subsection{The distribution of glitches}\label{sect:glDistrn}

In Figure~\ref{fig:glsize_hist}, the upper panel is a histogram of the
fractional glitch size $\Delta\nu_{\rm g}/\nu$. Our results reinforce
the bimodal distribution of the observed fractional glitch sizes
previously reported by numerous authors
\citep[e.g.,][]{wmp+00,ywml10,elsk11}. The first peak in this
distribution lies around $2\times10^{-9}$ and the second around
$10^{-6}$. Our observations mainly contribute to the second
peak. Because of our rather infrequent sampling, it is very difficult
for us to detect glitches with $\Delta\nu_{\rm
  g}/\nu\lesssim10^{-9}$. As noted by \citet{elsk11} and others, the
left edge of the distribution is strongly limited by observational
selection. The actual number of small glitches could be large and the
lower peak in Figure~\ref{fig:glsize_hist} may not even exist in the
intrinsic distribution. However the dip at $\Delta\nu_{\rm g}/\nu \sim
10^{-7}$ is clearly real and suggests that there may be two mechanisms
that can induce a glitch event. As previously mentioned in
\S\ref{sect:intro}, it has been proposed that starquakes caused by the
cracking of stellar crust may generate small glitches, whereas large
glitches may result from the sudden transfer of angular momentum from
a crustal superfluid to the rest of the star.

The fractional glitch size is affected both by the
size of the glitch and the pulse frequency of the pulsar. In the lower
panel of Figure~\ref{fig:glsize_hist}, we plot a histogram of the
frequency jump $\Delta\nu_{\rm g}$. As shown in the figure, the
distribution of $\Delta\nu_{\rm g}$ also has a bimodal distribution or
at least has a dip between the large and small glitches. It is
interesting that the peak for large glitches is much narrower in
$\Delta\nu_{\rm g}$ than it is in $\Delta\nu_{\rm g}/\nu$, whereas the
converse is true for the lower-frequency peak.  A large fraction of
the peak at the high end comes from just two pulsars, PSR J0537$-$6910
\citep{mmw+06} and the Vela pulsar. These two pulsars have frequent
glitches and most of them have $\Delta\nu_{\rm g}\sim 20\,\mu$Hz. The
pulse frequencies however differ by a factor of about six.

\begin{figure}
\begin{center}
\includegraphics[angle=-90,width=7cm]{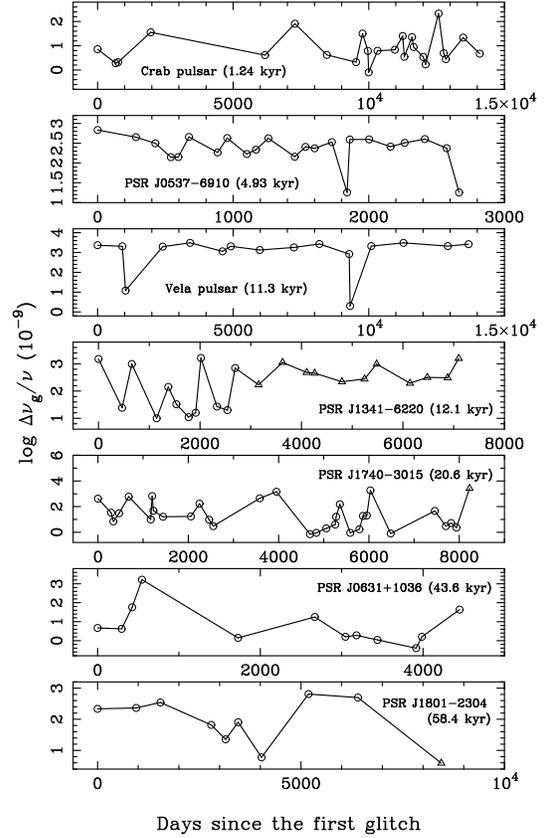}
\end{center}
\caption{The time series of glitch events for the seven pulsars where
    ten or more glitches were detected. Data from the ATNF Pulsar
  Catalogue glitch table are indicated by circles, while those from
  this work are indicated by triangles. For each pulsar, its
  characteristic age is shown in parentheses.}\label{fig:glseries}
\end{figure}

Figure \ref{fig:glseries} shows the time sequence of glitch fractional
sizes for seven pulsars where ten or more glitches were detected. This
figure shows that the bimodal distribution of glitch sizes may be seen
in individual pulsars as well. For example, most glitches in
PSR~J0537$-$6910 and the Vela pulsar (PSR~B0833$-$45) are large and
similar in amplitude, but much smaller glitches were occasionally
seen. Although not quite so clear, similar behaviour is seen in
PSRs~J1341$-$6220, J1740$-$3015, J0631$+$1036 (many small glitches and
only one large glitch) and J1801$-$2304. It is evident from Figure
\ref{fig:glseries} that we detected fewer small glitches in
PSR~J1341$-$6220 compared to earlier results. This may be because the
frequent occurrence of larger glitches obscured some small glitches in
our data set which is not as well sampled as the earlier ones. 

\begin{figure*}
\begin{center}
\begin{tabular}{cc}
\includegraphics[width=7cm,angle=-90]{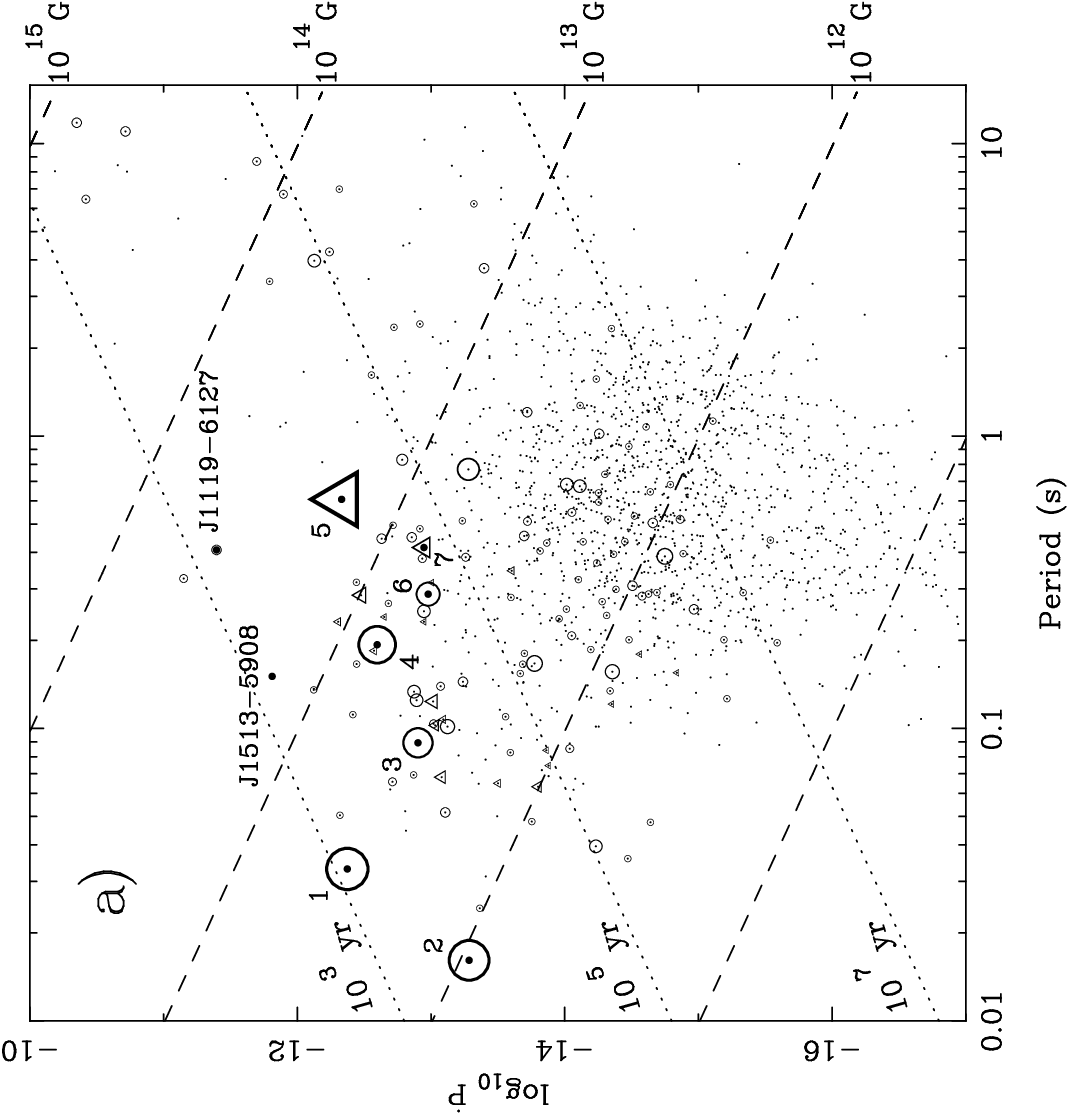} &
\includegraphics[width=7cm,angle=-90]{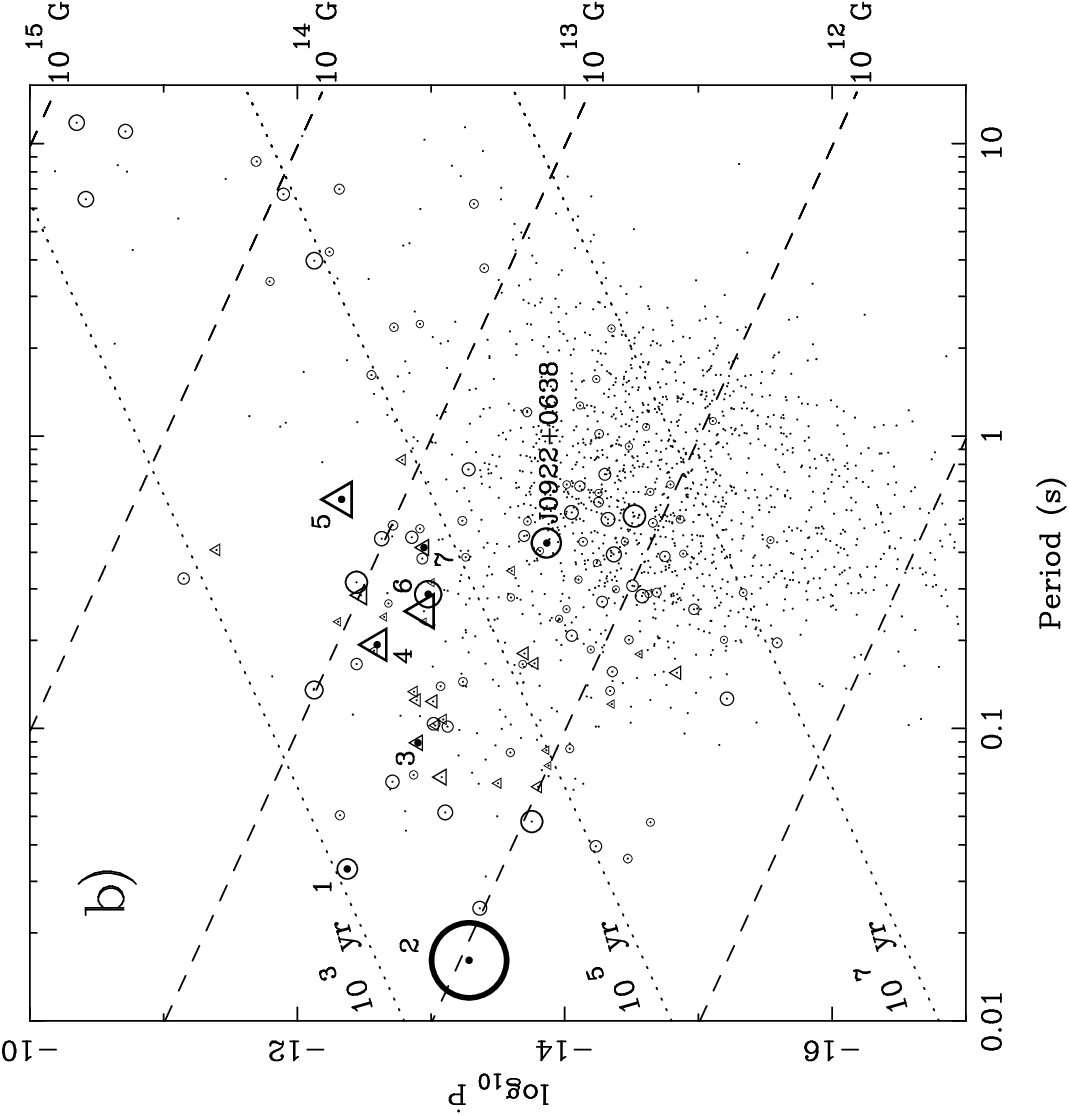} \\
\includegraphics[width=7cm,angle=-90]{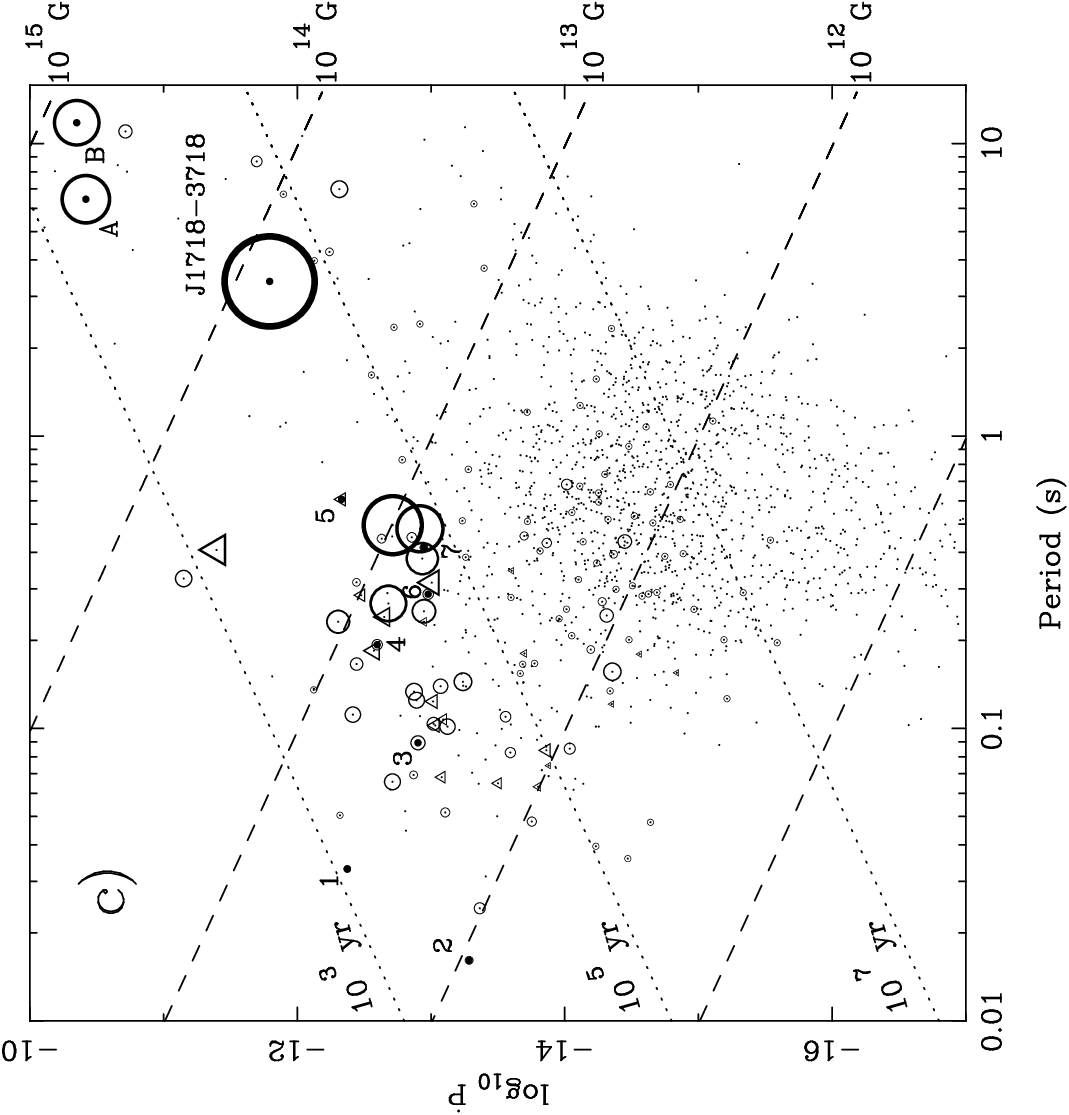} &
\includegraphics[width=7cm,angle=-90]{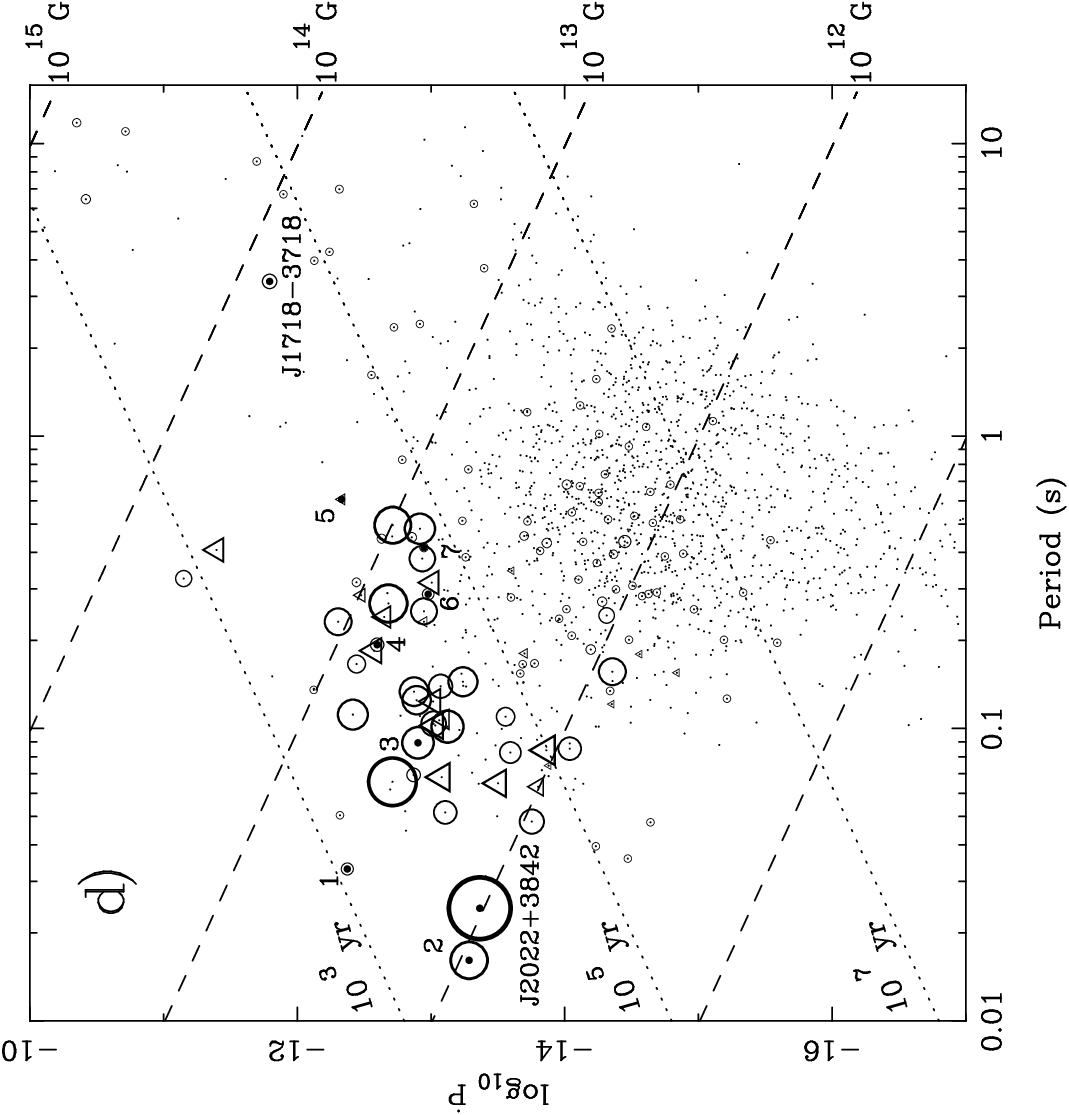} \\
\includegraphics[width=7cm,angle=-90]{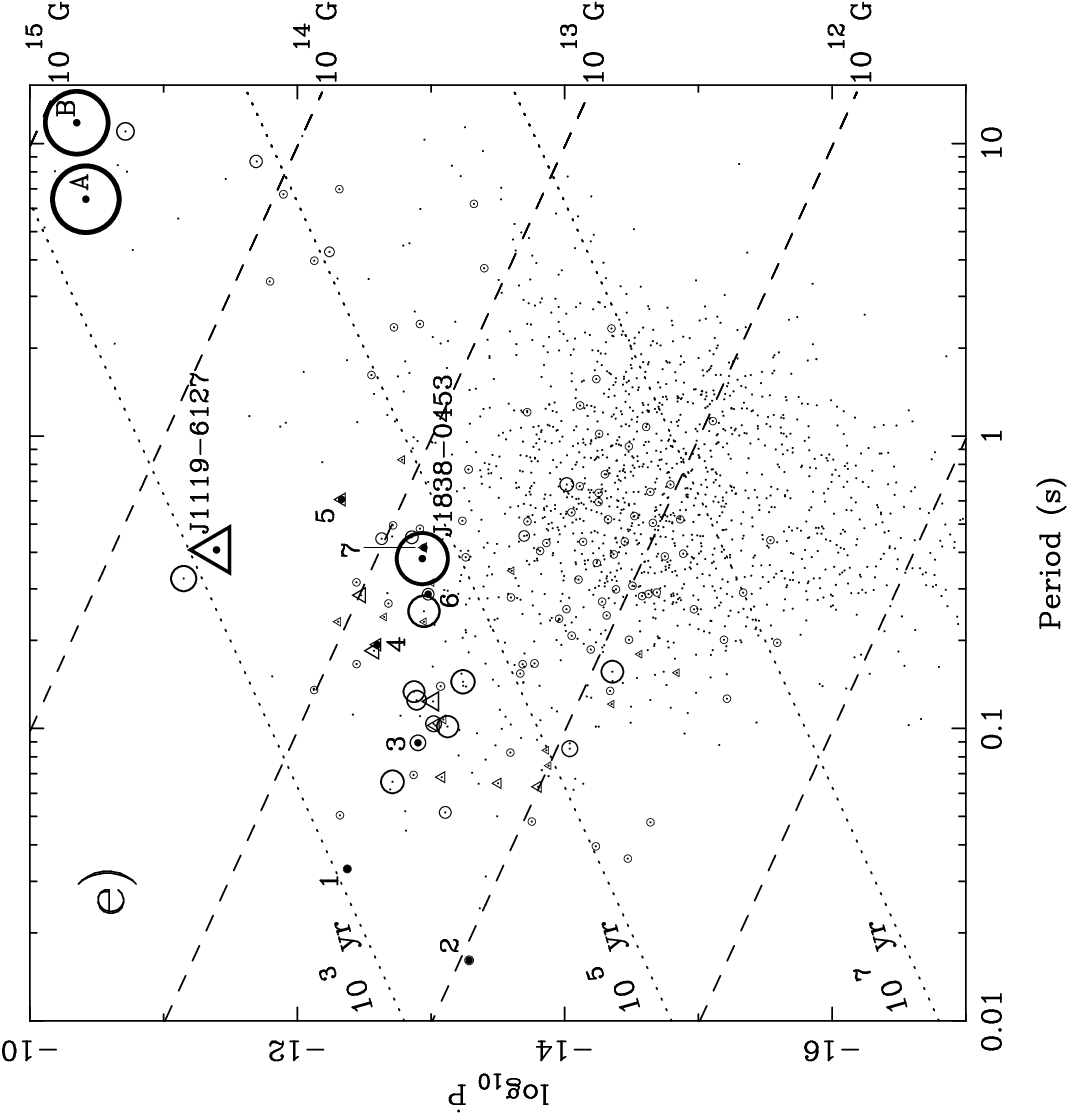} &
\includegraphics[width=7cm,angle=-90]{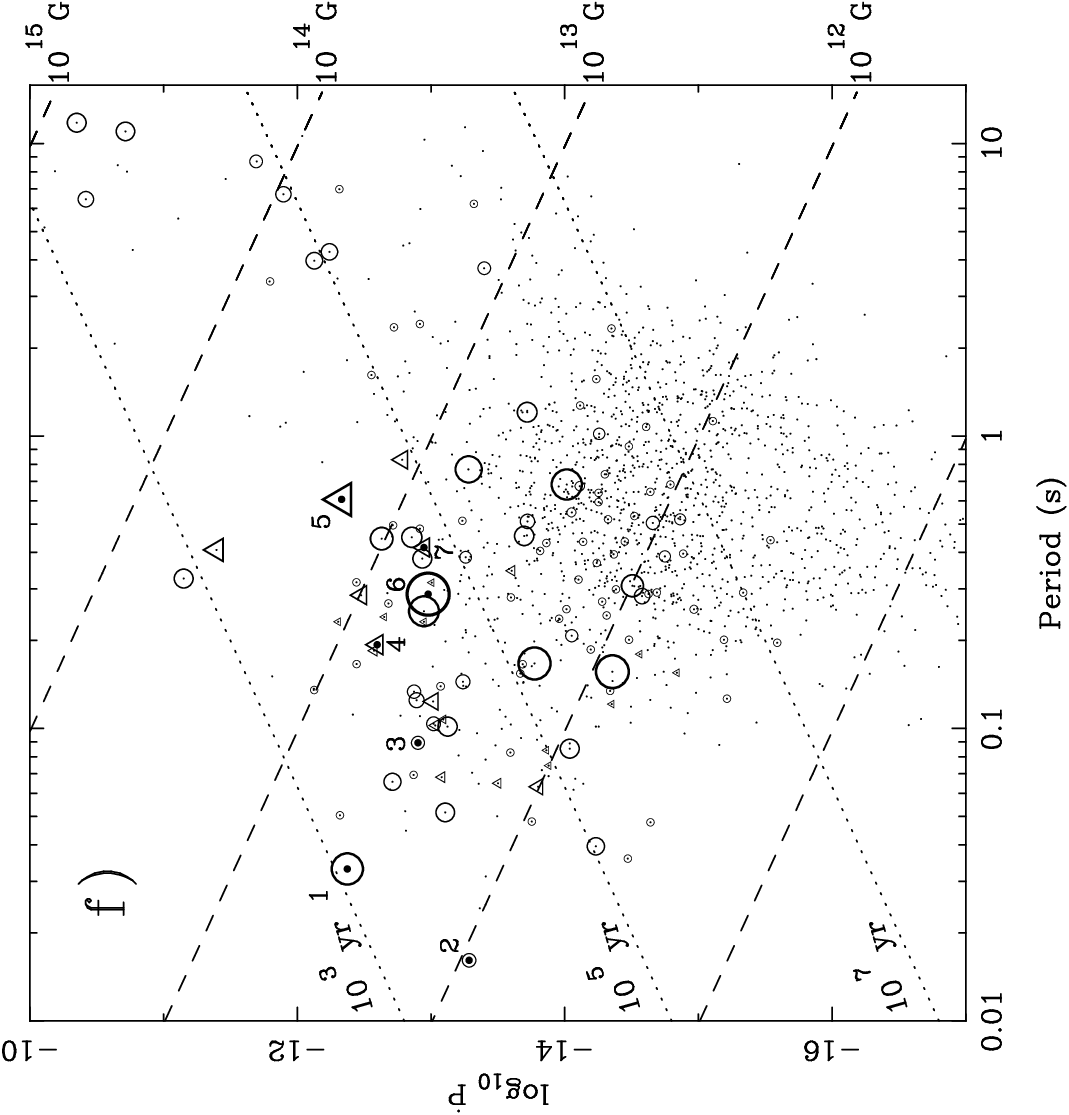} \\
\end{tabular}
\end{center}
\caption{$P$--$\dot{P}$ diagrams for glitch-related quantities of the
  a) number of detected glitches; b) average number of glitches per
  year; c) maximum fractional glitch size; d) maximum glitch size; e)
  rms fractional glitch size; and f) rms fractional size normalised by
  the mean. A circle indicates the parameter was obtained from the
  ATNF Pulsar Catalogue glitch table, whereas a triangle symbol
  indicates a parameter from this work. In the various plots, the
  seven pulsars exhibiting ten or more glitches are marked: 1 --
  PSR~B0531+21 (Crab pulsar); 2 -- PSR~J0537$-$6910; 3 --
  PSR~B0833$-$45 (Vela pulsar); 4 -- PSR~J1341$-$6220; 5 --
  PSR~J1740$-$3015; 6 -- PSR~J0631$+$1036; 7 -- PSR~J1801$-$2304; and
  two magnetars: A -- PSR~J1048$-$5937 (1E~1048.1$-$5937) and B --
  PSR~J1841$-$0456 (1E~1841$-$045).}
\label{fig:ppdot_gl}
\end{figure*}

In Figure~\ref{fig:ppdot_gl}, we show a set of
period--period-derivative ($P$--$\dot{P}$) diagrams to present six
quantities relevant to glitches. For previously published glitches, we
refer to the ATNF Pulsar Catalogue glitch table. The six plotted
quantities are a) the number of glitches detected in a given pulsar
$N_{\rm g}$, b) the average number of glitches per year $\dot{N}_{\rm
  g}$, c) the fractional glitch size $\Delta\nu_{\rm g}/\nu$, d) the
glitch size $\Delta\nu_{\rm g}$, e) the rms glitch fractional size and
f) the rms fractional size normalised by the mean value for that
pulsar. For sub-plots c) and d), if a pulsar has glitched more than
once, then the largest value is plotted. In each $P$--$\dot{P}$
diagram, the size of the symbols (circle or triangle) is a linear
function of the magnitude of the given quantity; we adjusted the
slopes and offsets for the different functions to give appropriate
sizes for the symbols.

Table~\ref{tab:glrate} gives the number of detected glitches $N_{\rm
  g}$, the observation data span and the rate of glitches for known
glitching pulsars $\dot{N}_{\rm g}$. The uncertainty of the glitch
rate was estimated as the square-root of $N_{\rm g}$ divided by the
data span. 

We discuss these results in the following subsections. 

\subsubsection{The number and rate of glitches}\label{sect:glnr}

Large numbers of glitches were observed in the pulsars with
characteristic ages between $10^3$ and $10^5$\,yr; the seven pulsars
that have been observed to show ten or more glitches are within this
age region. If magnetic-dipole radiation is assumed, then the inferred
surface magnetic field for the seven pulsars spans from
$\sim$10$^{12}$ to $\sim$10$^{13}$\,G. But there are some young
pulsars that have not been observed to glitch, although they also have
relatively long data spans. For instance, PSR~J1513$-$5908 has a
characteristic age of $\sim$1.5\,kyr; it is the youngest pulsar in our
sample of 165 pulsars. This pulsar has been observed for more than 28
years \citep{lk11} with no evidence for any glitch activity. On the
other hand, PSR~J1119$-$6127 has a similar characteristic age of
$\sim$1.6\,kyr and three glitches have been observed in its
$\sim$13-yr timing data span.

The 23 glitches observed for PSR~J0537$-$6910 occurred in
$\sim$7.6\,yr, resulting in a large glitch rate of
$\sim$3.0\,yr$^{-1}$; this pulsar is the most frequently glitching
pulsar known. The second most frequently glitching pulsar is
PSR~J1740$-$3015. This pulsar has 32 glitches in $\sim$25\,yr, giving
a glitch rate of $\sim$1.3\,yr$^{-1}$. As shown in sub-plot b) of
Figure \ref{fig:ppdot_gl}, the positions of these two pulsars on the
$P$--$\dot{P}$ diagram are quite different: the pulse period for
PSR~J0537$-$6910 is about a factor of 40 less than that of
PSR~J1740$-$3015 and $\dot{P}$ is about an order of magnitude
smaller. Hence the two pulsars are both relatively young, with
characteristic ages of 5\,kyr and 20\,kyr respectively. The dipole
magnetic field strength of PSR~J1740$-$3015 is more than an order of
magnitude stronger than that of PSR~J0537$-$6910. We found that
PSRs~J0729$-$1448, J1341$-$6220 and J0922$+$0638 also exhibit glitches
more than once per year. For PSR~J1341$-$6220, the Parkes observations
have detected 25 glitches during $\sim$20\,yr. PSR~J0729$-$1448 has
glitched four times during its data span of $\sim$3.3\,yr. For
PSR~J0922$+$0638, one glitch has been detected in a data span of just
one year \citep{sha10}, so the inferred high glitch rate is very
uncertain. On the $P$--$\dot{P}$ diagram, the characteristc-age lines
of $10^3$ and $10^5$\,yr together with the magnetic-field lines of
$10^{12}$ and $10^{13}$\,G define a region where pulsars are observed
to exhibit more glitches and large glitch rates.

\subsubsection{Glitch sizes}\label{sect:gl_size}

As shown in sub-plot c) in Figure~\ref{fig:ppdot_gl}, large fractional
glitch sizes $\Delta\nu_{\rm g}/\nu$ are generally observed in young
pulsars with long periods (i.e., small $\nu$). The largest known,
$\sim3.3\times10^{-5}$, was detected in PSR~J1718$-$3718 which has a
pulse period of 3.38\,s \citep{mh11} but a characteristic age of only
34\,kyr and an inferred dipole field strength of $7.4\times
10^{13}$\,G, one of the highest known for radio pulsars. Magnetars
detected at X-ray wavelengths are also known to suffer large glitches
and have long pulse periods and even higher inferred dipole
fields. For example, PSR~J1048$-$5937 (1E~1048.1$-$5937) and
PSR~J1841$-$0456 (1E~1841$-$045) are both X-ray detected magnetars
which have had glitches with $\Delta\nu_{\rm g}/\nu > 10^{-5}$
\citep{dkg08,dkg09}.

Not surprisingly, the absolute frequency jumps $\Delta\nu_{\rm g}$
tend to be larger for the shorter-period (larger $\nu$) pulsars since
they are not normalised by $\nu$. Sub-plot d) of Figure
\ref{fig:ppdot_gl} shows the distribution of $\Delta\nu_{\rm g}$
values on the $P$--$\dot{P}$ diagram confirming this expectation. The
largest frequency-jump was observed in the 24-ms pulsar J2022$+$3842;
the pulse frequency of this source gained $\sim$78\,$\mu$Hz in its MJD
$\sim$54675 glitch \citep{agr+11}. Sub-plot d) of Figure
\ref{fig:ppdot_gl} also shows that the glitch size $\Delta\nu_{\rm g}$
is also correlated with characteristic age. Pulsars with young ages
tend to have larger jumps, but this correlation breaks down for the
very young pulsars such as the Crab pulsar (PSR B0531$+$21) and PSR
J1513$-$5908.

\subsubsection{Glitch variability}\label{sect:gl_rms}

Sub-plot e) of Figure \ref{fig:ppdot_gl} shows the rms fluctuation in
fractional glitch sizes. The magnetars PSR~J1048$-$5937
(1E~1048.1$-$5937) and PSR~J1841$-$0456 (1E~1841$-$045) show a wide
range of glitch sizes, with $\Delta\nu_{\rm g}/\nu$ ranging from
$\sim1.4\times10^{-6}$ up to $\sim1.6\times10^{-5}$. Although most of
the glitches in the Vela pulsar are large, there have been two small
glitches (see Figure~\ref{fig:glseries}) and so the rms fluctuation of
glitch size is relatively large. After the magnetars, large rms
variations are found in the high-magnetic-field radio pulsar
J1119$-$6127 and PSR~J1838$-$0453. There seems a tendency for more
variability in glitch size in older pulsars. This is also seen in
sub-plot f) where the rms fluctuation has been normalised by the mean
value to give an effective ``modulation index'' for glitch size
fluctuations. However, this conclusion is not very certain as only a
small number of glitches were detected in PSR~J1838$-$0453 and some
other pulsars. Also, the results may be biased by the difficulty in
detecting small glitches in noisy pulsars. In sub-plot f) there is
more scatter in this plot though with the Crab pulsar being more
prominent because of its relatively small mean glitch size.

\subsubsection{Implications for neutron-star physics}\label{sect:gl_phy}

In the angular-momentum-transfer model, a glitch is understood as a
sudden transfer of angular momentum from a more rapidly rotating
interior superfluid to the neutron star crust
\citep[e.g.,][]{aaps81}. The angular momentum of a rotating superfluid
is carried by an array of vortices; each vortex contains a quantised
unit of angular momentum. In principle, if the rotation of a
superfluid is slowing down, then the surface density of the vortices
will be decreasing, or, in other words, the vortices will migrate away
from the rotation axis. However, in the solid crust of a neutron star,
the vortices of the neutron superfluid tend to pin onto the nuclear
lattice \citep{alp77,eb88}. The pinning between the vortices and the
crystal lattice will not be broken until the force induced by the
differential rotation (or the ``Magnus'' force) between the superfluid
and the crust reaches a critical value, beyond which an avalanche
process of unpinning and a substantial transfer of angular momentum
could be triggered.

Sub-plots c) and d) of Figure \ref{fig:ppdot_gl} shows that large
glitches, either relative or absolute, are mostly confined to the
pulsars with characteristic ages smaller than $10^5$\,yr. The largest
fractional glitch sizes are observed in long-period, low-$\nu$
pulsars. This suggests that a larger fraction of the excess angular
momentum of the superfluid is transferred to the crust in these
pulsars. For very young pulsars such as the Crab pulsar, glitch sizes
are much smaller and somewhat less frequent. For a very young neutron
star, as the result of the internal high temperature, the superfluid
vortices could creep against the pinning energy barrier, preventing
the formation of pinning zones and the sudden release of pinned
vortices \citep{aaps84a}. \citet{rzc98} have suggested that, because
of tangling of the vortices and magnetic flux tubes, as the neutron
star spins down the core flux-tubes are pulled by the vortex lines,
moving toward the stellar equator. The moving core flux-tubes lead to
a build-up of the shear stress in the stellar crust
\citep{sbmt90}. Once the stress grows to exceed the yield strength of
the crust, then the consequent cracking of the stellar surface may
cause small period jumps in the rotation of the neutron star
\citep{rud09}. This could be the mechanism for the observed small
glitches in the Crab and other pulsars such as Vela and
PSR~J1740$-$3015. It is also possible that such processes may trigger
the large-scale release of angular momentum needed for glitches with
large fractional glitch sizes.

\subsection{Glitch recoveries}\label{sect:glRec}

Pulsars show a variety of behaviours following a glitch. Part of the
step change in both $\nu$ and $\dot\nu$ often recovers exponentially,
in some cases with more than one identifiable time constant. Following
this exponential recovery, a linear increase in $\dot\nu$ (decrease in
slow-down rate $|\dot\nu|$) is often observed. Normally this continues
until the next glitch. Finally, apparently permanent changes in $\nu$,
$\dot\nu$ and/or $\ddot\nu$ are sometimes left after the transient
recoveries. We searched for evidence of these different types of
recovery in all 107 observed glitches.

\subsubsection{Exponential recoveries}\label{sect:glRec_exp}

\begin{figure}
\begin{center}
\begin{tabular}{c}
\includegraphics[width=5.5cm,angle=-90]{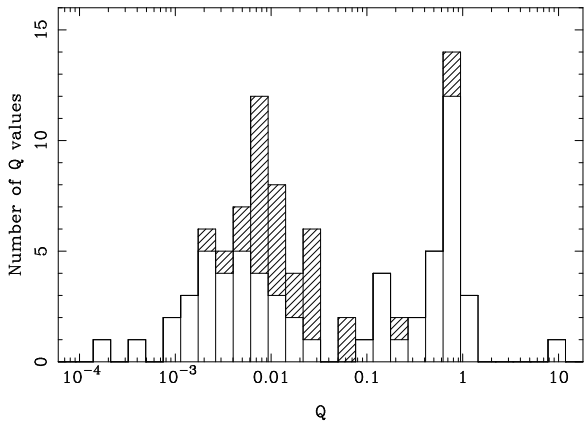} \\
\end{tabular}
\end{center}
\caption{Histogram for the recovery fraction $Q$ of exponential
  decays. Data for the blank bars are from the literature as given in
  Table \ref{tab:exp}, while the shaded bars show values from this
  work.}\label{fig:q_hist}
\end{figure}

In our sample, 27 glitches in 18 pulsars had an identifiable
exponential recovery. The observed fractional recovery $Q$ and
recovery time constant $\tau_d$ for these glitches are given in
Table~\ref{tab:glt}. All but two of the exponential recoveries are
well modelled by a single exponential term. Two exponential terms were
required for glitch 3 in PSR~J1119$-$6127 and glitch 1 in
PSR~J1803$-$2137. The largest $Q\sim0.84$ was detected in glitch 2 in
PSR~J1119$-$6127. Table \ref{tab:exp} summarises the parameters for
the previously reported exponential recoveries. Multiple decays were
observed for three pulsars, the Crab pulsar, the Vela pulsar and PSR
J2337+6151. However, the strong observational selection against
observing short-term recoveries in most pulsars needs to be
recognised.\footnote{Note that we do not consider the 1.2-min recovery
  observed after the Vela MJD 51559 glitch by \citet{dml02} for two
  reasons: a) the parameters of the recovery are very uncertain and b)
  it is unique in that such short timescale recoveries cannot be
  observed for any other known pulsar.}

Figure \ref{fig:q_hist} shows the histogram of the fractional
exponential recoveries $Q$. The observed values span a very wide range
from the smallest $\sim$0.00014 observed in PSR~J1841$-$0425
\citep{ywml10} to the largest $\sim$8.7 recently detected in the young
X-ray pulsar J1846$-$0258 \citep{lkg10}. The histogram is clearly
bimodal with a broad peak around 0.01 and another very close to
1.0. This strongly suggests that there are two different mechanisms
for the exponential recovery following a glitch.

\begin{figure*}
\begin{center}
\begin{tabular}{cc}
\includegraphics[width=5.5cm,angle=-90]{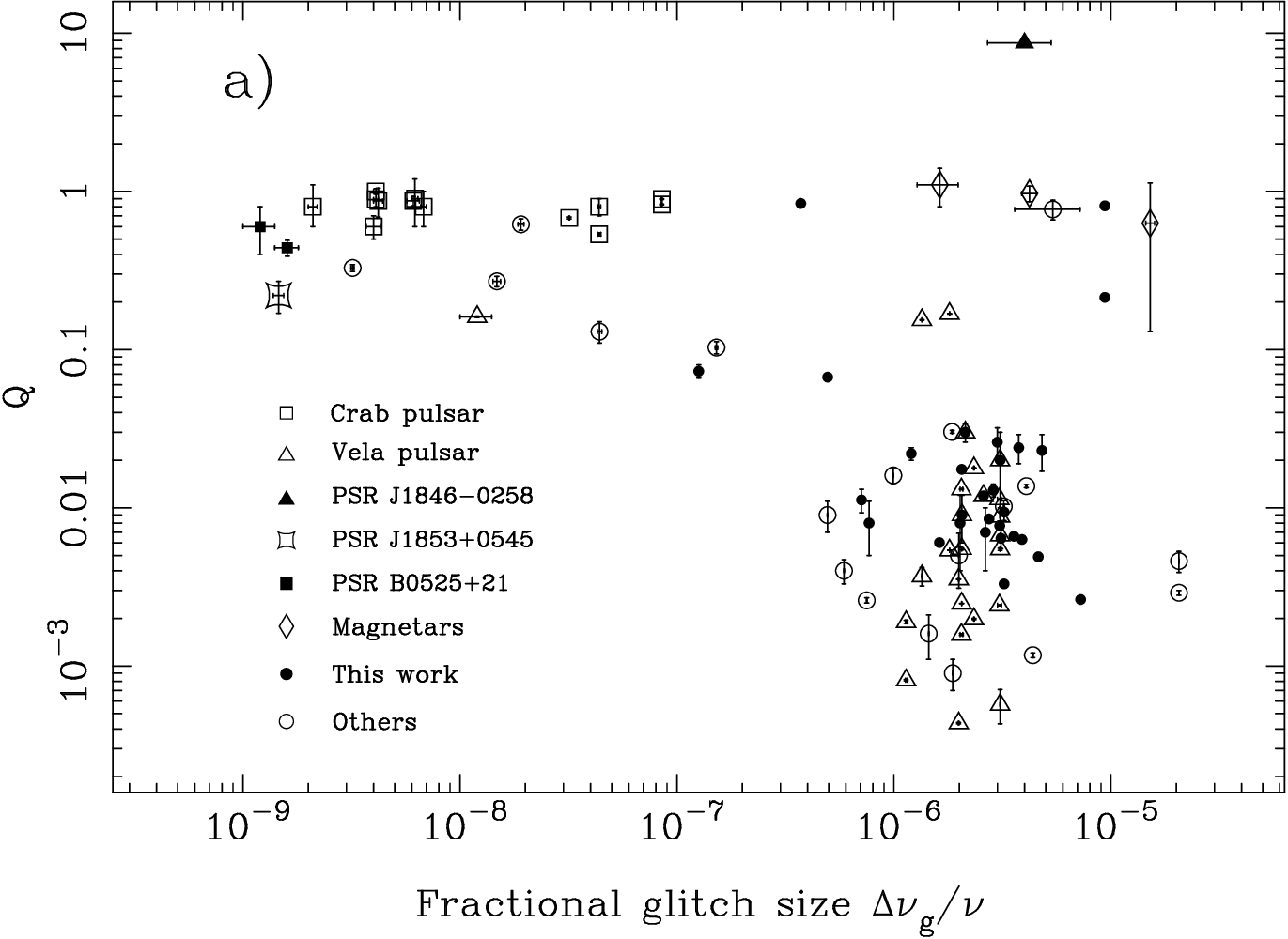} &
\includegraphics[width=5.5cm,angle=-90]{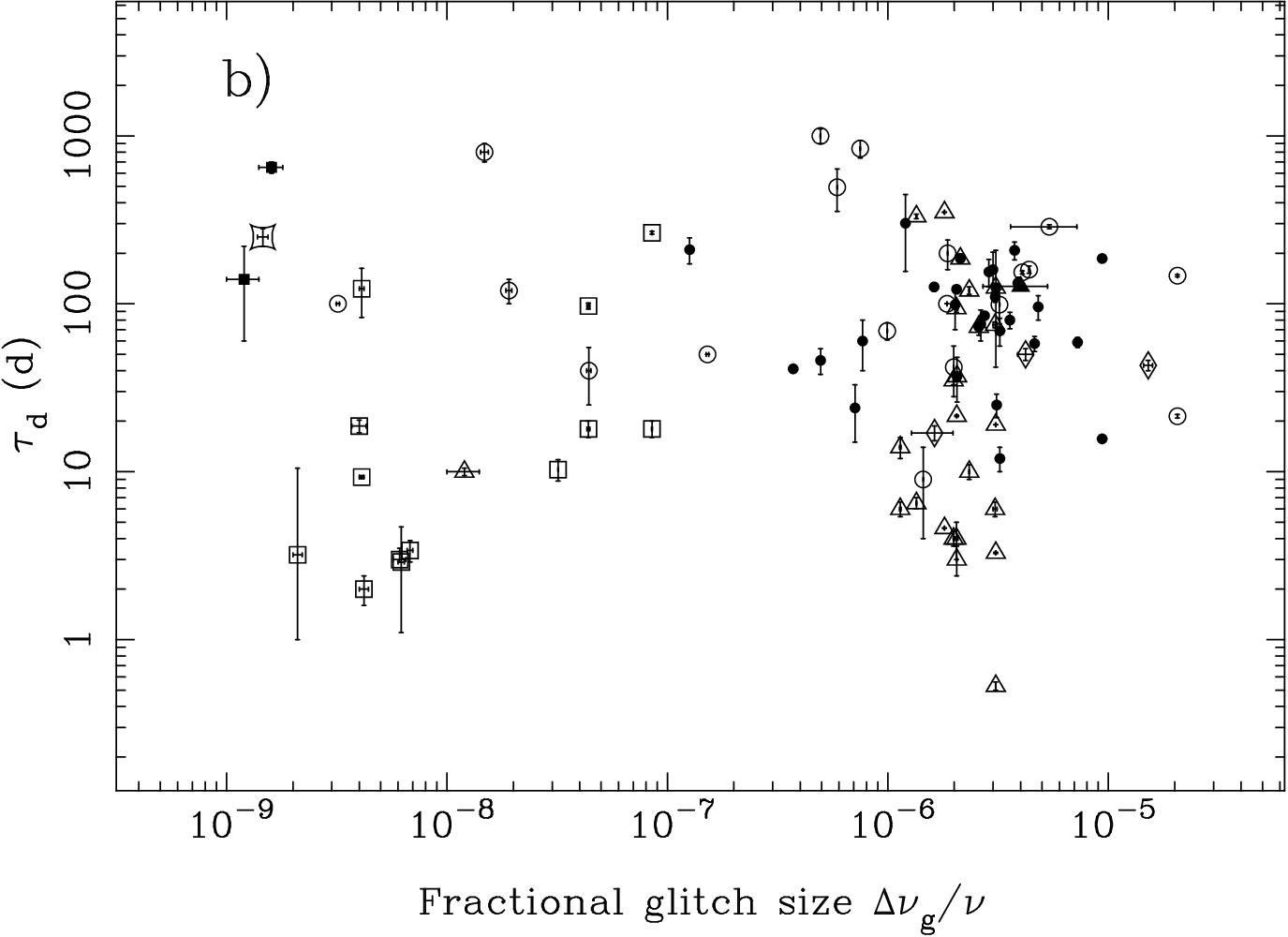} \\
\includegraphics[width=5.5cm,angle=-90]{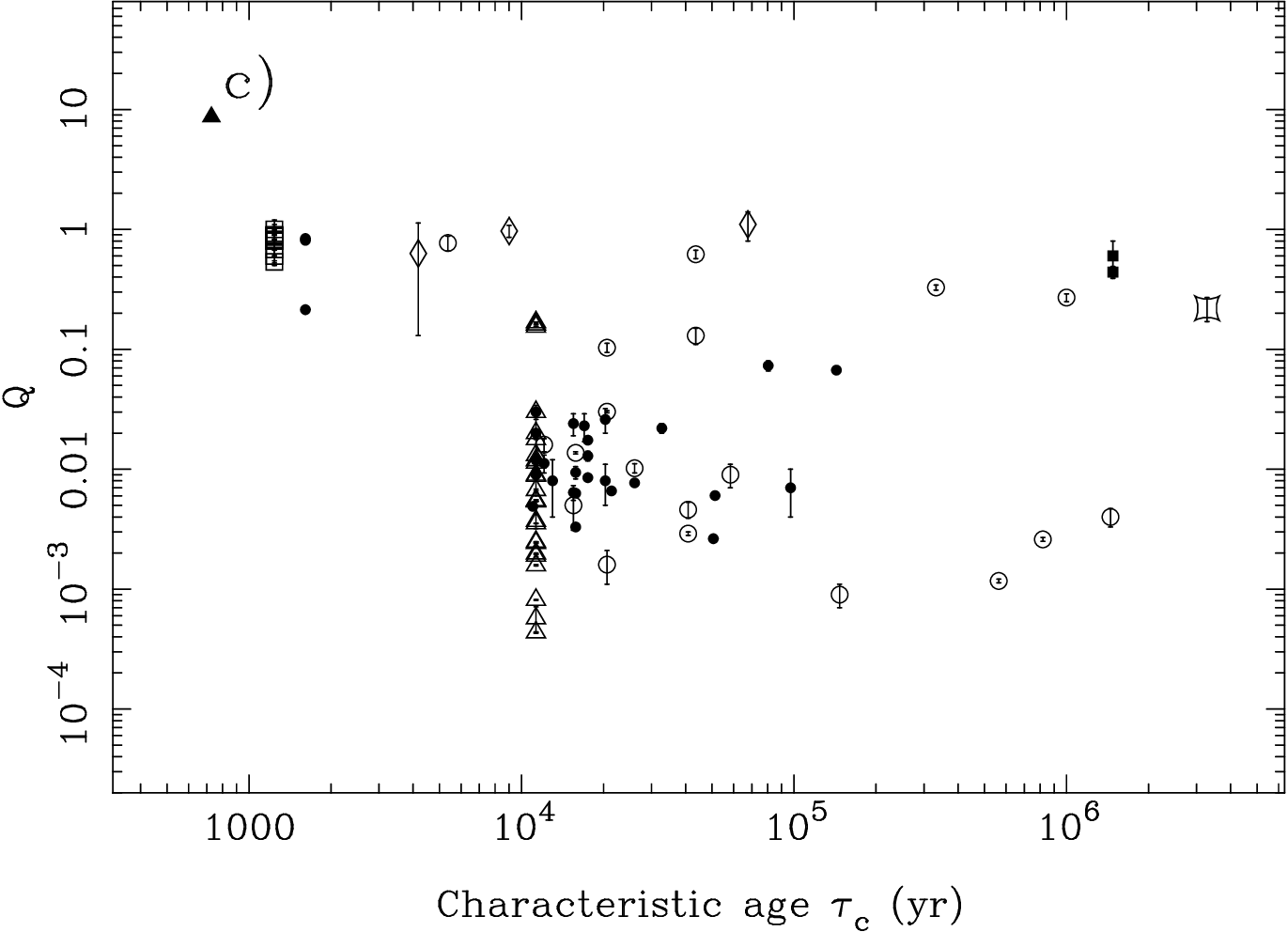} &
\includegraphics[width=5.5cm,angle=-90]{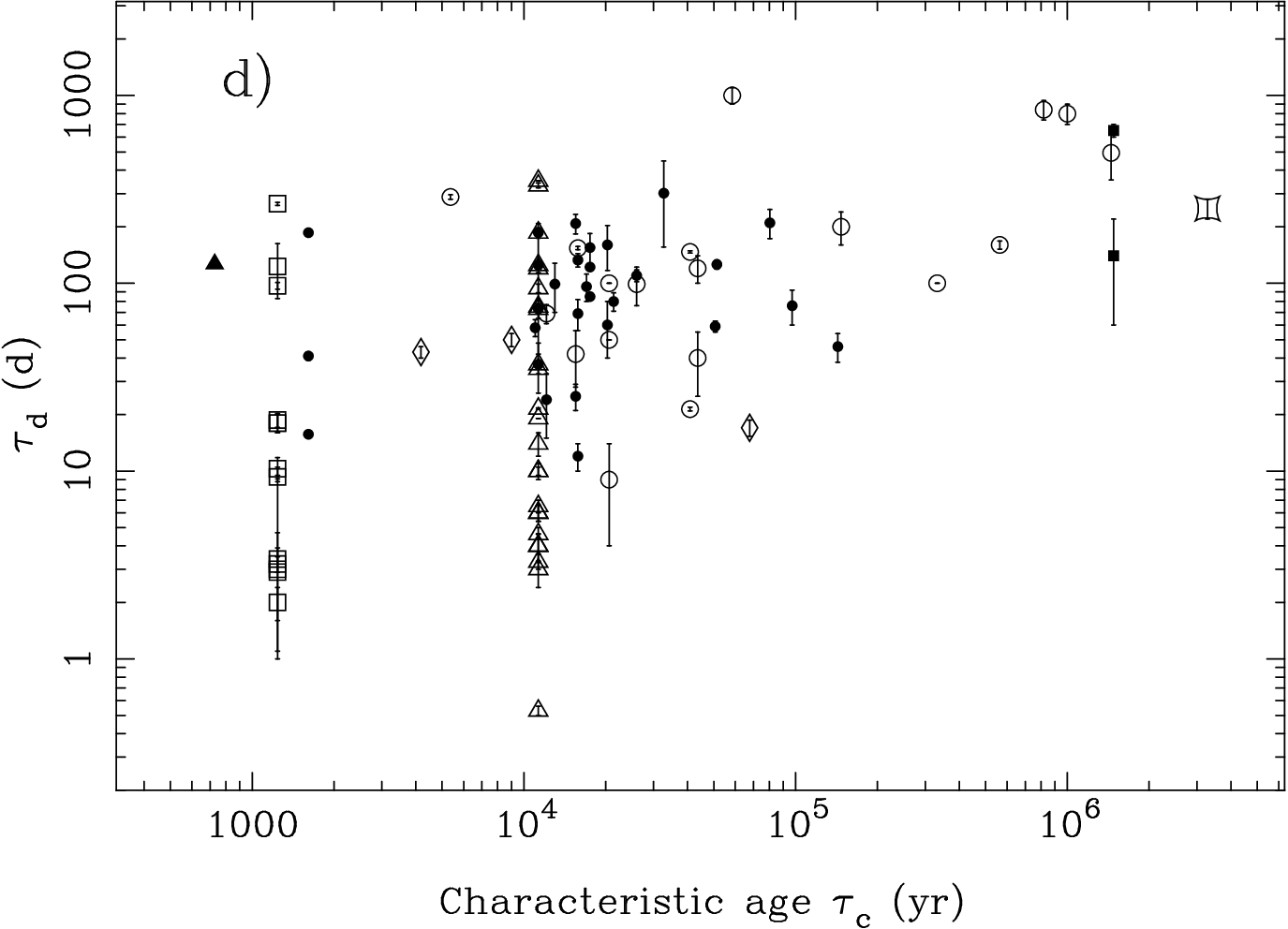} \\
\includegraphics[width=5.5cm,angle=-90]{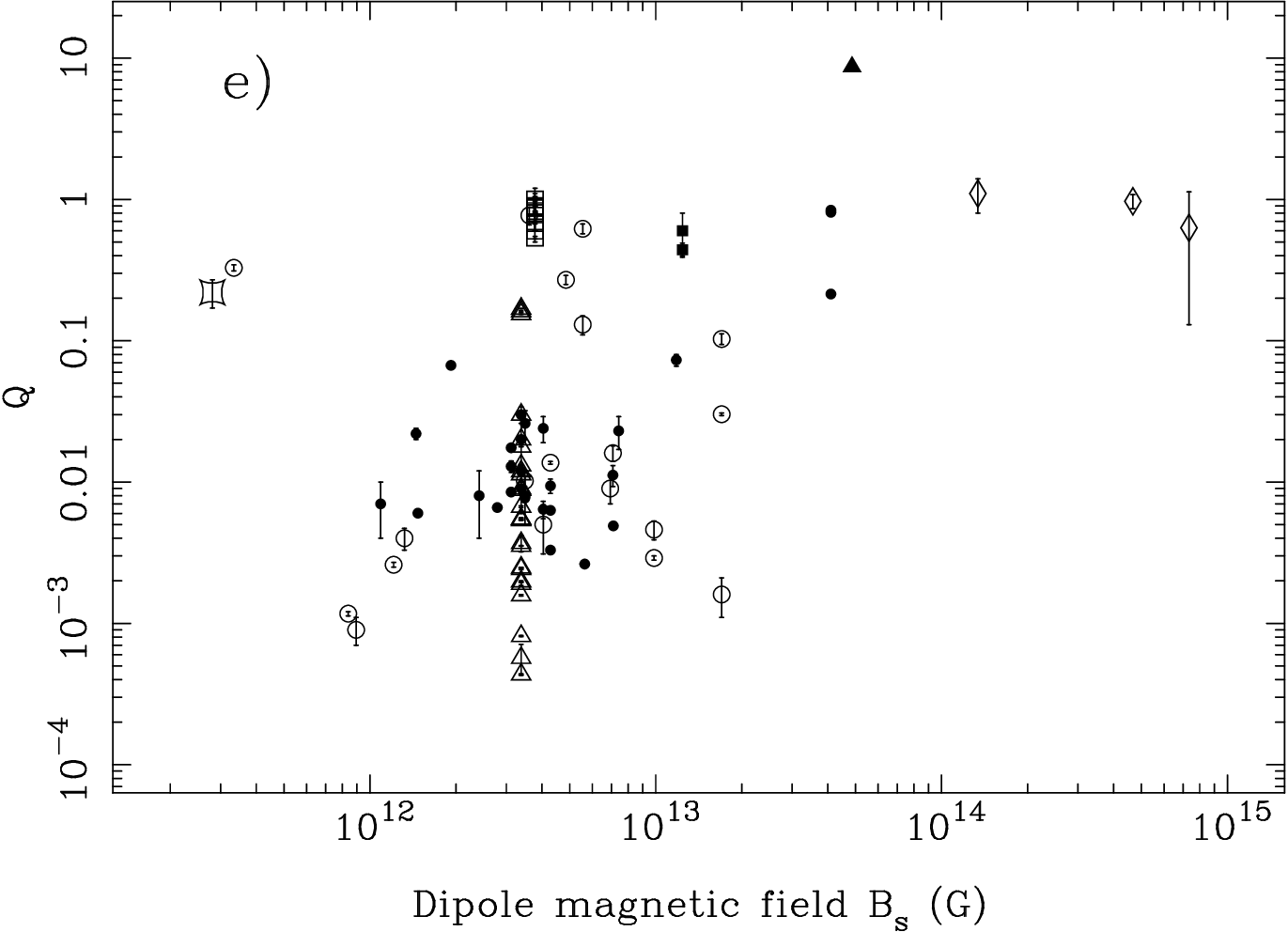} &
\includegraphics[width=5.5cm,angle=-90]{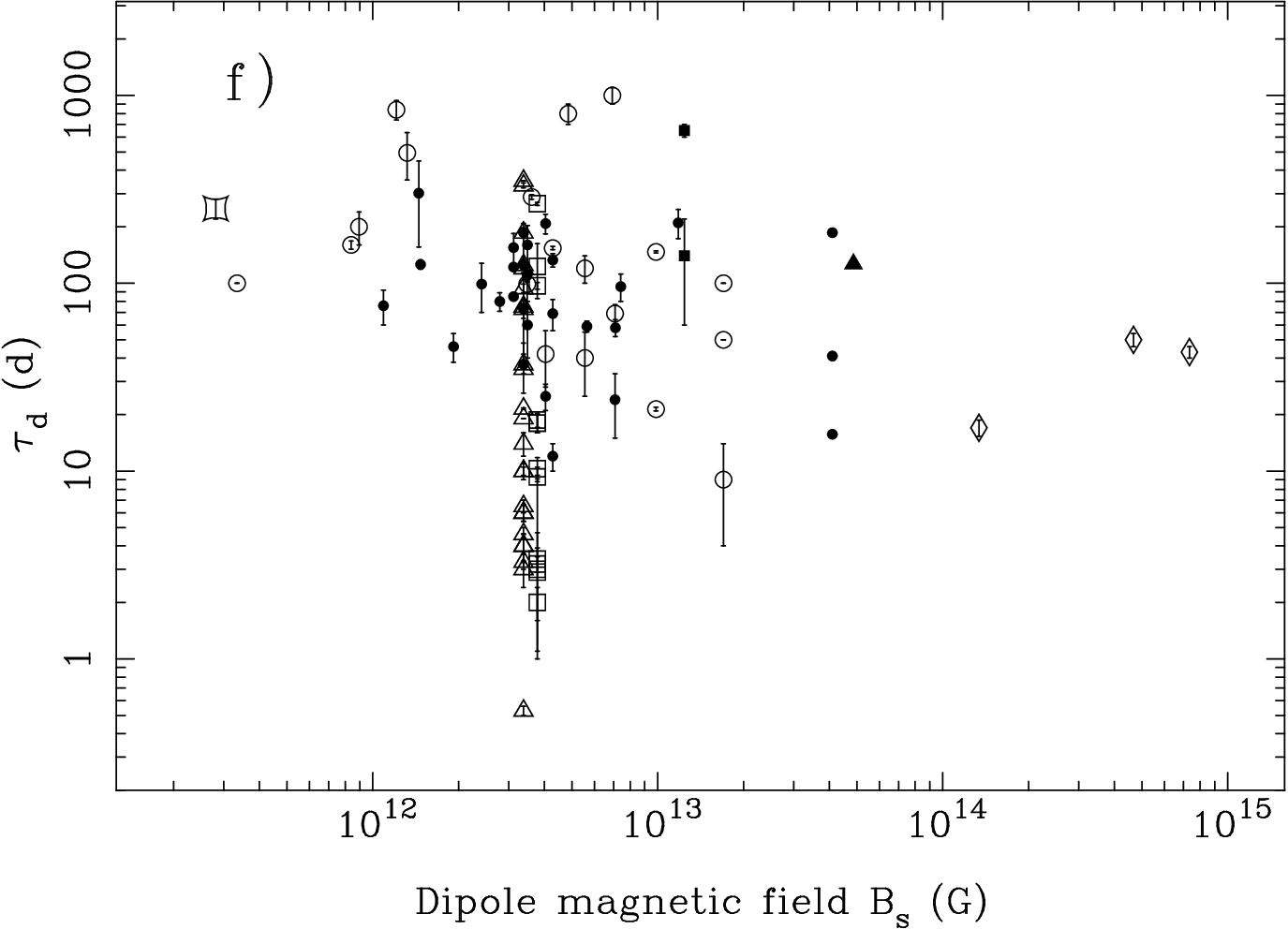} \\
\end{tabular}
\end{center}
\caption{Dependences of the post-glitch exponential recovery
  parameters $Q$ and $\tau_{\rm d}$ on the observed glitch fractional
  size, the pulsar characteristic age $\tau_{\rm c}$ and the dipole
  magnetic field $B_{\rm s}$: a) $Q$ versus $\Delta\nu_{\rm g}/\nu$,
  b) $\tau_{\rm d}$ versus $\Delta\nu_{\rm g}/\nu$, c) $Q$ versus
  $\tau_{\rm c}$, d) $\tau_{\rm d}$ versus $\tau_{\rm c}$, e) $Q$
  versus $B_{\rm s}$ and $\tau_{\rm d}$ versus $B_{\rm
    s}$.}\label{fig:exp_cmp}
\end{figure*}

To further explore the properties of the exponential recoveries, all
known values of $Q$ and $\tau_{\rm d}$ are shown in Figure
\ref{fig:exp_cmp} as a function of the fractional glitch size
$\Delta\nu_{\rm g}/\nu$, the pulsar characteristic age and the surface
magnetic-dipole field. In sub-plot a), it is striking that glitches
with $Q\sim 1$ are found over the whole range of fractional glitch
sizes, whereas those with small $Q$ are only found in the larger
glitches with $\Delta\nu_{\rm g}/\nu\gtrsim 10^{-6}$. Because of
random period irregularities, it will be difficult or impossible to
detect exponential recoveries in many small glitches with
$\Delta\nu_{\rm g}/\nu \lesssim 10^{-8}$. However, such decays should
be detectable in most glitches with $\Delta\nu_{\rm g}/\nu \sim
10^{-7}$. Therefore there appears to be a real absence of exponential
recoveries with $Q \sim 0.01$ for glitches with $\Delta\nu_{\rm g}/\nu
\lesssim 3\times10^{-7}$. It is also worth noting that, for many
glitches, both large and small, no exponential recovery is detected
(Table \ref{tab:glt}), so effectively $Q \sim 0$ in these cases. These
results suggest that the real physical distinction between pulsars
with large glitches and those with small glitches
(cf. Figure~\ref{fig:glsize_hist}) is more complex with the glitch
recovery parameter also being important.

Unlike $Q$, the distribution of the time constant $\tau_{\rm d}$
relative to $\Delta\nu_{\rm g}/\nu$ shown in sub-plot b) is more
uniform with decay timescales from a few days to a few hundred days
seen in glitches of all sizes.

Sub-plot c) of Figure \ref{fig:exp_cmp} shows that the large-$Q$
glitch recoveries are observed in both young and old pulsars, whereas
the large glitches, which have low $Q$, are not seen in the very young
pulsars. As discussed above, the large glitches primarily occur in
pulsars with characteristic ages between $10^4$ and $10^5$
years. Sub-plot d) gives some support to the suggestion of
\citet{ywml10} that there is a positive correlation of $\tau_{\rm d}$
with pulsar characteristic age. There is a clear absence of short-term
decays for the older pulsars that cannot be accounted for by
observational selection. Sub-plots e) and f) show the dependence of
$Q$ and $\tau_{\rm d}$ on surface dipole field strength. Since this is
derived from $P$ and $\dot{P}$, as is the characteristic age, it is
not surprising that these plots show basically the same dependences as
sub-plots c) and d). However, they do highlight the fact that the
magnetars, with $B_{\rm s} \gtrsim 10^{13}$~G, have high-$Q$ glitch
recoveries \citep{kg03,dkg08,gdk11}. The largest observed $Q\sim8.7$,
indicating a massive over-recovery, was detected in the young
X-ray-detected but spin-powered pulsar PSR~J1846$-$0258, which has an
intermediate period, $\sim$0.326\,s, a very high implied dipole
magnetic field, $B_{\rm s} \sim 5\times10^{13}$\,G and is located near
the centre of the supernova remnant Kes 75 \citep{lkg10}.

\subsubsection{Linear recovery}\label{sect:glRec_lnr}

\begin{figure}
\begin{center}
\includegraphics[width=5.5cm,angle=-90]{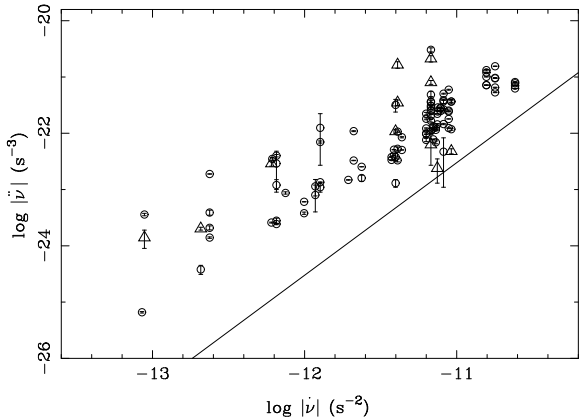}
\end{center}
\caption{Linear decay rate $\ddot\nu$ as a function of $|\dot\nu|$ for
  32 glitching pulsars. The symbols of a circle and triangle indicate
  positive and negative values, respectively. The solid line gives the
  $\ddot\nu_{\rm ext}$ resulting from external or magnetospheric
  braking with $n=3$ and for $\nu=10$\,Hz.}\label{fig:nudot_f2}
\end{figure}

It has been well recognised that the long-term recovery from a glitch
is generally dominated by a linear increase in $\dot\nu$, that is, a
linear decrease in the slow-down rate $|\dot\nu|$ or, equivalently, in
$\dot{P}$ \citep{dow81,lpgc96,lsg00,wmp+00,ywml10}. This effect is
clearly seen in most of the $\dot\nu$ plots given in
Figures~\ref{fig:glt1} -- \ref{fig:glt9}, especially for the larger
glitches. The linear trend normally becomes evident at the end of the
exponential recovery and persists until the next glitch event. In
cases where the exponential recovery is absent or has very low $Q$,
e.g., PSR J1301$-$6305 (Figure~\ref{fig:glt3}) the linear recovery
begins immediately after the glitch. The rate of increase in $\dot\nu$
is quantified by fitting for $\ddot\nu$ in the pre-, inter- and
post-glitch intervals following the decay of any exponential
recovery. Values of $\ddot\nu$ obtained in this way are given in
Table~\ref{tab:gltspin} and are plotted in Figure~\ref{fig:nudot_f2}
for 32 pulsars. No significant $\ddot\nu$ value was obtained for four
glitching pulsars, in most cases because available data spans were too
short. Of the 108 $\ddot\nu$ values plotted, 11 are negative. These
are for pulsars where timing noise is relatively strong and/or the
available data spans are short.

There is no doubt that these linear increases in $\dot\nu$ are related
to internal neutron-star dynamics and recovery from glitches. Firstly,
the value of $\ddot\nu$ often changes at the time of a glitch. Clear
examples of slope changes at glitches are seen for the Vela pulsar
(Figure~\ref{fig:glt1}), PSR~J1420$-$6048 (Figure~\ref{fig:glt4}) and
PSR~J1709$-$4429 (Figure~\ref{fig:glt6}). In other cases however,
e.g., PSR~J1301$-$6305 (Figure~\ref{fig:glt3}) and PSR~J1803$-$2137
(Figure~\ref{fig:glt8}), there is little or no slope change at a
glitch.
Secondly, although a positive value of $\ddot\nu$ is expected from
normal magnetospheric braking, the observed values are generally much
larger. Pulsar braking is normally described by the braking index $n$,
defined by
\begin{equation}
n = \frac{\ddot\nu \nu}{\dot\nu^2}
\end{equation}
where $n=3$ for magnetic-dipole braking.  This relation is shown in
Figure~\ref{fig:nudot_f2} assuming $n=3$ and for a typical young
pulsar with $\nu = 10$\,Hz. The magnetospheric or external
contribution $\ddot\nu_{\rm ext}$ is well below the observed
values. Observed braking indices attributable to magnetospheric
braking are generally less than 3.0 \citep[e.g.,][]{elk+11} which
increases the discrepancy. However, for young long-period pulsars, the
discrepancy is less. For example, PSR~J1119$-$6127 has a value of
$\ddot\nu$ for magnetic-dipole braking comparable to the minimum
observed values (Table~\ref{tab:gltspin}) and its braking index is
close to 3.0 \citep{wje11}. However, there are significant changes in
$\ddot\nu$ at the glitches in this pulsar, with the largest observed
value being more than a factor of two higher. So even in this case, it
is clear that glitch-related phenomena contribute to the $\ddot\nu$
value.

\begin{figure*}
\begin{center}
\begin{tabular}{cc}
\includegraphics[width=5.5cm,angle=-90]{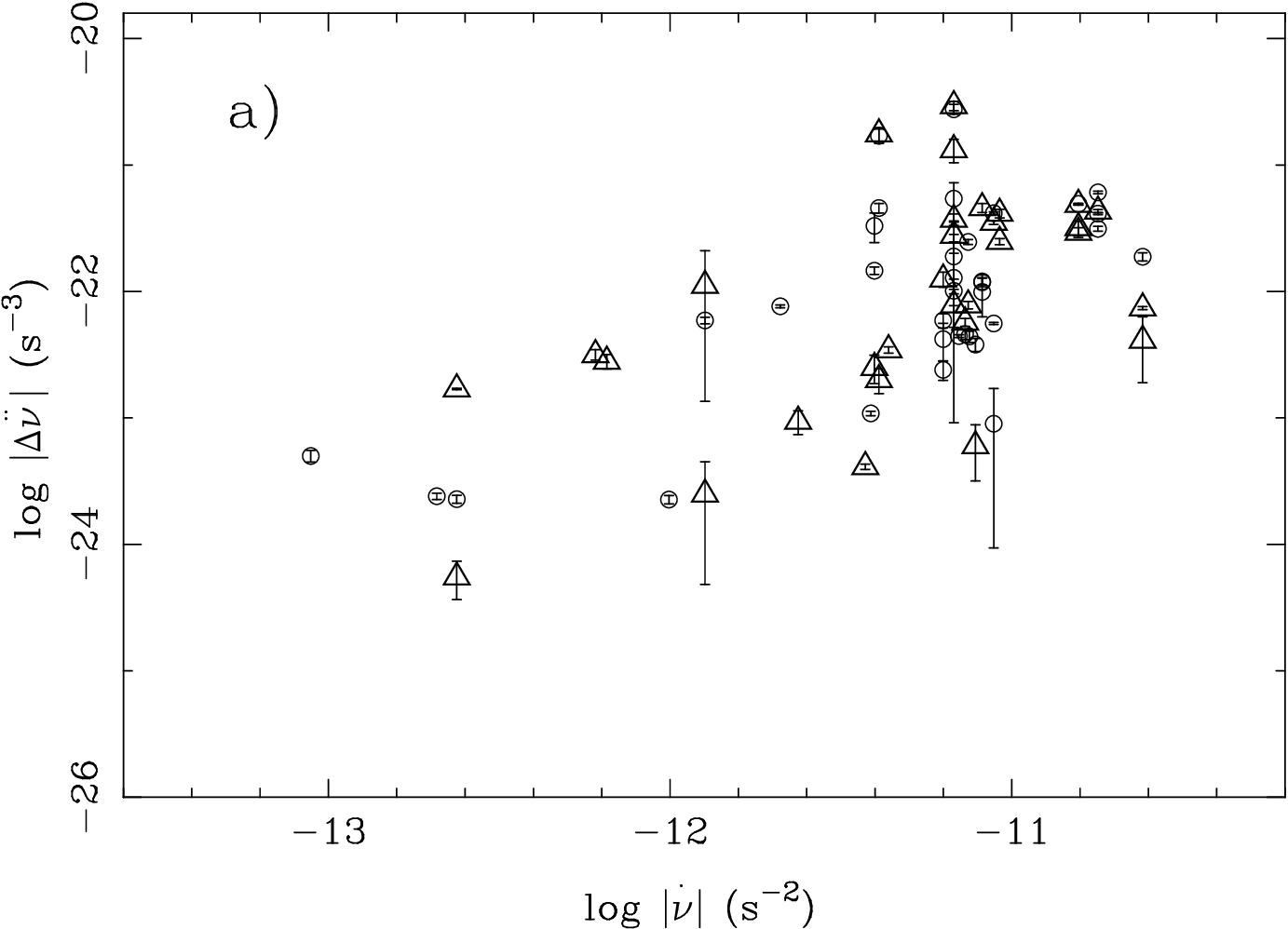} &
\includegraphics[width=5.5cm,angle=-90]{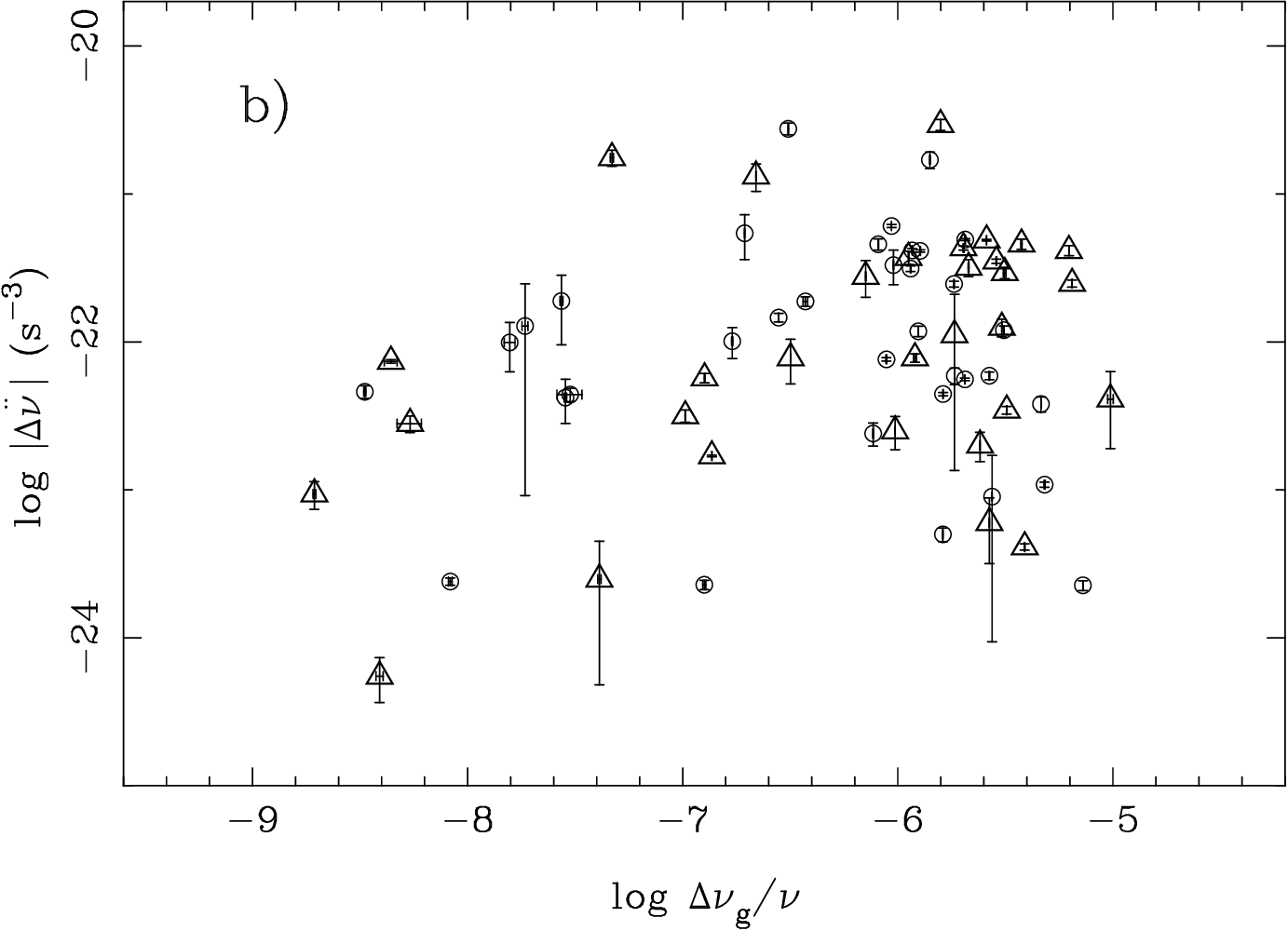}
\end{tabular}
\end{center}
\caption{The change in slope of linear $\dot\nu$ recoveries at
  glitches, $\Delta\ddot{\nu}$, as a function of a) $|\dot\nu|$ and b)
  glitch fractional size $\Delta\nu_{\rm g}/\nu$. Circles and
  triangles indicate the postive and negative values,
  respectively.}\label{fig:glf2}
\end{figure*}

Figure~\ref{fig:glf2} shows the change in slope of the linear
recoveries at glitches, $\Delta\ddot{\nu}$, as a function of slow-down
rate $|\dot\nu|$ and glitch fractional size. Sometimes
$\Delta\ddot{\nu}$ is solved for as part of the glitch fit
(Table~\ref{tab:glt}), but often the data spans used for this are too
short to define $\Delta\ddot{\nu}$ well. Therefore, in order to
enlarge the sample as much as possible, we took the difference in
$\ddot{\nu}$ between each pair of the pre- and post-glitch solutions
given in Table~\ref{tab:gltspin}. We found that, out of the available
66 values, 35 are positive (53\%) and 31 are negative (47\%) in
accordance with our expectation that the positive and negative changes
would be approximately evenly balanced. Panel a) further illustrates
that the changes are comparable in magnitude to the $\ddot\nu$ values
plotted in Figure~\ref{fig:nudot_f2}, implying that the amount of
additional braking can change dramatically at a glitch. Panel b) shows
that the amount of change is not strongly dependent on the glitch
size, with both large and small glitches inducing a wide range of
slope changes relative to the $\dot\nu$ value. Since large glitches
usually occur in relatively young pulsars, most of which have a large
$|\dot\nu|$ (cf. Figure~\ref{fig:ppdot_gl}, panel d), it is not
surprising that the largest values of $\Delta\ddot{\nu}$ occur in
glitches with large $\Delta\nu_{\rm g}/\nu$.

\subsubsection{Implications for neutron-star physics}\label{sect:glRec_phy}

\citet{aaps84a,aaps84b,accp93} have suggested that the observed
different types of glitch recoveries are a manifestation of
``vortex-capacitive'' and ``vortex-resistive'' regions in the interior
superfluid. For the former, the vortices decouple from the normal spin
of the crust in a glitch but only share angular momentum at the time
of the glitch, whereas for the latter, the vortices drift via a
continuous pinning and unpinning, continuously coupling their angular
momentum to the crust. At a glitch, the vortices in the resistive
zones unpin and then gradually repin until the next glitch, resulting
in the observed exponential and linear recoveries in $\dot\nu$.  The
resistive region may contain several sub-layers, some of which may
have a linear response and others a non-linear response to the
glitch. In this context, a linear response means that the amplitude of
the associated exponential recovery toward the steady state is
proportional to the perturbation in the angular velocity of a
particular sub-layer at the time of a glitch. This occurs when the
internal termperature is high compared to the pinning energy and the
differential angular velocity between the superfluid layer and the
crust is small \citep{accp93}. On the other hand, in outer layers
where the temperature is low compared to the pinning energy, the
equilibrium lag is large and the response is not linear with the
angular-velocity perturbation. Equilibrium is generally not reached
before the occurence of the next glitch and there is an effectively
linear increase in $\dot\nu$. 

Following \citet{accp93}, we label the linear-response layers that are
responsible for the exponential recoveries, $i=1,2,...$, with moments
of inertia $I_{\rm i}$, and the nonlinear response layer, responsible
for the linear recovery in $\dot\nu$, layer A with moment of inertia
$I_{\rm A}$.  For the exponential recoveries, the decaying part of
$\Delta\dot\nu$ is
\begin{equation}
\Delta\dot\nu_{\rm d,i} = -Q_{\rm i} \frac{\Delta\nu_{\rm g}}
{\tau_{\rm d,i}} = -\frac{I_{\rm i}}{I}\frac{\Delta\nu_{\rm g}}{\tau_{\rm d,i}}
\end{equation}
where $I$ is the total moment of interia of the star. For the large
glitches, the median observed value of $Q$ is about 0.01 (Figure
\ref{fig:exp_cmp}a) showing that the linear-response superfluid layers
contain about 1\% of the total moment of inertia. 

The situation is clearly different for the pulsars for which $Q\sim
1$. In the simplest two-component models with a solid crust and a
superfluid interior, where the superfluid is weakly coupled to the
crust, $Q \approx I_{\rm n}/I$, where $I_{\rm n}$ is the total moment
of inertia of the superfluid neutrons \citep[e.g.,][]{bppr69}. So one
explanation for the high-$Q$ events is that a large fraction of the
total stellar moment of inertia is in the form of superfluid neutrons
that are weakly coupled to the crust. The degree of coupling is
evidently quite variable, with decay time constants ranging from a few
days for the Crab pulsar to tens or hundreds of days for other
pulsars. This explanation for high-$Q$ decays can apply regardless of
the mechanism for the glitch itself; this may be related to the fact
that these decays are observed to follow both large and small
glitches. The observed overshoot in the very young pulsar
J1846$-$0258, with an apparent $Q$ of $\sim$8.7 \citep{lkg10}, is
clearly anomalous with no other similar overshoots being observed. As
suggested by \citet{lkg10} it is likely that this behaviour resulted
from a glitch-induced change in the external torque.

\begin{figure}
\begin{center}
\includegraphics[width=5.5cm,angle=-90]{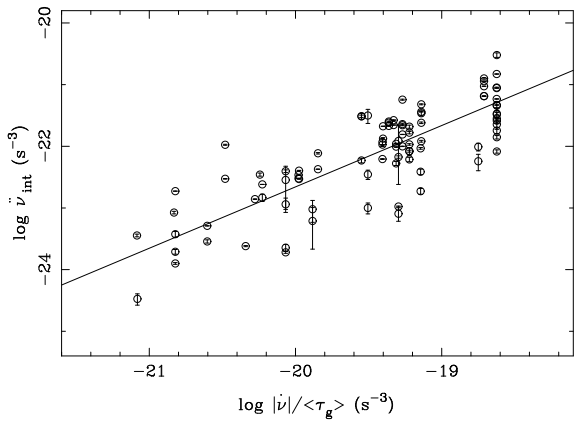}
\end{center}
\caption{The observed glitch-related $\ddot\nu_{\rm int}$ as a
  function of the ratio of $|\dot\nu|$ and the mean interglitch
  interval $\langle\tau_{\rm g}\rangle$. The solid line shows a fitted
  straight line which has a slope very close to 1.0 (see
  text). }\label{fig:f2int}
\end{figure}

In the Alpar et al. models, the observed approximately linear
recoveries in $\dot\nu$ are related to the properties of an outer
superfluid layer in which the coupling is very weak so that the
rotational lag between the crust and the superfluid is very large. For
this case, \citet{ab06} show that the $\ddot\nu_{\rm int}$ related to
glitch recovery is given by
\begin{equation}\label{eq:ddnu_int}
\ddot\nu_{\rm int} = \frac{I_{\rm A}}{I} \frac{|\dot\nu|}{\tau_{\rm
    g}}
\end{equation}
where $\tau_{\rm g}$ is the interglitch interval and $I_{\rm A}$ is
the inertial moment of layer A. Figure \ref{fig:f2int} shows
$\ddot{\nu}_{\rm int}=\ddot\nu_{\rm obs}-3\dot\nu^2/\nu$ versus
$|\dot\nu|/\langle\tau_{\rm g}\rangle$, where $\langle\tau_{\rm
  g}\rangle$ is the mean interglitch interval
(Table~\ref{tab:glrate}).  A correlation is clearly seen, with an
unweighted-least-squares fit giving
\begin{equation}
\ddot{\nu}_{\rm int} = 10^{-2.8(1.4)} (|\dot\nu|/\langle\tau_{\rm
  g}\rangle)^{1.00(7)}. \label{eq:f2int}
\end{equation}
Remarkably, the slope of the fitted line is equal to the expected
value of 1.0 based on Equation~(\ref{eq:ddnu_int}). Furthermore, the
proportionality constant, $\sim 10^{-2.8} \approx 0.0016$ is
comparable to the median value of $\Delta\dot\nu_{\rm g}/\dot\nu$,
0.0034 (Table~\ref{tab:glt}), although this quantity has a large
scatter. This is consistent with the non-linear response models where
both of these quantities are equal to $I_{\rm A}/I$
\citep{accp93,ab06}.

\subsection{Slow glitches}\label{sect:J1539}

\begin{figure}
\begin{center}
\includegraphics[width=6cm,angle=-90]{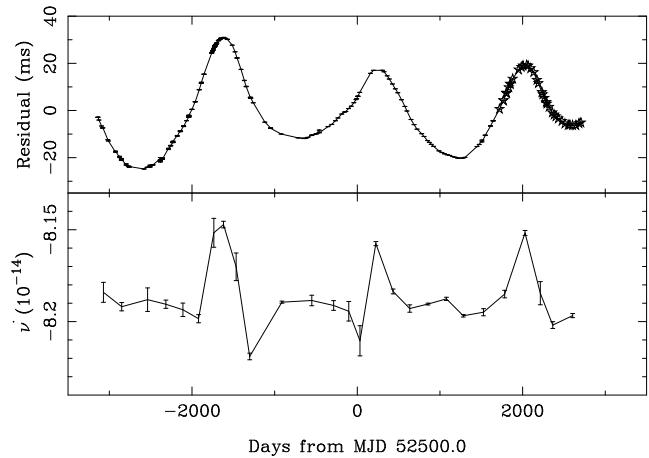}
\caption{Timing residuals for PSR~J1539$-$5626 after fitting up to
  $\ddot\nu$ are shown in the top panel. Points marked with $\star$
  are from digital filterbank observations. The bottom panel shows the
  variations $\dot\nu$.}\label{fig:J1539}
\end{center}
\end{figure}  

So-called ``slow glitches'' have been observed in a number of
pulsars. In these events, $\dot\nu$ rises sharply and declines over
the next few hundred days roughly to its pre-glitch state. This
results in a persistent increase in $\nu$ relative to the pre-glitch
variation. Slow glitches were first observed in PSR J1825$-$0935 (PSR
B1822$-$09) \citep{zww+04,sha07,ywml10} and have been reported in
several other pulsars by \citet{ywml10}. \citet{hlk10} suggested that
these slow glitches are a manifestation of the discrete states in
spin-down rate first seen in PSR B1931+24
\citep{klo+06}. \citet{lhk+10} further showed that the discrete
spin-down states were correlated with pulse shape changes, implying
that the slow-glitch glitch phenomenon has a different origin to
normal glitches for which the pulse shape changes are not
expected. Never-the-less they are discrete events which result in a
step change in spin frequency, so it is reasonable to label them as
``slow glitches''.

Our data for PSR~J1825$-$0935 (Figure~\ref{fig:glt9}) do not clearly
show the slow glitches since there were insufficient observations at
the relevant times. However, for PSR~J1539$-$5626, we see slow-glitch
features as shown in Figure~\ref{fig:J1539}. The top panel shows the
timing residuals for this pulsar showing the quasi-sinusoidal features
as observed in many pulsars \citep{hlk10}. The bottom panel shows the
episodic increases in $\dot{\nu}$ that characterise the slow-glitch
phenomenon. The two states have $\dot{\nu}\sim-8.19\times
10^{-14}$\,s$^{-2}$ and $\dot{\nu}\sim -8.15\times10^{-14}$\,s$^{-2}$,
a relative variation of just 0.5\% in $\dot{\nu}$. \citet{lhk+10}
found values of $\Delta\dot{\nu}/\dot{\nu}$ of between 0.3\% and 45\%
in different pulsars. Furthermore, they found a relationship between
the size of the change in $\dot{\nu}$ and the size of the pulse-shape
change between the two states. The small percentage change in
$\dot{\nu}$ for PSR~J1539$-$5626 therefore suggests that any
correlated pulse shape changes will also be small and so far we have
been unable to find convincing evidence for the changes in either
pulse shape or polarisation propertities.

\section{Conclusion}\label{sect:conc}

In this paper, we reported the results of a search for glitch events
in the timing residuals of 165 pulsars, covering a total data span of
1911\,yr. A total of 107 glitches was detected in 36 pulsars of which
13 were not previously known to glitch and 46 glitches are new
discoveries.  As constrained by our observational sampling, glitches
with $\Delta\nu_{\rm g}/\nu\lesssim10^{-9}$ are difficult for us to
detect, and 22 events that have previously been reported were missed.
Similar difficulties also occur for the detection of post-glitch
exponential recoveries with timescales $\lesssim20$\,d.  However, our
observations do reveal exponential recoveries mostly with timescales
of a few tens of days for 27 glitches in 18 pulsars.  A linear
increase in $\dot\nu$ is clearly observed following most
glitches. Linear increases, presumably related to a previous unseen
glitch, were also seen before the first observed glitch in most
pulsars. To quantify the linear recoveries as accurately as possible,
the solutions for $\ddot\nu$ (see Table \ref{tab:gltspin}) were
measured after any observed exponential recoveries were essentially
complete.  Of the 108 $\ddot\nu$ measurements obtained, 97 are
positive; the 11 negative ones are generally from short data spans
and/or for pulsars with strong timing noise.

With the contribution of 46 new glitches, the observed bimodal
distribution of the glitch fractional size has been reinforced,
implying that there are two different glitch mechanisms, possibly the
starquake and the vortex pinning -- unpinning theories. Post-glitch
exponential recoveries have been observed over a wide range of
fractional glitch size and pulsar age. Large recovery fractions $Q$
have been seen in small glitches in both young (e.g. the Crab) and old
(e.g. PSR~B0525$+$21) pulsars and also in large glitches in young
pulsars (e.g. PSR~J1119$-$6127) and in magnetars. Small values of $Q$,
typically around 0.01, are more commonly observed in large glitches. A
bimodal distribution shown in the histogram for $Q$ has also clearly
been seen. Moreover, decay timescales $\tau_{\rm d}$ have been
observed to show some positive correlation with pulsar age. Figure
\ref{fig:nudot_f2} shows that the inter-glitch $\ddot\nu$ has a strong
correlation with the slow-down rate $|\dot\nu|$ and is generally much
larger than the expectation from a magnetic-dipole braking. The excess
decay in braking is clearly related to glitches and is consistent with
the predictions of a model based on the properties of a weakly coupled
superfluid in the outer layers of the neutron star \citep{ab06}.

It is very clear that the true distribution of the glitch fractional
size at the low end is not well determined (cf. Figure
\ref{fig:glsize_hist}). Glitches with $\Delta\nu_{\rm
  g}/\nu\lesssim10^{-9}$ are at the lower limit of the detectability
for most timing programs. The observational sampling, timing accuracy
and intrinsic timing noise also hamper the detection of exponential
recoveries with very short timescales, leaving an incomplete sample
for short time constants. Further studies on these issues require
intensive timing observations for the glitching pulsar population
supported by simulations to better reveal the true distributions of
glitch-related phenomena.

\section*{Acknowledgements}

The Parkes radio telescope is part of the Australia Telescope, which
is funded by the Commonwealth of Australia for operation as a National
Facility managed by the Commonwealth Scientific and Industrial
Research Organisation (CSIRO). MY is funded by the National Basic
Research Program of China (2012CB821800) and China Scholarship Council
(No. 2009601129). GH is the recipient of an Australian Research
Council QEII Fellowship (\#DP0878388). VK is funded by NSERC, CIFAR,
FQRNT, CRC and is the recipient of the killam Fellowship and the Lorne
Trottier Chair. VR is a recipient of a John Stocker Postgraduate
Scholarship from the Science and Industry Endowment Fund. RX and GJ
acknowledge the NSFC (10935001, 10973002 and 10833003) and the
National Basic Research Program of China (2012CB821800 and
2009CB824800).


\appendix
\section{Timing solutions and glitch parameters}\label{sect:tim_solu}

Timing solutions incorporating pre- and post-glitch pulse parameters
and glitch parameters, obtained from \textsc{Tempo2} fits for the 36
glitching pulsars analysed in this paper are provided here in Tables
\ref{tab:gltspin} and \ref{tab:glt}.

To support the discussion in \S\ref{sect:dis}, Table \ref{tab:glrate}
summarises the number of observed glitches, $N_{\rm g}$, the observing
range and the derived glitching rates $\dot{N}_{\rm g}$ for the known
glitching pulsars. Table \ref{tab:exp} gives previously reported
exponential-recovery parameters.

ASCII machine-readable versions of the four tables in this Appendix
are provided as an on-line supplement.

\begin{table*}
\caption{Pre- and post-glitch timing solutions for 36 glitching pulsars.}\label{tab:gltspin}
\begin{center}
{\tiny
\begin{tabular}{cD{-}{-}{1.1}D{.}{.}{16}D{.}{.}{9}cccD{.}{.}{1}D{.}{.}{3}l}
\hline\\
PSR J & \multicolumn{1}{c}{Int.} & \multicolumn{1}{c}{$\nu$} & \multicolumn{1}{c}{$\dot{\nu}$} & \multicolumn{1}{c}{$\ddot{\nu}$} & Epoch & Data range & \multicolumn{1}{c}{No. of} & \multicolumn{1}{c}{Rms res.} & \multicolumn{1}{c}{$\chi_r^2$ [d.o.f]} \\
 & & \multicolumn{1}{c}{(s$^{-1}$)} & \multicolumn{1}{c}{(10$^{-12}$ s$^{-2}$)} & \multicolumn{1}{c}{(10$^{-24}$ s$^{-3}$)} & (MJD) & (MJD) & \multicolumn{1}{c}{ToAs} & \multicolumn{1}{c}{(ms)} & \\
\hline\\
J0729$-$1448 & -1 & 3.97318845714(15) & -1.7840(4) & - & 54263 & 54218 --- 54309 & 11 & 0.32 & 5.23[8] \\
 & 1-2 & 3.9731688132(5) & -1.7824(4) & - & 54391 & 54333 --- 54450 & 13 & 1.39 & 34.0[10] \\
 & 2-3 & 3.97314824736(14) & -1.7821(2) & - & 54525 & 54485 --- 54565 & 8 & 0.21 & 2.34[5] \\
 & 3-4 & 3.9731315102(3) & -1.7837(4) & - & 54634 & 54597 --- 54673 & 4 & 0.17 & 3.41[1] \\
 & 4- & 3.9730918334(6) & -1.79905(8) & - & 55059 & 54690 --- 55429 & 49 & 13.2 & 18000[46] \\\\
J0742$-$2822 & -1 & 5.9964098032(3) & -0.604567(3) & 2.59(5) & 52188 & 49364 --- 55014 & 402 & 32.5 & 16900000[398] \\
 & 1- & 5.99624713923(16) & -0.605276(8) & $-$29(3) & 55315 & 55051 --- 55579 & 79 & 0.97 & 26700[75] \\\\
J0834$-$4159 & -1 & 8.256502162607(19) & -0.2918552(9) & - & 52347 & 51299 --- 53395 & 64 & 1.00 & 23.5[61] \\
 & 1- & 8.25645335747(3) & -0.2918531(12) & - & 54283 & 53423 --- 55145 & 56 & 0.887 & 27.9[53] \\\\
J0835$-$4510 & -1 & 11.196712768(4) & -15.5997(2) & 1326(33) & 49985 & 49608 --- 50364 & 59 & 14.9 & 47200[55] \\
 & 1-2 & 11.195158486(3) & -15.58872(12) & 1008(26) & 51155 & 50819 --- 51493 & 54 & 8.25 & 24500[50] \\
 & 2-3 & 11.1932878578(11) & -15.59077(3) & 715(3) & 52568 & 51945 --- 53191 & 73 & 9.68 & 21300[69] \\
 & 3-4 & 11.1918728700(4) & -15.61044(5) & 1206(6) & 53635 & 53323 --- 53948 & 33 & 1.37 & 240[29] \\
 & 4- & 11.1904829031(5) & -15.587931(19) & 719(3) & 54687 & 54202 --- 55172 & 209 & 4.88 & 7980[205] \\\\
J0905$-$5127 & -1 & 2.88776598305(12) & -0.20879(7) & - & 49425 & 49364 --- 49488 & 6 & 0.25 & 5.64[3] \\
 & 1- & 2.88776198459(6) & -0.20845(3) & - & 49649 & 49560 --- 49739 & 9 & 0.26 & 2.36[6] \\
 & -2 & 2.88771636557(6) & -0.2071686(11) & $-$2.02(12) & 52195 & 51526 --- 52865 & 21 & 0.95 & 355[17] \\
 & 2- & 2.88768278363(6) & -0.2073986(12) & 0.38(7) & 54071 & 52998 --- 55145 & 65 & 3.70 & 2790[61] \\\\
J1016$-$5857 & -1 & 9.3126315989(4) & -6.990896(13) & 78(1) & 51913 & 51299 --- 52527 & 97 & 3.28 & 1830[93] \\
 & 1-2 & 9.3115113267(6) & -6.994765(11) & 122.4(5) & 53791 & 52571 --- 55011 & 134 & 15.3 & 27700[130] \\
 & 2- & 9.3106478766(6) & -7.00431(5) & 133(18) & 55250 & 55072 --- 55429 & 19 & 0.78 & 64.8[15] \\\\
J1048$-$5832 & -1 & 8.0873027360(12) & -6.27372(4) & 76(5) & 48418 & 47910 --- 48928 & 60 & 10.6 & 52500[56] \\
 & 1-2 & 8.08699243910(13) & -6.27060(14) & - & 48991 & 48957 --- 49025 & 9 & 0.09 & 3.85[6] \\
 & 2-3 & 8.0863750399(8) & -6.27954(3) & 98(3) & 50172 & 49559 --- 50786 & 61 & 7.22 & 148000[57] \\
 & 3-4 & 8.0854473241(13) & -6.26780(3) & 122(3) & 51894 & 51093 --- 52696 & 61 & 16.4 & 1000000[57] \\
 & 4-5 & 8.0847483076(11) & -6.27386(5) & 181(7) & 53212 & 52771 --- 53653 & 35 & 6.29 & 96600[31] \\
 & 5-6 & 8.0842772583(18) & -6.26424(9) & 223(12) & 54082 & 53680 --- 54486 & 34 & 7.37 & 393000[30] \\
 & 6- & 8.0838894263(12) & -6.28529(6) & 98(12) & 54843 & 54505 --- 55183 & 34 & 4.42 & 214000[30] \\\\
J1052$-$5954 & -1 & 5.5372526159(4) & -0.61341(11) & - & 54352 & 54220 --- 54485 & 12 & 1.25 & 8.34[9] \\
 & 1- & 5.53721546844(16) & -0.617277(7) & 35(2) & 55096 & 54733 --- 55461 & 27 & 0.98 & 3.37[23] \\\\
J1105$-$6107 & -1 & 15.8248209436(12) & -3.95900(6) & $-$108(8) & 49995 & 49589 --- 50402 & 41 & 2.54 & 2590[37] \\
 & 1-2 & 15.8245012491(16) & -3.96232(4) & 38(6) & 50942 & 50434 --- 51451 & 108 & 6.19 & 17400[104] \\
 & 2-3 & 15.823737653(2) & -3.96361(4) & 13(2) & 53217 & 51744 --- 54690 & 99 & 28.0 & 661000[95] \\
 & 3-4 & 15.8231268298(20) & -3.96587(13) & $-$11(30) & 55002 & 54733 --- 55272 & 38 & 3.38 & 13200[34] \\
 & 4- & 15.8230116650(6) & -3.96652(9) & 319(81) & 55382 & 55304 --- 55461 & 11 & 0.13 & 35.9[7] \\\\
J1112$-$6103 & -1 & 15.3936495872(5) & -7.45479(4) & $-$24(11) & 51055 & 50850 --- 51261 & 50 & 0.67 & 6.61[46] \\
 & 1-2 & 15.3927981288(7) & -7.471152(13) & 222(1) & 52417 & 51529 --- 53307 & 58 & 5.26 & 904[54] \\
 & 2- & 15.3913344780(9) & -7.46853(4) & 144(5) & 54712 & 54220 --- 55206 & 42 & 3.21 & 344[38] \\\\
J1119$-$6127 & -1 & 2.45326884390(8) & -24.211160(7) & 700(2) & 51121 & 50852 --- 51392 & 57 & 1.08 & 7.46[53] \\
 & 1-2 & 2.4507184171(3) & -24.142246(5) & 626.2(4) & 52342 & 51405 --- 53279 & 149 & 21.04 & 3800[145] \\
 & 2-3 & 2.447638860(2) & -24.0540(1) & 814(14) & 53821 & 53423 --- 54220 & 33 & 19.5 & 3230[29] \\
 & 3- & 2.444874218(3) & -23.95777(11) & 773(17) & 55154 & 54733 --- 55576 & 56 & 64.8 & 11000[52] \\\\
J1301$-$6305 & -1 & 5.4190840865(3) & -7.82942(1) & 248(2) & 51420 & 50941 --- 51901 & 62 & 3.61 & 17.1[58] \\
 & 1-2 & 5.4182027029(12) & -7.84022(4) & 286(4) & 52758 & 52145 --- 53371 & 40 & 17.3 & 434[36] \\
 & 2- & 5.417207268(1) & -7.833490(18) & 276(2) & 54249 & 53395 --- 55104 & 64 & 20.2 & 580[60] \\\\
J1341$-$6220 & -1 & 5.1729360602(5) & -6.76765(6) & 253(38) & 49651 & 49540 --- 49763 & 10 & 0.47 & 6.36[6] \\
 & 1-2 & 5.1728273829(3) & -6.76580(19) & - & 49837 & 49787 --- 49888 & 6 & 0.30 & 11.3[3] \\
 & 2-3 & 5.1727579534(9) & -6.7529(14) & - & 49956 & 49920 --- 49992 & 7 & 0.65 & 8.79[4] \\
 & 3-4 & 5.17263893435(17) & -6.77137(3) & 167(9) & 50174 & 50026 --- 50322 & 16 & 0.36 & 4.75[12] \\
 & 4-5 & 5.1724951732(7) & -6.76951(12) & 356(93) & 50420 & 50341 --- 50501 & 13 & 0.42 & 9.74[9] \\
 & 5-6 & 5.1723882419(4) & -6.76865(7) & 484(74) & 50603 & 50537 --- 50670 & 17 & 0.31 & 8.02[13] \\
 & 6-7 & 5.1721912539(7) & -6.76959(5) & 207(20) & 50946 & 50760 --- 51133 & 30 & 1.29 & 158[26] \\
 & 7-8 & 5.1719833589(3) & -6.76796(4) & 308(13) & 51303 & 51155 --- 51451 & 8 & 0.41 & 10.9[4] \\
 & 8-9 & 5.1716336852(6) & -6.76774(8) & $-$63(36) & 51911 & 51782 --- 52041 & 16 & 0.93 & 68.5[12] \\
 & 9-10 & 5.1714729568(9) & -6.7725(9) & - & 52190 & 52145 --- 52235 & 4 & 0.48 & 57.6[1] \\
 & 10-11 & 5.171283363(4) & -6.7866(3) & $-$803(63) & 52518 & 52266 --- 52771 & 17 & 11.8 & 11700[13] \\
 & 11-12 & 5.170994398(12) & -6.7534(8) & $-$2120(269) & 53013 & 52804 --- 53222 & 15 & 32.0 & 188000[11] \\
 & 12-13 & 5.1708006160(7) & -6.7484(3) & - & 53347 & 53241 --- 53455 & 9 & 2.01 & 640[6] \\
 & 13-14 & 5.170541851(19) & -6.7618(8) & $-$177(181) & 53799 & 53488 --- 54112 & 19 & 71.7 & 26.4[11] \\
 & 14-15 & 5.1702517491(3) & -6.76414(4) & 365(14) & 54297 & 54144 --- 54451 & 15 & 0.58 & 189[11]  \\
 & 15-16 & 5.1700342469(7) & -6.76214(7) & 287(22) & 54672 & 54486 --- 54860 & 15 & 1.71 & 189[11] \\
 & 16-17 & 5.169858206(3) & -6.7592(4) & 3044(251) & 54976 & 54881 --- 55072 & 10 & 1.90 & 362[6] \\
 & 17- & 5.16968689489(14) & -6.764590(16) & 109(5) & 55282 & 55104 --- 55461 & 20 & 0.45 & 13.5[16] \\\\
J1412$-$6145 & -1 & 3.17232313958(4) & -0.9928240(13) & 3.80(17) & 51353 & 50850 --- 51858 & 45 & 0.79 & 2.59[41] \\
 & 1- & 3.17212789172(5) & -0.9969084(6) & 6.06(3) & 53887 & 52314 --- 55461 & 101 & 4.73 & 110[97] \\\\
J1413$-$6141 & -1 & 3.5012521433(5) & -4.08114(8) & - & 51055 & 50850 --- 51261 & 26 & 7.20 & 36.7[23] \\
 & 1-2 & 3.5011397601(19) & -4.0842(8) & - & 51374 & 51294 --- 51454 & 9 & 5.52 & 21.8[6] \\
 & 2-3 & 3.50105387384(14) & -4.08410(5) & - & 51627 & 51472 --- 51782 & 24 & 0.90 & 3.02[21] \\
 & 3-4 & 3.5009433012(4) & -4.08332(7) & $-$350(40) & 51941 & 51844 --- 52039 & 16 & 0.79 & 3.21[12] \\
 & 4-5 & 3.5007434868(4) & -4.084389(14) & 107(3) & 52515 & 52145 --- 52886 & 27 & 3.05 & 60.9[23] \\
 & 5-6 & 3.5005683237(18) & -4.0871(3) & $-$1648(218) & 53012 & 52925 --- 53101 & 6 & 1.61 & 35.9[2] \\
 & 6-7 & 3.5003276299(11) & -4.08149(4) & 53(4) & 53708 & 53150 --- 54268 & 36 & 17.5 & 1470[32] \\
 & 7- & 3.4999222810(7) & -4.077751(19) & 33(2) & 54882 & 54303 --- 55461 & 54 & 14.0 & 630[50] \\\\
J1420$-$6048 & -1 & 14.6675125451(3) & -17.83946(3) & 651(7) & 51311 & 51100 --- 51523 & 26 & 0.37 & 15.5[22] \\
 & 1-2 & 14.666147261(2) & -17.83559(11) & 964(12) & 52207 & 51678 --- 52737 & 80 & 13.9 & 17000[76] \\
 & 2-3 & 14.6644360883(7) & -17.85639(4) & 533(6) & 53336 & 52957 --- 53716 & 58 & 2.51 & 618[54] \\
 & 3-4 & 14.6631475381(9) & -17.85577(3) & 945(5) & 54183 & 53734 --- 54633 & 38 & 2.20 & 599[34] \\
 & 4-5 & 14.6618531329(16) & -17.84922(8) & 1552(14) & 55031 & 54672 --- 55391 & 46 & 4.95 & 2.91[42] \\
 & 5- & 14.6612348736(20) & -17.90(12) & - & 55445 & 55429 --- 55461 & 5 & 0.15 & 4.40[2] \\\\
\end{tabular}}
\end{center}
\end{table*}

\addtocounter{table}{-1}

\begin{table*}
\begin{center}
\caption{--- {\it continued}}
{\tiny
\begin{tabular}{cD{-}{-}{1.1}D{.}{.}{16}D{.}{.}{9}cccD{.}{.}{1}D{.}{.}{3}l}
\hline\\
PSR J & \multicolumn{1}{c}{Int.} & \multicolumn{1}{c}{$\nu$} & \multicolumn{1}{c}{$\dot{\nu}$} & \multicolumn{1}{c}{$\ddot{\nu}$} & Epoch & Data range & \multicolumn{1}{c}{No. of} & \multicolumn{1}{c}{Rms res.} & \multicolumn{1}{c}{$\chi_r^2$ [d.o.f]} \\
 & & \multicolumn{1}{c}{(s$^{-1}$)} & \multicolumn{1}{c}{(10$^{-12}$ s$^{-2}$)} & \multicolumn{1}{c}{(10$^{-24}$ s$^{-3}$)} & (MJD) & (MJD) & \multicolumn{1}{c}{ToAs} & \multicolumn{1}{c}{(ms)} & \\
\hline\\
J1452$-$6036 & -1 & 6.451956870952(9) & -0.060343(1) & - & 54635 & 54220 --- 55051 & 35 & 0.16 & 52.2[32] \\
 & 1- & 6.451953766111(19) & -0.060447(5) & - & 55266 & 55072 --- 55461 & 17 & 0.12 & 33.4[14] \\\\
J1453$-$6413 & -1 & 5.571461007545(8) & -0.08520253(5) & 0.0659(19) & 52608 & 50669 --- 54548 & 117 & 0.29 & 643[113] \\
 & 1- & 5.571444247574(7) & -0.0852032(9) & - & 54885 & 54566 --- 55206 & 28 & 0.09 & 237[25] \\\\
J1531$-$5610 & -1 & 11.87624700867(4) & -1.938066(6) & - & 51448 & 51215 --- 51680 & 31 & 0.15 & 4.98[28] \\
 & 1- & 11.8758384521(3) & -1.946724(4) & 14.85(15) & 54060 & 52659 --- 55461 & 99 & 5.11 & 6700[95] \\\\
J1614$-$5048 & -1 & 4.31760554(1) & -9.1754(2) & 366(13) & 48848 & 47910 --- 49784 & 100 & 447 & 2740000[96] \\ 
 & 1-2 & 4.316214218(3) & -9.22347(7) & 118(5) & 50635 & 49819 --- 51451 & 171 & 95.5 & 632000[167] \\
 & & 4.314824505(11) & -9.1601(3) & 366(32) & 52385 & 51782 --- 52989 & 43 & 159 & 3800000[39] \\
 & 2- & 4.313373849(7) & -9.17841(7) & $-$48(5) & 54248 & 53036 --- 55461 & 100 & 282 & 4330000[96] \\\\
J1646$-$4346 & -1 & 4.3173283630(18) & -2.095392(11) & 32.8(3) & 50857 & 47913 --- 53803 & 258 & 294 & 1080000[254] \\
 & 1- & 4.3166537681(7) & -2.091929(20) & 109(2) & 54610 & 53949 --- 55273 & 47  & 13.1 & 718[43] \\\\
J1702$-$4310 & -1 & 4.15729055112(16) & -3.864815(4) & 40.25(14) & 52498 & 51223 --- 53774 & 76 & 7.49 & 1100[72] \\
 & 1- & 4.15649409771(9) & -3.868451(3) & 51.1(4) & 54941 & 54421 --- 55461 & 36 & 1.15 & 10.1[32] \\\\
J1709$-$4429 & -1 & 9.76127199222(13) & -8.863764(4) & 123.9(8) & 48327 & 47910 --- 48746 & 46 & 0.56 & 487[42] \\
 & 1-2 & 9.7597767546(8) & -8.853391(18) & 179.7(7) & 50305 & 49160 --- 51451 & 168 & 17.5 & 385000[164] \\
 & 2-3 & 9.758422208(3) & -8.8516(1) & 595(11) & 52091 & 51524 --- 52659 & 36 & 17.1 & 1280000[32] \\
 & 3-4 & 9.7570513620(17) & -8.85369(4) & 244(4) & 53919 & 53150 --- 54689 & 60 & 18.7 & 1120000[56] \\
 & 4- & 9.7560834373(5) & -8.86295(3) & 253(7) & 55219 & 54932 --- 55507 & 19 & 1.21 & 3890[15] \\\\
J1718$-$3825 & -1 & 13.39192401122(9) & -2.3679369(7) & 25.38(3) & 52890 & 50878 --- 54903 & 146 & 2.85 & 1500[142] \\
 & 1- & 13.39144806554(9) & -2.362796(5) & 16(2) & 55219 & 54931 --- 55507 & 18 & 0.14 & 7.09[14] \\\\
J1730$-$3350 & -1 & 7.170087711(1) & -4.35574(2) & 85(2) & 51303 & 50539 --- 52069 & 69 & 11.5 & 35900[65] \\
 & 1- & 7.1690354505(5) & -4.361361(6) & 50.5(3) & 54155 & 52805 --- 55507 & 90 & 14.2 & 36200[86] \\\\
J1731$-$4744 & -1 & 1.20511216181(5) & -0.2382630(12) & 18.89(12) & 48773 & 48184 --- 49363 & 21 & 0.41 & 8.58[17] \\
 & 1-2 & 1.20508590674(6) & -0.2376701(20) & 2.09(17) & 50059 & 49415 --- 50703 & 47 & 3.18 & 6330[43] \\
 & 2-3 & 1.20505448010(3) & -0.2375479(7) & 1.40(4) & 51590 & 50722 --- 52458 & 60 & 1.53 & 2230[56] \\
 & 3-4 & 1.20502077952(3) & -0.2375618(16) & 3.9(4) & 53239 & 52925 --- 53554 & 24 & 0.50 & 108[20] \\
 & 4- & 1.204993924876(15) & -0.2374356(6) & - & 54548 & 53589 --- 55507 & 64 & 1.96 & 23500[61] \\\\
J1737$-$3137 & -1 & 2.21990926991(9) & -0.68317(5) & - & 54286 & 54221 --- 54351 & 10 & 0.33 & 0.75[7] \\
 & 1- & 2.21987416781(9) & -0.684329(8) & - & 54930 & 54353 --- 55507 & 49 & 8.16 & 141[46] \\\\
J1740$-$3015 & -1 & 1.648184042812(13) & -1.265524(4) & - & 50798 & 50669 --- 50927 & 42 & 0.26 & 82.3[39] \\
 & 1-2 & 1.64809155129(19) & -1.266547(13) & - & 51665 & 50987 --- 52344 & 27 & 19.8 & 399000[24] \\
 & 2-3 & 1.6479812921(9) & -1.26564(5) & 125(98) & 52675 & 52361 --- 52989 & 19 & 13.0 & 179000[15] \\
 & 3-4 & 1.64786917225(7) & -1.266266(2) & 13.49(18) & 53728 & 53036 --- 54420 & 48 & 2.16 & 26600[44] \\
 & 4-5 & 1.64774895802(16) & -1.264817(7) & 11(2) & 54828 & 54450 --- 55206 & 37 & 3.65 & 134000[33] \\
 & 5- & 1.64769409872(4) & -1.266147(5) & 70(3) & 55370 & 55234 --- 55507 & 17 & 0.24 & 625[13] \\\\
J1801$-$2304 & -1 & 2.4051703923(3) & -0.65365(4) & - & 48176 & 47911 --- 48442 & 16 & 3.19 & 23.0[13] \\
 & 1-2 & 2.40512164856(9) & -0.653541(3) & 2.8(3) & 49054 & 48465 --- 49643 & 61 & 3.26 & 21.5[57] \\
 & 2-3 & 2.40507528156(11) & -0.65348(5) & - & 49878 & 49730 --- 50026 & 11 & 0.93 & 5.07[8] \\
 & 3-4 & 2.4050548970(3) & -0.65345(4) & 29(19) & 50240 & 50117 --- 50363 & 25 & 1.19 & 7.83[21] \\
 & 4-5 & 2.40502945211(8) & -0.653426(7) & 40(2) & 50694 & 50462 --- 50927 & 83 & 1.32 & 8.79[79] \\
 & 5-6 & 2.4049822461(6) & -0.65254(3) & 12(3) & 51531 & 51021 --- 52041 & 20 & 7.93 & 464[16] \\
 & 6-7 & 2.40491876991(6) & -0.653179(5) & - & 52684 & 52145 --- 53223 & 17 & 2.22 & 45.8[14] \\
 & 7-8 & 2.40482774696(6) & -0.653049(1) & 2.44(7) & 54318 & 53307 --- 55330 & 73 & 3.72 & 37.5[69] \\
 & 8- & 2.40476474326(9) & -0.65292(5) & - & 55435 & 55364 --- 55507 & 5 & 0.22 & 0.24[2] \\\\
J1801$-$2451 & -1 & 8.0071549008(15) & -8.17840(11) & 369(32) & 49171 & 48957 --- 49386 & 17 & 2.71 & 1020[13] \\
 & 1-2 & 8.006538717(1) & -8.19043(3) & 387(3) & 50064 & 49482 --- 50646 & 81 & 8.59 & 8550[77] \\
 & 2-3 & 8.005640187(4) & -8.18049(8) & 505(9) & 51348 & 50656 --- 52041 & 97 & 41.5 & 234000[93] \\
 & 3-4 & 8.004687982(3) & -8.19041(14) & 47(36) & 52737 & 52484 --- 52990 & 19 & 4.77 & 4330[15] \\
 & 4-5 & 8.0039126298(19) & -8.17244(5) & 146(4) & 53834 & 53036 --- 54633 & 58 & 26.5 & 73400[54] \\
 & 5- & 8.0029848506(6) & -8.19035(3) & 266(7) & 55182 & 54857 --- 55507 & 35 & 2.16 & 343[31] \\\\
J1803$-$2137 & -1 & 7.48329875057(11) & -7.48842(10) & - & 50709 & 50669 --- 50750 & 16 & 0.12 & 9.17[13] \\
 & 1-2 & 7.4820940765(8) & -7.4904029(17) & 244(2) & 52602 & 51782 --- 53423 & 58 & 11.8 & 101000[54] \\
 & 2- & 7.4807378694(16) & -7.50335(4) & 288(4) & 54739 & 53949 --- 55530 & 58 & 25.4 & 192000[54] \\\\
J1809$-$1917 & -1 & 12.08488375958(18) & -3.727797(2) & 37.74(12) & 52012 & 50782 --- 53242 & 57 & 1.99 & 365[53] \\
 & 1- & 12.08405907802(11) & -3.7299474(3) & 33.63(16) & 54632 & 53734 --- 55530 & 62 & 1.04 & 62.6[58] \\\\
J1825$-$0935 & -1 & 1.3003898596(5) & -0.088534(7) & $-$1.4(5) & 52978 & 51844 --- 54112 & 69 & 46.3 & 578000[65] \\
 & 1- & 1.30037753992(5) & -0.0888195(16) & 3.60(18) & 54608 & 54144 --- 55073 & 16 & 0.32 & 485[12] \\\\
J1826$-$1334 & -1 & 9.8548791982(16) & -7.277563(15) & 126(1) & 51967 & 50749 --- 53187 & 86 & 24.9 & 180000[82] \\
 & 1-2 & 9.85391251222(15) & -7.264712(11) & 69(4) & 53506 & 53279 --- 53734 & 17 & 0.21 & 23.4[13] \\
 & 2- & 9.8531185307(7) & -7.296496(18) & 115(2) & 54821 & 54112 --- 55530 & 62 & 7.17 & 7620[58] \\\\
J1835$-$1106 & -1 & 6.02730411248(17) & -0.74791(6) & - & 52068 & 51945 --- 52191 & 9 & 0.47 & 395[6] \\
 & 1- & 6.0271868732(8) & -0.74881(1) & 8.7(4) & 53882 & 52234 --- 55530 & 96 & 39.5 & 1210000[92] \\\\
J1841$-$0524 & -1 & 2.243267777344(15) & -1.175696(4) & - & 53813 & 53619 --- 54008 & 8 & 0.19 & 0.09[5] \\
 & 1-2 & 2.24322183389(19) & -1.175543(14) & 8(4) & 54266 & 54048 --- 54485 & 18 & 1.34 & 2.06[14] \\
 & 2- & 2.24314896191(9) & -1.175947(3) & 11.4(37) & 55006 & 54506 --- 55507 & 37 & 1.96 & 4.40[33] \\\\
\hline
\end{tabular}}
\end{center}
\end{table*}

\begin{landscape}
\begin{table}
\caption{Observed glitch parameters.}
\label{tab:glt}
\begin{center}
\begin{threeparttable}
{\tiny
\begin{tabular}{cclcccccccD{.}{.}{1}D{.}{.}{0}D{.}{.}{4}l}
\hline\\ \multicolumn{1}{c}{PSR J} & Gl. No. &
\multicolumn{1}{c}{Gl. Epoch} & \multicolumn{1}{c}{New?} &
\multicolumn{1}{c}{$\Delta\nu_{\rm g}/\nu$} &
\multicolumn{1}{c}{$\Delta\dot{\nu}_{\rm g}/\dot{\nu}$} &
\multicolumn{1}{c}{$\Delta\dot{\nu}_{\rm p}$} &
\multicolumn{1}{c}{$\Delta\ddot{\nu}_{\rm p}$} &
\multicolumn{1}{c}{$Q$} & \multicolumn{1}{c}{$\tau_{\rm d}$} &
\multicolumn{1}{c}{No. of} & \multicolumn{1}{c}{Data span} &
\multicolumn{1}{c}{Rms res.} & \multicolumn{1}{c}{$\chi_r^2$ [d.o.f]}
\\ & & \multicolumn{1}{c}{(MJD)} & \multicolumn{1}{c}{(N/P)} &
\multicolumn{1}{c}{($10^{-9}$)} & \multicolumn{1}{c}{($10^{-3}$)} &
\multicolumn{1}{c}{($10^{-15}$ s$^{-2}$)} &
\multicolumn{1}{c}{($10^{-24}$ s$^{-3}$)} & & \multicolumn{1}{c}{(d)}
& \multicolumn{1}{c}{ToAs} & \multicolumn{1}{c}{(MJD)} &
\multicolumn{1}{c}{(ms)} & \\ \hline\\
J0729$-$1448 & 1 &54316.8(3)$^*$ & P & 21.2(7) & - & - & - & - & - & 21 & 54220 - 54380 & 0.42 & 6.25[16] \\
 & 2 & 54479.5(9)$^*$ & P & 15.4(7) & - & - & - & - & - & 15 & 54351 - 54565 & 0.55 & 11.4[10] \\
 & 3 & 54589.8(9)$^*$ & P & 13.0(9) & - & - & - & - & - & 12 & 54485 - 54673 & 0.44 & 10.8[7] \\ 
 & 4 & 54681(9) & P & 6651.6(8) & - & - & - & - & - & 15 & 54597 - 54763 & 0.58 & 16.0[9] \\\\
J0742$-$2822 & 1 & 55020.66(9)$^*$ & P & 102.73(11) & 2.1(5) & $-$1.3(3) & - & - & - & 90 & 54772 - 55309 & 0.60 & 8820[83] \\\\
J0834$-$4159 & 1 & 53415(2)$^*$ & N & 1.85(4) & 0.26(4) & $-$0.07(1) & - & - & - & 25 & 52822 - 54144 & 0.30 & 2.43[19] \\\\
J0835$-$4510 & 1 & 50369.394$^1$ & P & 2133(10) & 7.7(4) & $-$76(3) & - & 0.030(4) & 186(12) & 159 & 50214 - 50669 & 0.17 & 11.3[150] \\
 & 2 & 51559.3190(5)$^2$ & P & 3140(46) & 8(4) & $-$57(11) & - & 0.02(1) & 125(83) & 58 & 51093 - 51900 & 4.04 & 4600[49] \\
 & 3 & 53193.09$^3$ & P & 2059(6) & 11(2) & $-$104.1(8) & 304(23) & 0.009(3) & 37(11) & 78 & 52666 - 53803 & 0.99 & 245[68] \\
 & 4 & 53959.93$^4$ & P & 2585(3) & 8.1(5) & $-$72(2) & - & 0.0119(6) & 73(8) & 103 & 53523 - 54390 & 0.73 & 124[94] \\\\
J0905$-$5127 & 1 & 49552(2)$^*$ & N & 13.6(4) & $-$1.8(9) & 0.38(18) & - & - & - & 11 & 49364 - 49644 & 0.23 & 4.89[5] \\
 & 2 & 52931(67) & N & 8.31(16)(53) & 1.2(1) & $-$0.258(20) & - & - & - & 16 & 52463 - 53367 & 0.55 & 165[10] \\\\
J1016$-$5857 & 1 & 52549(22) & N & 1622.6(3)(51) & 3.69(5) & $-$25.8(3)(4) & 69(7) & - & - & 97 & 51941 - 53244 & 1.89 & 717[89] \\
 & 2 & 55041(30) & N & 1912.4(3) & 4.4(3) & $-$31(2) & - & - & - & 13 & 54858 - 55236 & 0.24 & 7.14[6] \\\\
J1048$-$5832 & 1 & 48946.9(2)$^*$ & P & 17.95(19) & - & - & - & - & - & 17 & 48814 - 49025 & 0.17 & 17.5[11] \\
 & 2 & 49034(9) & P & 2995(26) & - & - & - & 0.026(6)(7) & 160(43) & 28 & 48957 - 49236 & 0.39 & 84.3[20] \\
 & 3 & 50791.485(5)$^5$ & P & 768(3) & 3.7(8) & $-$13.5(1) & - & 0.008(3) & 60(20) & 43 & 50703 - 50939 & 0.12 & 67[35] \\
 & 4 & 52733(37) & N & 1838.4(5)(90) & 3.7(3) & $-$23(2) & - & - & - & 33 & 52312 - 53151 & 2.72 & 25200[26] \\
 & 5 & 53673.0(8)$^*$ & N & 28.5(4) & 0.19(14) & $-$1.2(9) & - & - & - & 23 & 53349 - 53959 & 0.88 & 2002[16] \\ 
 & 6 & 54495(10) & P & 3042.56(14)(340) & 5.6(1) & $-$35.2(7) & - & - & - & 32 & 54303 - 54690 & 0.28 & 1050[25] \\\\
J1052$-$5954 & 1 & 54495(10) & P & 495(3)(7) & 86(14)(19) & $-$6.4(5) & - & 0.067(4)(13) & 46(8) & 32 & 54220 - 54787 & 1.05 & 4.26[24] \\\\
J1105$-$6107 & 1 & 50417(16) & P & 279.20(7)(36) & 1.07(20) & $-$4.2(8) & - & - & - & 38 & 50267 - 50576 & 0.086 & 5.02[31] \\
 & 2 & 51598(147) & N & 971.7(2)(5) & 0.13(9) & $-$0.5(4) & - & - & - & 32 & 50986 - 52004 & 0.36 & 49.5[25] \\
 & 3 & 54711(21) & P & 29.5(3)(15) & 3.4(6) & $-$13(3) & - & - & - & 26 & 54504 - 54903 & 0.40 & 169[19] \\
 & 4 & 55288(16) & N & 954.42(7) & - & - & - & - & - & 21 & 55145 - 55461 & 0.13 & 40.2[15] \\\\
J1112$-$6103 & 1 & 51395(134) & N & 1825(2)(25) & 4.66(11)(38) & $-$34.8(8)(29) & 242(20) & - & - & 81 & 50850 - 52397 & 1.04 & 21.3[73] \\
 & 2 & 53337(30) & N & 1202(20)(21) & 7(2) & $-$35(3) & - & 0.022(2) & 302(146) & 44 & 52661 - 54008 & 1.37 & 78.0[35] \\\\
J1119$-$6127 & 1 & 51399(3)$^*$ & P & 4.4(3) & 0.036(5) & $-$0.86(10) & - & - & - & 29 & 51258 - 51557 & 0.72 & 4.81[23] \\
 & 2 & 53293(13) & P & 372(9)(80) & 8.9(4)(23) & 2(2) & - & 0.84(3)(28) & 41(2) & 31 & 53148 - 53427 & 0.23 & 0.77[22] \\
 & 3 & 54244(24) & P & 9400(300)(5900) & 580(14)(440) & 24.9(9) & - & 0.81(4)(81) & 15.7(3) & 65 & 53651 - 54820 & 3.31 & 50.4[54] \\
 & & & & & & & & 0.214(7)(136) & 186(3) & & & \\\\
J1301$-$6305 & 1 & 51923(23) & N & 4630(2)(17) & 8.6(4)(11) & $-$42.9(3) & - & 0.0049(3)(11) & 58(6) & 59 & 51370 - 52507 & 1.39 & 2.97[50] \\
 & 2 & 53383(12) & N & 2664(2)(6) & 3.92(11) & $-$30.7(8) & - & - & - & 17 & 53185 - 53523 & 1.63 & 5.94[11] \\\\
J1341$-$6220 & 1 & 49775(12) & P & 12.2(3) & - & - & - & - & - & 14 & 49589 - 49888 & 0.30 & 5.91[8] \\
 & 2 & 49904(16) & P & 14(1) & - & - & - & - & - & 13 & 49787 - 49992 & 0.28 & 5.91[7] \\
 & 3 & 50008(16) & P & 1634(1)(5) & 2.87(17) & $-$19(2) & - & - & - & 16 & 49920 - 50190 & 0.49 & 6.17[10] \\
 & 4 & 50332(10) & P & 27.3(4) & 0.06(6) & $-$0.4(4) & - & - & - & 14 & 50213 - 50434 & 0.35 & 5.96[8] \\
 & 5 & 50532.1(8)$^*$ & P & 18.5(6) & 0.18(8) & $-$1.2(5) & - & - & - & 22 & 50399 - 50621 & 0.40 & 12.4[16] \\
 & 6 & 50683(13) & P & 709(2)(4) & - & - & - & 0.0112(19)(49) & 24(9) & 43 & 50576 - 50799 & 0.22 & 4.49[35] \\
 & 7 & 51144(11) & N & 170(1) & - & - & - & - & - & 8 & 50987 - 51241 & 0.51 & 30.1[2] \\
 & 8 & 51617(165) & N & 1121.5(7)(8) & - & - & - & - & - & 14 & 51241 - 51901 & 0.35 & 8.77[8] \\
 & 9 & 52093(53) & N & 480(4) & - & - & - & - & - & 9 & 52001 - 52235 & 1.06 & 161[4] \\
 & 10 & 52250(16) & N & 454.5(7) & - & - & - & - & - & 9 & 52145 - 52360 & 0.33 & 19.0[3] \\
 & 11 & 52788(17) & N & 219.2(4)(150) & $-$8.3(3) & 57(2) & - & - & - & 9 & 52659 - 52886 & 0.05 & 0.80[2] \\
 & 12 & 53232(10) & N & 277(3) & - & - & - & - & - & 8 & 53101 - 53307 & 0.77 & 488[2] \\ 
 & 13 & 53471(17) & N & 985(6) & - & - & - & - & - & 8 & 53371 - 53553 & 1.65 & 541.9[2] \\
 & 14 & 54128(16) & N & 194.0(4)(84) & 4.97(12) & $-$33.5(8) & - & - & - & 15 & 54008 - 54306 & 0.18 & 4.11[8] \\
 & 15 & 54468(18) & P & 317.2(4)(18) & 0.90(4) & $-$6.1(3) & - & - & - & 11 & 54335 - 54597 & 0.27 & 6.67[5] \\
 & 16 & 54871(11) & P & 309.6(6) & - & - & - & - & - & 11 & 54763 - 54975 & 0.42 & 16.5[5] \\
 & 17 & 55088(16) & N & 1579(2) & - & - & - & - & - & 8 & 54975 - 55183 & 0.57 & 67.5[2] \\\\
J1412$-$6145 & 1 & 51868(10) & N & 7253.0(7)(31) & 17.5(8)(19) & $-$5.67(3) & - & 0.00263(8)(38) & 59(4) & 33 & 51370 - 52348 & 0.58 & 1.94[25] \\\\
J1413$-$6141 & 1 & 51290(3)$^*$ & N & 39(4) & - & - & - & - & - & 19 & 51146 - 51405 & 4.26 & 15.9[14] \\
 & 2 & 51463(9) & N & 970(2) & - & - & - & - & - & 30 & 51370 - 51556 & 1.53 & 7.33[25] \\
 & 3 & 51796.3(4)$^*$ & N & 59.7(4) & $-$0.33(3) & 1.35(12) & - & - & - & 39 & 51472 - 52004 & 0.94 & 3.66[3] \\
 & 4 & 52092(53) & N & 811(2) & 0.28(20)(58) & $-$1.1(8)(24) & 491(42) & - & - & 27 & 51844 - 52426 & 0.69 & 3.15[19] \\
 & 5 & 52899.4(3)$^*$ & N & 46.9(8) & - & - & - & - & - & 11 & 52771 - 53036 & 0.64 & 4.47[6] \\
 & 6 & 53125(24) & N & 1410(5) & - & - & - & - & - & 8 & 52990 - 53279 & 2.87  & 222[2] \\
 & 7 & 54286(18) & N & 2409.8(7)(11) & 0.4(4) & $-$2(2) & - & - & - & 18 & 54112 - 54421 & 0.71 & 2.64[11] \\\\
J1420$-$6048 & 1 & 51600(77) & N & 1146.2(6)(300) & 3.83(8)(14) & $-$68(2)(24) & 281(62) & - & - & 24 & 51333 - 51901 & 0.17 & 2.98[16] \\
 & 2 & 52754(16) & N & 2019(10)(13) & 6.6(8)(9) & $-$90(6) & $-$269(152) & 0.008(4) & 99(29) & 47 & 52526 - 53105 & 0.16 & 2.83[37] \\
 & 3 & 53725(9) & N & 1270(3)(4) & 3.9(3) & $-$70(5) & 371(234) & - & - & 32 & 53488 - 54144 & 1.68 & 179[24] \\
 & 4 & 54653(20) & P & 934.5(4)(97) & 4.84(6)(8) & $-$86(1)(2) & 633(50) & - & - & 41 & 54335 - 55011 & 0.62 & 48.0[33] \\
 & 5 & 55410(19) & N & 1346.00(18) & - & - & - & - & - & 11 & 55304 - 55461 & 0.12 & 2.35[6] \\
\end{tabular}}
\end{threeparttable}
\end{center}
\end{table}
\end{landscape}

\addtocounter{table}{-1}

\begin{landscape}
\begin{table}
\caption{--- {\it continued}}
\begin{center}
\begin{threeparttable}
{\tiny
\begin{tabular}{cclcccccccD{.}{.}{1}D{.}{.}{0}D{.}{.}{4}l}
\hline\\ \multicolumn{1}{c}{PSR J} & Gl. No. &
\multicolumn{1}{c}{Gl. Epoch} & \multicolumn{1}{c}{New?} &
\multicolumn{1}{c}{$\Delta\nu_{\rm g}/\nu$} &
\multicolumn{1}{c}{$\Delta\dot{\nu}_{\rm g}/\dot{\nu}$} &
\multicolumn{1}{c}{$\Delta\dot{\nu}_{\rm p}$} &
\multicolumn{1}{c}{$\Delta\ddot{\nu}_{\rm p}$} &
\multicolumn{1}{c}{$Q$} & \multicolumn{1}{c}{$\tau_{\rm d}$} &
\multicolumn{1}{c}{No. of} & \multicolumn{1}{c}{Data span} &
\multicolumn{1}{c}{Rms res.} & \multicolumn{1}{c}{$\chi_r^2$ [d.o.f]}
\\ & & \multicolumn{1}{c}{(MJD)} & \multicolumn{1}{c}{(N/P)} &
\multicolumn{1}{c}{($10^{-9}$)} & \multicolumn{1}{c}{($10^{-3}$)} &
\multicolumn{1}{c}{($10^{-15}$ s$^{-2}$)} &
\multicolumn{1}{c}{($10^{-24}$ s$^{-3}$)} & & \multicolumn{1}{c}{(d)}
& \multicolumn{1}{c}{ToAs} & \multicolumn{1}{c}{(MJD)} &
\multicolumn{1}{c}{(ms)} & \\ \hline\\
J1452$-$6036 & 1 & 55055.22(4)$^*$ & N & 28.95(20) & 2.37(20) & $-$0.143(12) & - & - & - & 20 & 54820 - 55304 & 0.054 & 8.18[14] \\\\
J1453$-$6413 & 1 & 54552(4)$^*$ & N & 0.299(18) & 0.55(11) & $-$0.047(9) & - & - & - & 27 & 54335 - 54794 & 0.054 & 78.3[21] \\\\ 
J1531$-$5610 & 1 & 51731(51) & N & 2637(2)(11) & 25(4)(10) & $-$15.1(8) & - & 0.007(3)(4) & 76(16) & 29 & 51453 - 52042 & 0.20 & 11.3[21] \\\\
J1614$-$5048 & 1 & 49803(16) & P & 6456.7(8)(257) & 9.3(3) & $-$85(2) & - & - & - & 20 & 49589 - 49985 & 1.44 & 57.8[13] \\
 & 2 & 53013(24) & N & 6242.4(6)(370) & 9.14(4) & $-$83.8(4) & - & - & - & 12 & 52886 - 53150 & 0.40 & 27.0[6] \\\\
J1646$-$4346 & 1 & 53876(73) & N & 885(3)(5) & 1.5(3)(4) & $-$3.1(8) & - & - & - & 16 & 53524 - 54112 & 2.09 & 32.6[10] \\\\
J1702$-$4310 & 1 & 53943(169) & N & 4810(27)(104) & 17(4)(13) & $-$13.3(4)(6) & 30(6) & 0.023(6)(20) & 96(16) & 73 & 52883 - 54932 & 0.92 & 16.3[63] \\\\
J1709$-$4429 & 1 & 48779(33) & P & 2050.6(4)(99) & 5.86(8)(87) & $-$18.72(11)(22) & 70(2) & 0.01748(8)(404) & 122(3) & 106 & 47910 - 50026 & 0.50 & 336[96] \\
 & 2 & 51488(37) & N & 1166.73(17)(1680) & 6.22(3)(17) & $-$55.0(3)(15) & 470(8) & - & - & 30 & 50987 - 51901 & 0.23 & 150[22] \\
 & 3 & 52716(57) & N & 2872(7)(26) & 8.0(7)(12) & $-$44(2)(20) & $-$248(18)(20) & 0.0129(12)(42) & 155(29)(32) & 67 & 51946 - 53734 & 2.10 & 22900[57] \\
 & 4 & 54711(22) & P & 2743.9(4)(96) & 8.41(8)(78) & $-$43.27(11) & - & 0.00849(7)(187) & 85(2) & 49 & 54221 - 55104 & 0.14 & 49.2[40] \\\\
J1718$-$3825 & 1 & 54911(2)$^*$ & N & 1.94(3) & $-$0.12(4) & 0.28(9) & - & - & - & 32 & 54451 - 55391 & 0.15 & 10.0[25] \\\\
J1730$-$3350 & 1 & 52107(38) & P & 3208(6)(15) & 11(2)(3) & $-$20(1) & - & 0.0102(9)(34) & 99(23) & 30 & 51844 - 52458 & 0.60 & 127[22] \\\\
J1731$-$4744 & 1 & 49386.72(12)$^*$ & P & 136.37(20) & 1.25(11) & $-$0.30(3) & - & - & - & 40 & 49044 - 50026 & 1.03 & 125[33] \\
 & 2 & 50715.8(9)$^*$ & P & 3.90(15) & 0.46(14) & $-$0.11(4) & - & - & - & 20 & 50588 - 50819 & 0.16 & 62.1[14] \\
 & 3 & 52472.65(10)$^*$ & P & 126(2) & 2.7(5) & $-$0.041(19) & - & 0.073(7) & 210(37) & 49 & 51782 - 53243 & 0.84 & 142[41] \\
 & 4 & 53582(6)$^*$ & N & 2.69(12) & 0.2(2) & $-$0.06(6) & - & - & - & 23 & 53223 - 53949 & 0.36 & 51.4[16] \\\\
J1737$-$3137 & 1 & 54352.334(8)$^*$ & P & 1342.2(3) & 3.01(15) & $-$2.0(1) & - & - & - & 26 & 54221 - 54505 & 0.66 & 1.65[20] \\\\
J1740$-$3015 & 1 & 50936.803(4)$^5$ & P & 1440.7(3) & 0.97(3) & $-$1.23(4) & - & - & - & 34 & 50750 - 51155 & 0.46 & 241[28] \\
 & 2 & 52348.1(8)$^*$ & P & 151(2) & 5.4(7) & $-$6.8(8) & - & - & - & 23 & 52002 - 52659 & 4.78 & 26800[16] \\
 & 3 & 53023.5190(4)$^6$ & P & 1835(2) & 4.69(18) & $-$5.9(3) & - & - & - & 14 & 52805 - 53223 & 2.66 & 10200[8] \\
 & 4 & 54449(1)$^*$ & P & 41.0(7) & 0.19(8) & $-$0.24(10) & - & - & - & 27 & 54267 - 54634 & 0.95 & 9940[21] \\
 & 5 & 55220(14) & N & 2664.50(15)(120) & 1.35(3) & $-$1.71(3) & - & - & - & 19 & 55052 - 55364 & 0.23 & 546[13] \\\\
J1801$-$2304 & 1 & 48453.68(6)$^*$ & P & 348.3(3) & $-$0.14(4) & 0.09(3) & - & - & - & 53 & 47911 - 48896 & 2.07 & 8.10[47] \\
 & 2 & 49702.1(4)$^*$ & P & 63.4(4) & 0.3(1) & $-$0.20(6) & - & - & - & 19 & 49364 - 50026 & 1.01 & 3.72[13] \\
 & 3 & 50054(2)$^*$ & P & 22.5(5) & $-$0.06(11) & 0.04(7) & - & - & - & 36 & 49730 - 50363 & 1.25 & 7.43[30] \\
 & 4 & 50363.414(4)$^7$ & P & 80.2(4) & 0.61(13) & $-$0.40(9) & - & - & - & 58 & 50191 - 50647 & 0.98 & 5.17[52] \\
 & 5 & 50936(9)$^*$ & P & 5.4(7) & - & - & - & - & - & 47 & 50696 - 51211 & 1.64 & 12.7[41] \\
 & 6 & 52093(53) & P & 649.1(3)(4) & $-$0.1(1) & 0.09(7) & - & - & - & 25 & 51783 - 52823 & 0.94 & 7.90[18] \\
 & 7 & 53306.98(1)$^6$ & P & 493.29(10) & 0.19(5) & $-$0.12(3) & - & - & - & 61 & 52145 - 54380 & 2.41 & 28.8[54] \\
 & 8 & 55356(3)$^*$ & N & 3.9(3) & - & - & - & - & - & 12 & 55145 - 55507 & 0.48 & 0.64[7] \\\\
J1801$-$2451 & 1 & 49475.95(3)$^6$ & P & 1989(1) & 3.95(18) & $-$32(2) & - & - & - & 20 & 49160 - 49643 & 0.62 & 50.3[13] \\
 & 2 & 50651.44(3)$^6$ & P & 1245.27(12) & 3.37(7) & $-$27.5(6) & - & - & - & 80 & 50462 - 50799 & 0.39 & 20.9[73] \\
 & 3 & 52054.74(7)$^6$ & P & 3757(26) & 8(2) & $-$25(5) & - & 0.024(5) & 208(25) & 25 & 51879 - 52427 & 0.27 & 16.3[16] \\
 & 4 & 53032(4)$^*$ & P & 15.7(9) & $-$1.0(5) & 8(4) & - & - & - & 20 & 52805 - 53243 & 1.45 & 529[13] \\
 & 5 & 54653(19) & P & 3113(3)(15) & 15(2)(6) & $-$49(2) & - & 0.0064(9)(35) & 25(4) & 20 & 54451 - 54860 & 0.25 & 5.92[11] \\\\
J1803$-$2137 & 1 & 50765(15) & P & 3220(4)(22) & 37.36(18)(2012) & $-$51.8(5) & - & 0.0094(11)(65) & 12(2) & 54 & 50669 - 51451 & 0.16 & 12.2[43] \\
 & & & & & & & & 0.00330(17)(64) & 69(13) & & & & \\ 
 & 2 & 53473(50) & P & 3889(2)(30) & 8.96(18)(69) & $-$50.97(20) & - & 0.00630(16)(196) & 133(11) & 46 & 52804 - 54145 & 0.67 & 325[37] \\\\
J1809$-$1917 & 1 & 53261(18) & P & 1620.8(3)(19) & 5.84(17)(41) & $-$10.90(7) & 3.5(8) & 0.00602(9)(77) & 126(7) & 100 & 51783 - 54690 & 0.49 & 22.6[90] \\\\
J1825$-$0935 & 1 & 54115.5(3)$^*$ & P & 126.4(3) & - & - & - & - & - & 23 & 53949 - 55073 & 0.38 & 471[17] \\\\
J1826$-$1334 & 1 & 53236(2)$^*$ & P & 3.33(4) & 0.054(18) & $-$0.39(13) & - & - & - & 31 & 52805 - 53619 & 0.18 & 15.8[24] \\
 & 2 & 53752(18) & P & 3575(2)(9) & 11.1(6)(11) & $-$47.5(6) & - & 0.0066(3)(13) & 80(9) & 20 & 53423 - 54048 & 0.06 & 1.80[11] \\\\
J1835$-$1106 & 1 & 52222.1(7)$^*$ & P & 18.33(12) & 1.0(5) & $-$0.7(4) & - & - & - & 20 & 51945 - 52505 & 0.33 & 202[13] \\\\
J1841$-$0524 & 1 & 54011.3(5)$^*$ & P & 30.89(17) & $-$0.131(18) & 0.15(2) & - & - & - & 26 & 53619 - 54485 & 1.11 & 1.86[20] \\
 & 2 & 54495(10) & P & 1032.5(4)(6) & 1.00(7) & $-$1.18(8) & - & - & - & 27 & 54269 - 54762 & 1.13 & 1.49[21] \\\\
\hline
\end{tabular}}
\begin{tablenotes}
\item[$^*$] Glitch epoch determined by phase fit in this
  work. References are given for glitch epochs adopted from previously
  published work. Other values are determined from our data sets. \\
  References: 1 -- \citet{fla96}; 2 -- \citet{dml02}; 3 --
  \citet{dbr+04}; 4 -- \citet{fb06}; 5 -- \citet{ura02}; 6 --
  \citet{elsk11}; 7 -- \citet{klgj03}.
\end{tablenotes}
\end{threeparttable}
\end{center}
\end{table}
\end{landscape}

\begin{table*}
\caption{Number of observed glitches and their mean rate for known
  glitching pulsars.}\label{tab:glrate}
\begin{center}
\begin{threeparttable}
{\scriptsize
\begin{tabular}{clcD{.}{.}{3.1}ll}
\hline\\
PSR J & \multicolumn{1}{c}{Name} & Data span & \multicolumn{1}{c}{$N_{\rm g}$} & \multicolumn{1}{c}{$\dot{N}_{\rm g}$} & \multicolumn{1}{c}{References} \\
 & & (MJD) & & \multicolumn{1}{c}{(yr$^{-1}$)} & \\
\hline\\
J0007$+$7303 & J0007$+$7303 & 54682 --- 55222 & 1 & 0.7(7) & \citet{rkp+11} \\
J0146$+$6145 & 4U 0142+61 & 49613 --- 54239 & 2 & 0.16(12) & \citet{mks05,gdk11} \\
J0147$+$5922 & B0144$+$59 & 52486 --- 54831 & 1 & 0.1(2) & \citet{ywml10} \\
J0157$+$6212 & B0154$+$61 & 46866 --- 50496 & 1 & 0.1(1) & \citet{klgj03} \\
J0205$+$6449 & J0205$+$6449 & 52327 --- 54669 & 2 & 0.3(3) & \citet{lrc+09} \\\\
J0358$+$5413 & B0355$+$54 & 41808 --- 54946 & 6 & 0.17(7) & \citet{elsk11} \\
J0406$+$6138 & B0402$+$61 & 52469 --- 54830 & 1 & 0.2(2) & \citet{ywml10} \\ 
J0502$+$4654 & B0458$+$46 & 46238 --- 54946 & 1 & 0.04(5) & \citet{elsk11} \\
J0528$+$2200 & B0525$+$21 & 45010 --- 54947 & 3 & 0.11(7) & \citet{elsk11} \\
J0534$+$2200 & B0531$+$21 & 40491 --- 54947 & 24 & 0.61(13) & \citet{elsk11} \\\\
J0537$-$6910 & J0537$-$6910 & 51197 --- 53968 & 23 & 3.0(7) & \citet{mmw+06} \\
J0540$-$6919 & B0540$-$09 & 50150 --- 52935 & 1 & 0.13(14) & \citet{lkg05} \\
J0601$-$0527 & B0559$-$05 & 44815 --- 54948 & 1 & 0.04(4) & \citet{elsk11} \\
J0631$+$1036 & J0631$+$1036 & 49994 --- 54942 & 12 & 0.9(3) & \citet{elsk11} \\
J0633$+$1746 & J0633$+$1746 & 41725 --- 51673 & 1 & 0.04(4) & \citet{jhgm02} \\\\
J0659$+$1414 & B0656$+$14 & 43955 --- 54949 & 2 & 0.07(5) & \citet{elsk11} \\
J0729$-$1836 & B0727$-$18 & 43584 --- 54949 & 2 & 0.06(5) & \citet{elsk11} \\
J0729$-$1448 & J0729$-$1448 & 54218 --- 55429 & 4 & 1.2(6) & this work \\
J0742$-$2822 & B0740$-$28 & 44838 --- 55579 & 7 & 0.24(9) & \citet{elsk11}, this work \\
J0758$-$1528 & B0756$-$15 & 47133 --- 54939 & 1 & 0.05(5) & \citet{elsk11} \\\\
J0834$-$4159 & J0834$-$4159 & 51299 --- 55145 & 1 & 0.09(10) & this work \\
J0835$-$4510 & B0833$-$45 & 40276 --- 55172 & 16 & 0.4(1) & \citet{cdk88}, this work \\
J0905$-$5127 & J0905$-$5127 & 49363 --- 55145 & 2 & 0.13(9) & this work \\
J0922$+$0638 & B0919$+$06 & 54892 --- 55254 & 1 & 1(1) & \citet{sha10} \\
J1016$-$5857 & J1016$-$5857 & 51299 --- 55429 & 2 & 0.18(13) & this work \\\\
J1048$-$5832 & B1046$-$58 & 47910 --- 55183 & 6 & 0.30(13) & this work \\
J1048$-$5937 & 1E 1048.1$-$5937 & 52386 --- 54202 & 2 & 0.4(3) & \citet{dkg09} \\
J1052$-$5954 & J1052$-$5954 & 54220 --- 55460 & 1 & 0.3(3) & this work \\
J1105$-$6107 & J1105$-$6107 & 49589 --- 55461 & 4 & 0.25(13) & this work \\
J1112$-$6103 & J1112$-$6103 & 50850 --- 55207 & 2 & 0.17(12) & this work \\\\
J1119$-$6127 & J1119$-$6127 & 50852 --- 55576 & 3 & 0.23(14) & this work \\
J1123$-$6259 & J1123$-$6259 & 49316 --- 51155 & 1 & 0.2(2) & \citet{wmp+00} \\
J1124$-$5916 & J1124$-$5916 & 54682 --- 55415 & 1 & 0.5(5) & \citet{rkp+11} \\
J1141$-$3322 & J1141$-$3322 & 49420 --- 54940 & 1 & 0.07(7) & \citet{elsk11} \\
J1141$-$6545 & J1141$-$6545 & 53834 --- 54785 & 1 & 0.4(4) & \citet{mks+10} \\\\
J1301$-$6305 & J1301$-$6305 & 50941 --- 55104 & 2 & 0.18(13) & this work \\
J1302$-$6350 & B1259$-$63 & 47900 --- 52900 & 1 & 0.07(8) & \citet{wjm04} \\
J1328$-$4357 & B1325$-$43 & 43566 --- 44098 & 1 & 0.7(7) & \citet{nmc81} \\
J1341$-$6220 & B1338$-$62 & 47915 --- 55461 & 23 & 1.1(3) & \citet{wmp+00}, this work \\
J1357$-$6429 & J1357$-$6429 & 51458 --- 53104 & 1 & 0.2(3) & \citet{cml+04} \\\\
J1412$-$6145 & J1412$-$6145 & 50850 --- 55461 & 1 & 0.08(8) & this work \\
J1413$-$6141 & J1413$-$6141 & 50850 --- 55461 & 7 & 0.6(2) & this work \\
J1420$-$6048 & J1420$-$6048 & 51100 --- 55461 & 5 & 0.42(19) & this work \\
J1452$-$6036 & J1452$-$6036 & 54220 --- 55461 & 1 & 0.3(3) & this work \\
J1453$-$6413 & J1453$-$6413 & 50669 --- 55205 & 1 & 0.08(8) & this work \\\\
J1509$+$5531 & B1508$+$55 & 40500 --- 42000 & 1 & 0.2(3) & \citet{mt74} \\
J1531$-$5610 & J1531$-$5610 & 51215 --- 55461 & 1 & 0.09(9) & this work \\
J1532$+$2745 & B1530$+$27 & 45109 --- 54946 & 1 & 0.04(4) & \citet{elsk11} \\
J1614$-$5048 & B1610$-$50 & 47910 --- 55461 & 2 & 0.10(7) & this work \\
J1617$-$5055 & J1617$-$5055 & 47590 --- 51434 & 1 & 0.1(1) & \citet{tgv+00} \\\\
J1644$-$4559 & B1641$-$45 & 42563 --- 47888 & 3 & 0.20(12) & \citet{mngh78,fla93} \\
J1645$-$0317 & B1642$-$03 & 40000 --- 54000 & 7 & 0.18(7) & \citet{sha09} \\
J1646$-$4346 & B1643$-$43 & 47913 --- 55273 & 1 & 0.05(5) & this work \\
J1702$-$4310 & J1702$-$4310 & 51223 --- 55461 & 1 & 0.09(9) & this work \\
J1705$-$1906 & B1702$-$19 & 43587 --- 54935 & 1 & 0.03(4) & \citet{elsk11} \\\\
J1705$-$3423 & J1705$-$3423 & 49086 --- 54936 & 3 & 0.19(11) & \citet{elsk11} \\
J1708$-$4009 & 1RXS J1708$-$4009 & 50826 --- 54015 & 3 & 0.34(20) & \citet{dkg08} \\
J1709$-$4429 & B1706$-$44 & 47910 --- 55507 & 4 & 0.19(10) & this work \\
J1718$-$3718 & J1718$-$3718 & 51383 --- 55649 & 1 & 0.09(9) & \citet{mh11} \\
J1718$-$3825 & J1718$-$3825 & 50878 --- 55507 & 1 & 0.08(8) & this work \\\\
J1720$-$1633 & B1717$-$16 & 46718 --- 54945 & 1 & 0.04(5) & \citet{elsk11} \\
J1721$-$3532 & B1718$-$35 & 47907 --- 54934 & 1 & 0.05(6) & \citet{elsk11} \\
J1730$-$3350 & B1727$-$33 & 47880 --- 54946 & 2 & 0.10(8) & \citet{elsk11} \\
J1731$-$4744 & B1727$-$47 & 48184 --- 55507 & 4 & 0.2(1) & this work \\
J1739$-$2903 & B1736$-$29 & 46270 --- 54947 & 1 & 0.04(5) & \citet{elsk11} \\\\
\end{tabular}}
\end{threeparttable}
\end{center}
\end{table*}

\addtocounter{table}{-1}

\begin{table*}
\caption{--- {\it continued}}
\begin{center}
\begin{threeparttable}
{\scriptsize
\begin{tabular}{clcD{.}{.}{3.1}ll}
\hline\\
PSR J & \multicolumn{1}{c}{Name} & Data span & \multicolumn{1}{c}{$N_{\rm g}$} & \multicolumn{1}{c}{$\dot{N}_{\rm g}$} & \multicolumn{1}{c}{References} \\
 & & (MJD) & & \multicolumn{1}{c}{(yr$^{-1}$)} & \\
\hline\\
J1740$-$3015 & B1737$-$30 & 46270 --- 55507 & 32 & 1.3(3) & \citet{elsk11}, this work \\
J1737$-$3137 & J1737$-$3137 & 50759 --- 54925 & 3 & 0.26(16) & \citet{elsk11} \\
J1743$-$3150 & B1740$-$31 & 47880 --- 54926 & 1 & 0.05(6) & \citet{elsk11} \\
J1751$-$3323 & J1751$-$3323 & 52496 --- 54714 & 2 & 0.3(3) & \citet{ywml10} \\
J1801$-$2451 & B1757$-$24 & 48957 --- 55507 & 5 & 0.28(13) & this work \\\\
J1801$-$0357 & B1758$-$03 & 46719 --- 54935 & 1 & 0.04(5) & \citet{elsk11} \\
J1801$-$2304 & B1758$-$23 & 46694 --- 55507 & 10 & 0.41(14) & \citet{elsk11}, this work \\
J1803$-$2137 & B1800$-$21 & 46270 --- 55530 & 5 & 0.20(9) & \citet{elsk11}, this work \\
J1806$-$2125 & J1806$-$2125 & 50802 --- 54940 & 1 & 0.09(9) & \citet{elsk11} \\
J1809$-$1917 & J1809$-$1917 & 50821 --- 54939 & 1 & 0.09(9) & \citet{elsk11} \\\\
J1809$-$2004 & J1809$-$2004 & 51510 --- 54945 & 1 & 0.11(11) & \citet{elsk11} \\
J1812$-$1718 & B1809$-$173 & 46271 --- 54936 & 3 & 0.13(8) & \citet{elsk11} \\
J1813$-$1246 & J1813$-$1246 & 54682 --- 55226 & 1 & 0.7(7) & \citet{rkp+11} \\
J1814$-$1744 & J1814$-$1744 & 50833 --- 54945 & 5 & 0.44(20) & \citet{elsk11} \\
J1818$-$1422 & B1815$-$14 & 51512 --- 54831 & 1 & 0.11(11) & \citet{ywml10} \\\\
J1819$-$1458 & J1819$-$1458 & 51031 --- 54938 & 1 & 0.1(1) & \citet{lmk+09} \\
J1824$-$1118 & B1821$-$11 & 46612 --- 54936 & 1 & 0.04(5) & \citet{elsk11} \\
J1824$-$2452 & B1821$-$24 & 47800 --- 52800 & 1 & 0.07(8) & \citet{cb04} \\
J1825$-$0935 & B1822$-$09 & 45008 --- 54948 & 8 & 0.29(11) & \citet{elsk11} \\
J1826$-$1334 & B1823$-$13 & 46302 --- 54944 & 5 & 0.2(1) & \citet{elsk11} \\\\
J1833$-$0827 & B1830$-$08 & 46449 --- 54944 & 2 & 0.08(6) & \citet{elsk11} \\
J1830$-$1135 & J1830$-$1135 & 51816 --- 54945 & 1 & 0.12(12) & \citet{elsk11} \\
J1834$-$0731 & J1834$-$0731 & 51833 --- 54945 & 1 & 0.12(12) & \citet{elsk11} \\
J1835$-$1106 & J1835$-$1106 & 49071 --- 54940 & 1 & 0.06(7) & \citet{elsk11} \\
J1837$-$0559 & J1837$-$0559 & 51153 --- 54945 & 1 & 0.1(1) & \citet{elsk11} \\\\
J1838$-$0453 & J1838$-$0453 & 51251 --- 54948 & 2 & 0.20(14) & \citet{elsk11} \\
J1841$-$0425 & B1838$-$04 & 46270 --- 54936 & 1 & 0.04(5) & \citet{elsk11} \\
J1844$-$0538 & B1841$-$05 & 46270 --- 54936 & 1 & 0.04(5) & \citet{elsk11} \\
J1841$-$0456 & 1E 1841$-$045 & 51224 --- 53970 & 3 & 0.4(3) & \citet{dkg08} \\
J1841$-$0524 & J1841$-$0524 & 51816 --- 54939 & 3 & 0.4(2) & \citet{elsk11} \\\\
J1845$-$0316 & J1845$-$0316 & 51609 --- 54942 & 2 & 0.22(16) & \citet{elsk11} \\
J1846$-$0258 & J1846$-$0258 & 51574 --- 54800 & 2 & 0.23(16) & \citet{lkgk06,lkg10} \\
J1847$-$0130 & J1847$-$0130 & 52135 --- 54942 & 2 & 0.26(19) & \citet{elsk11} \\
J1851$-$0029 & J1851$-$0029 & 53817 --- 54948 & 1 & 0.3(4) & \citet{elsk11} \\
J1853$+$0545 & J1853$+$0545 & 52493 --- 54830 & 1 & 0.16(16) & \citet{ywml10} \\\\
J1856$+$0113 & B1853$+$01 & 47577 --- 54948 & 1 & 0.05(5) & \citet{elsk11} \\
J1901$+$0156 & B1859$+$01 & 46724 --- 54936 & 1 & 0.04(5) & \citet{elsk11} \\
J1901$+$0716 & B1859$+$07 & 46564 --- 54938 & 1 & 0.04(5) & \citet{elsk11} \\
J1902$+$0615 & B1900$+$06 & 44817 --- 54938 & 5 & 0.18(8) & \citet{elsk11} \\
J1909$+$0007 & B1907$+$00 & 44818 --- 54936 & 3 & 0.11(7) & \citet{elsk11} \\\\
J1909$+$1102 & B1907$+$10 & 52470 --- 54821 & 2 & 0.3(3) & \citet{ywml10} \\
J1910$-$0309 & B1907$-$03 & 44817 --- 54938 & 3 & 0.11(7) & \citet{elsk11} \\
J1910$+$0358 & B1907$+$03 & 47389 --- 54936 & 1 & 0.05(5) & \citet{elsk11} \\
J1913$+$0446 & J1913$+$0446 & 51832 --- 54939 & 1 & 0.12(12) & \citet{elsk11} \\
J1913$+$0832 & J1913$+$0832 & 51643 --- 54939 & 1 & 0.11(11) & \citet{elsk11} \\\\
J1913$+$1011 & J1913$+$1011 & 51465 --- 54935 & 1 & 0.11(11) & \citet{elsk11} \\
J1915$+$1009 & B1913$+$10 & 45279 --- 54948 & 1 & 0.04(4) & \citet{elsk11} \\
J1915$+$1606 & B1913$+$16 & 46671 --- 54929 & 1 & 0.04(5) & \citet{elsk11} \\
J1919$+$0021 & B1917$+$00 & 46001 --- 54948 & 1 & 0.04(4) & \citet{elsk11} \\
J1926$+$0431 & B1923$+$04 & 44819 --- 54948 & 1 & 0.04(4) & \citet{elsk11} \\\\
J1932$+$2220 & B1930$+$22 & 44816 --- 54947 & 3 & 0.11(7) & \citet{elsk11} \\
J1937$+$2544 & B1935$+$25 & 46786 --- 54937 & 1 & 0.04(5) & \citet{elsk11} \\
J1952$+$3252 & B1951$+$32 & 47029 --- 54945 & 5 & 0.23(11) & \citet{elsk11} \\
J1955$+$5059 & B1953$+$50 & 43960 --- 54938 & 2 & 0.07(5) & \citet{elsk11} \\
J1957$+$2831 & J1957$+$2831 & 50239 --- 54938 & 3 & 0.23(14) & \citet{elsk11} \\\\
J2021$+$3651 & J2021$+$3651 & 52305 --- 54948 & 2 & 0.28(20) & \citet{hrr+04,elsk11} \\
J2022$+$3842 & J2022$+$3842 & 54400 --- 55500 & 1 & 0.3(4) & \citet{agr+11} \\
J2116$+$1414 & B2113$+$14 & 44329 --- 54934 & 1 & 0.03(4) & \citet{elsk11}\\
J2225$+$6535 & B2224$+$65 & 42000 --- 54831 & 5 & 0.14(7) & \citet{sl96,ywml10} \\
J2229$+$6114 & J2229$+$6114 & 51977 --- 54946 & 3 & 0.4(3) & \citet{elsk11} \\\\
J2257$+$5909 & B2255$+$58 & 44817 --- 54935 & 1 & 0.04(4) & \citet{elsk11} \\
J2301$+$5852 & 1E 2259$+$586 & 50356 --- 52575 & 1 & 0.16(17) & \citet{kgw+03} \\
J2337$+$6151 & B2334$+$61 & 52486 --- 55045 & 1 & 0.14(15) & \citet{ymw+10} \\\\
\hline\\
\end{tabular}}
\end{threeparttable}
\end{center}
\end{table*}

\begin{table*}
\caption{Previously reported parameters for the exponential recoveries that are not covered by our observations.}\label{tab:exp}
\begin{center}{\tiny
\begin{tabular}{clrrlccccl}
\hline\\
PSR J & \multicolumn{1}{c}{Name} & \multicolumn{1}{c}{Age} & \multicolumn{1}{c}{$B_{\rm s}$} & \multicolumn{1}{c}{Gl. Epoch} & $\Delta\nu_{\rm g}/\nu$ & $\Delta\dot{\nu}_{\rm g}/\dot{\nu}$ & $Q$ & $\tau_{\rm d}$ & \multicolumn{1}{c}{References} \\
 & & \multicolumn{1}{c}{(kyr)} & \multicolumn{1}{c}{($10^{12}$\,G)} & \multicolumn{1}{c}{(MJD)} & ($10^{-9}$) & ($10^{-3}$) & & (d) & \\ 
\hline\\
J0146$+$6145 & 4U 0142$+$61 & 67.7 & 134 & 53809.185840 & 1630(350) & 5100(1100) & 1.1(3) & 17.0(1.7) & \citet{gdk11} \\\\
J0205$+$6449 & J0205$+$6449 & 5.37 & 3.61 & 52920(144) & 5400(1800) & 52(1) & 0.77(11) & 288(8) & \citet{lrc+09} \\\\
J0358$+$5413 & B0355$+$54 & 564 & 0.839 & 46497(8) & 4368(2) & 96(17) & 0.00117(4) & 160(8) & \citet{sl96} \\\\
J0528$+$2200 & B0525$+$21 & 1480 & 12.4 & 42057(14) & 1.2(2) & 2(2) & 0.6(2) & 140(80) & \citet{sl96} \\
 & & & & 52280(4) & 1.6(2) & 1.1(1) & 0.44(5) & 650(50) & \citet{ywml10} \\\\
J0534$+$2200 & B0531$+$21 & 1.24 & 3.78 & 40494 & 4.0(3) & 0.116(19) & 0.6(1) & 18.7(1.6) & \citet{lps93} \\
 & & & & 42447.5 & 43.8(7) & 2.15(19) & 0.8(1) & 18(2) & \citet{lps93} \\
 & & & & & & & 0.536(12) & 97(4) & \\
 & & & & 46664.4 & 4.1(1) & 2.5(2) & 1.00(4) & 9.3(2) & \citet{lps93} \\
 & & & & &  & & 0.89(9) & 123(40) & \\
 & & & & 47767.4 & 85.1(4) & 4.5(5) & 0.894(6) & 18(2) & \citet{lps93} \\
 & & & & & & & 0.827(5) & 265(5) &  \\
 & & & & 48947.0(2) & 4.2(2) & 0.32(3) & 0.87(18) & 2.0(4) & \citet{wbl01} \\
 & & & & 50020.6(3) & 2.1(1) & 0.20(1) & 0.8$^{+0.3}_{-0.2}$ & 3.2$^{+7.3}_{-2.2}$ & \citet{wbl01} \\
 & & & & 50259.93$^{+0.25}_{-0.01}$ & 31.9(1) & 1.73(3) & 0.680(10) & 10.3(1.5) & \citet{wbl01} \\
 & & & & 50459.15(5) & 6.1(4) & 1.1(1) & 0.87(6) & 3.0$^{+0.5}_{-0.1}$ & \citet{wbl01} \\
 & & & & 50812.9$^{+0.3}_{-1.5}$ & 6.2(2) & 0.62(4) & 0.9(3) & 2.9(1.8) & \citet{wbl01} \\
 & & & & 51452.3$^{+1.2}_{-1.6}$ & 6.8(2) & 0.7(1) & 0.8(2) & 3.4(5) & \citet{wbl01} \\\\
J0631$+$1036 & J0631$+$1036 & 43.6 & 5.55 & 52852.0(2) & 19.1(6) & 3.1(6) & 0.62(5) & 120(20) & \citet{ywml10} \\
 & & & & 54632.41(14) & 44(1) & 4(2) & 0.13(2) & 40(15) & \citet{ywml10} \\\\
J0835$-$4510 & B0833$-$45 & 11.3 & 3.38 & 40280(4) & 2338(9) & 10.1(3) & 0.001980(18) & 10(1) & \citet{cdk88} \\
 & & & & & & & 0.01782(5) & 120(6) & \\
 & & & & 41192(8) & 2047(30) & 14.8(2) & 0.00158(2) & 4(1) & \citet{cdk88} \\
 & & & & & & & 0.01311(9) & 94(5) & \\
 & & & & 41312(4) & 12(2) & 1.9(2) & 0.1612(15) & 10.0(5) & \citet{cdk88} \\
 & & & & 42683(3) & 1987(8) & 11(1) & 0.000435(5) & 4.0(4) & \citet{cdk88} \\
 & & & & & & & 0.003534(16) & 35(2) & \\
 & & & & 43693(12) & 3063(65) & 18.3(2) & 0.00242(2) & 6.0(6) & \citet{cdk88} \\
 & & & & & & & 0.01134(2) & 75(3) & \\
 & & & & 44888.4(4) & 1138(9) & 8.43(6) & 0.000813(8) & 6.0(6) & \citet{cdk88} \\
 & & & & & & & 0.00190(4) & 14(2) & \\
 & & & & 45192.1(5) & 2051(3) & 23.1(3) & 0.002483(7) & 3.0(6) & \citet{cdk88} \\
 & & & & & & & 0.00550(8) & 21.5(2.0) & \\
 & & & & 46259(2) & 1346(5) & 6.16(3) & 0.0037(5) & 6.5(5) & \citet{mkhr87} \\
 & & & & & & & 0.1541(6) & 332(10) & \\
 & & & & 47519.80360(8) & 1805.2(8) & 77(6) & 0.005385(10) & 4.62(2) & \citet{mhmk90} \\
 & & & & & & & 0.1684(4) & 351(1) & \\
 & & & & 51559.3190(5) & 3152(2) & 495(37) & 0.0088(6) & 0.53(3) & \citet{dml02} \\
 & & & & & & & 0.00547(6) & 3.29(3) & \\
 & & & & & & & 0.006691(7) & 19.07(2) & \\\\
J1123$-$6259 & J1123$-$6259 & 819 & 1.21 & 49705.87(1) & 749.12(12) & 1.0(4) & 0.0026(1) & 840(100) & \citet{wmp+00} \\\\
J1141$-$6545 & J1141$-$6545 & 1450 & 1.32 & 54277(20) & 589.0(6) & 5.0(9) & 0.0040(7) & 495(140) & \citet{mks+10} \\\\
J1302$-$6350 & B1259$-$63 & 332 & 0.334 & 50690.7(7) & 3.20(5) & 2.5(1) & 0.328(16) & 100 & \citet{wjm04} \\\\
J1341$-$6220 & B1338$-$62 & 12.1 & 7.08 & 48645(10) & 993(2) & 0.7(5) & 0.016(2) & 69(8) & \citet{sl96} \\\\
J1708$-$4009 & 1RXS J1708$-$4009 & 901 & 467 & 52014.77 & 4210(330) & 546(62) & 0.97(11) & 50(4) & \citet{kg03} \\\\
J1730$-$3350 & B1727$-$33 & 26 & 3.48 & 47990(20) & 3070(10) & 9.7(7) & 0.0077(5) & 110(8) & \citet{sl96} \\\\
J1740$-$3015 & B1737$-$30 & 20.6 & 17 & 50936.803(4) & 1445.5(3) & 2.6(8) & 0.0016(5) & 9(5) & \citet{ura02} \\
& & & & 52347.66(6) & 152(2) & 0.1(7) & 0.103(9) & 50 & \citet{zwm+08} \\
& & & & 53036(13) & 1853.6(14) & 3.0(2) & 0.0302(6) & 100 & \citet{zwm+08} \\\\
J1801$-$2451 & B1757$-$24 & 15.5 & 4.04 & 49476(6) & 1990.1(9) & 5.6(3) & 0.0050(19) & 42(14) & \citet{lkb+96} \\\\
J1801$-$2304 & B1758$-$23 & 58.4 & 6.93 & 53309(18) & 494(1) & 0.19(3) & 0.009(2) & 1000(100) & \citet{ywml10} \\\\
J1803$-$2137 & B1800$-$21 & 15.8 & 4.28 & 48245(20) & 4073(16) & 9.1(2) & 0.0137(3) & 154(3) & \citet{sl96} \\\\
J1812$-$1718 & B1809$-$173 & 1000 & 4.85 & 53105(2) & 14.8(6) & 3.6(5) & 0.27(2) & 800(100) & \citet{ywml10} \\\\
J1833$-$0827 & B1830$-$08 & 147 & 0.895 & 48041(20) & 1865.9(4) & 1.8(5) & 0.0009(2) & 200(40) & \citet{sl96} \\\\
J1841$-$0425 & B1838$-$04 & 461 & 1.1 & 53408(21) & 578.8(1) & 1.4(6) & 0.00014(20) & 80(20) & \citet{ywml10} \\\\
J1841$-$0456 & 1E 1841$-$045 & 418 & 734 & 52464.00448 & 15170(711) & 848(76) & 0.63(5) & 43(3) & \citet{dkg08} \\\\
J1846$-$0258 & J1846$-$0258 & 0.728 & 48.6 & 53883.0(3.0) & 4000(1300) & 4.1(2) & 8.7(2.5) & 127(5) & \citet{lkg10} \\\\
J1853$+$0545 & J1853$+$0545 & 3280 & 0.281 & 53450(2) & 1.46(8) & 3.5(7) & 0.22(5) & 250(30) & \citet{ywml10} \\\\
J2337$+$6151 & B2334$+$61 & 40.9 & 9.86 & 53615(6) & 20579.4(12) & 156(4) & 0.0046(7) & 21.4(5) & \citet{ymw+10} \\
& & & & & & & 0.0029(1) & 147(2) & \\\\
\hline\\
\end{tabular}}
\end{center}
\end{table*}

\end{document}